\documentclass[12pt,preprint]{aastex}
\shorttitle{A multiwavelength study of star formation in the vicinity of
 Galactic \hii region Sh 2-100}
\shortauthors{Samal et al.}
\newcommand{\msun}{\mbox{\rm $M_{\odot}$}}

\newcommand{\arcs}{\hbox{$^{\prime\prime}$}}
\newcommand{\arcm}{\mbox{$^{\prime}$}}
\newcommand{\into}{\mbox{$\times$~}}

\newcommand{\jhk}{\mbox{$JHK_{\rm s}$}~}
\newcommand{\ks}{\mbox{$K_{\rm s}$~}}
\newcommand{\ksb}{\mbox{$K_{\rm s}$-band~}}

\newcommand{\av}{\mbox{$A_{\rm V}$~}}
\newcommand{\mv}{\mbox{$M_{\rm V}$~}}

\newcommand{\hii}{\mbox{H~{\sc ii}~}}
\newcommand{\hi}{\mbox{H~{\sc i}~}}
\newcommand{\uchii}{\mbox{UCH~{\sc ii}~}}
\newcommand{\chii}{\mbox{CH~{\sc ii}~}}
\newcommand{\sii}{\mbox{[S~{\sc ii}]}~}

\newcommand{\nii}{\mbox{[N~{\sc ii}]}~}

\newcommand{\hei}{\mbox{He~{\sc i}~}}
\newcommand{\heii}{\mbox{He~{\sc ii}~}}
\newcommand{\mgii}{\mbox{Mg~{\sc ii}~}}
\newcommand{\siii}{\mbox{Si~{\sc ii}~}}

\newcommand{\hco}{\mbox{HCO$^+$}~}

\newcommand{\app}{\mbox{$\approx$}~}
\usepackage{txfonts}

\begin{document}
\title{A multiwavelength study of star formation in the vicinity of
 Galactic \hii region Sh 2-100}
\author{M.R. Samal\altaffilmark{1},
A.K. Pandey\altaffilmark{1},
D.K. Ojha\altaffilmark{2},
S.K. Ghosh\altaffilmark{2},
V.K. Kulkarni\altaffilmark{3},
N. Kusakabe\altaffilmark{4},
M. Tamura\altaffilmark{4},
B.C. Bhatt\altaffilmark{5},
M.A. Thompson\altaffilmark{6},
R. Sagar\altaffilmark{1}
}
\email{manash@aries.res.in}
\altaffiltext{1}{Aryabhatta Research Institute of Observational Sciences, Nainital, 263 129, India}
\altaffiltext{2}{Tata Institute of Fundamental Research, Mumbai (Bombay), 400 005, India}
\altaffiltext{3}{National Center for Radio Astrophysics, Post Bag 3, Ganeshkhind, Pune 411007, India}
\altaffiltext{4}{National Astronomical Observatory of Japan, Mitaka, Tokyo 181-8588, Japan}
\altaffiltext{5}{Indian Institute of Astrophysics, Koramangala, Bangalore 560 034, India}
\altaffiltext{6}{School of Physics, Astronomy and Mathematics, University of Hertfordshire, College Lane, Hatfield, Herts AL10 9AB, UK}
\begin{abstract}
We present multiwavelength investigation of morphology, physical-environment, stellar contents and star formation activity
in the vicinity of star-forming region Sh 2-100. It is found that the Sh 2-100 region contains seven \hii regions 
of  ultracompact and compact nature. The present estimation of distance for three \hii regions, along with the kinematic 
distance for others, suggests that all of them belong to the same molecular cloud complex. Using near-infrared 
photometry, we identified the most probable ionizing sources of  six \hii regions.  Their approximate 
 photometric spectral type estimates suggest that they are massive early-B to mid-O zero-age-main-sequence stars and 
agree well with radio continuum observations at 1280 MHz, for sources whose 
emissions are optically thin at this frequency. 
The morphology of the complex shows a non-uniform distribution of warm and hot dust, well mixed with the 
ionized gas, which correlates well with the variation of  average visual extinction ($\sim$ 4.2 - 97 mag) across 
the region. We estimated the physical parameters of ionized gas
with the help of radio continuum observations. 
 We detected an optically visible compact nebula located to the
south of the 850 $\mu$m emission associated with one of the \hii regions  and the diagnostic of the optical emission 
line ratios gives electron density and electron temperature of $\sim$ 0.67 $\times$ 10$^{3}$ cm$^{-3}$ 
and $\sim$  10$^{4}$ K, respectively. The physical parameters suggest that all the \hii regions 
are in different stages of evolution, which correlate well with the probable ages in the range $\sim$ 0.01 - 2 Myr 
of the ionizing sources. The spatial distribution of infrared excess 
stars, selected from near-infrared and IRAC color-color diagrams, correlates well with the association of gas and dust. 
 The positions of infrared excess stars, ultracompact and compact \hii regions at the 
periphery of an \hi shell, possibly created by a WR star, indicate that  star formation  in Sh 2-100 
region might have been induced by an expanding \hi shell.
\end{abstract}


\section{Introduction}
Massive star formation is poorly understood as compared to low-mass stars,
because their formation takes place in the dense core of a molecular
cloud of high visual extinction, usually observable at far-infrared (FIR) to millimeter 
wavelengths and their  evolutionary time scales are much
shorter ($\leq$ 10$^{5}$ yr) (e.g., Bernasconi \& Maeder 1996). 
The radio free-free emission in terms of ultracompact \hii (\uchii) regions,
represents a later evolutionary sequence of a massive protostar outlined
by Beuther et al. (2007),
where the associated high mass protostar may still be in the accretion phase  
(e.g, Shepherd \& Churchwell 1996) or has already ceased the accretion
(Garay \& Lizano 1999). The \uchii region can further evolve into the less 
obscure stage of compact \hii (\chii) and classical \hii  regions (Garay \& Lizano 1999) 
by the disruption of
associated gas and dust, revealing both, the embedded high-mass and lower
mass stellar population at shorter wavelengths.  
Most of the  high-mass star-forming regions in the Galaxy lie at 
a distance of more than 1 kpc.  
As a result, the problem in interpreting observations of massive star
formation originates in source confusion at the core of distant
molecular clouds due to their cluster mode formation.
 Statistically, it has been 
shown that the most luminous protostars form in molecular clouds 
associated with evolved \hii regions (Dobashi et al. 2001), where the interplay between the 
stars' radiation field and associated gas and dust makes the environment even 
more complex. Therefore, a detailed study of the star-forming region hosting 
young massive stars, using optical, near-infrared (NIR), mid-infrared (MIR) 
and radio bands, is necessary to understand the
high-mass star formation scenario in  these complexes. 

 In    this   paper, we   have    studied a  young star-forming region (SFR)
K3-50 shown in Fig. 1, which contains a group of \hii regions, namely A, B, C, D, E, and F
(cf. Israel 1976). The \hii region C itself consists of 
two \uchii regions, C1 and C2 (Harris 1975). 
The kinematic distance towards K3-50 ranges from 7.9 to
9.3 kpc (Harris  1975; Bronfman et al. 1996; Araya et al. 2002). In the present
study we have adopted a widely used value of distance of $\sim$ 8.7 kpc for our analysis.
 We also derived distances of three \hii regions that come out to be close to the value of
8.7 kpc.
The group of \hii regions is located 
within an area of radius  $\sim$ 3\arcm.5  centered on $\alpha_{2000} = 20^{h}01^{m}42^{s}$,
$\delta_{2000} = +33^{\circ}33^{\prime}49^{\prime\prime}$,
 in the proximity of an evolved diffuse nebula Sh 2-100 of $\sim$ 4$^{\prime}$ in size (Sharpless 1959). 
Sh 2-100 is  
itself a part of the large molecular cloud complex W58, which consists of 
several classes of \hii regions with widely varying physical parameters. 
The observed radio luminosity of the complex can be explained by the 
presence of OB association (cf. Israel 1976). 
In Fig. 1 one can notice that K3-50D, K3-50E, and K3-50F 
are associated with  nebulosity. A faint nebulosity is also seen 
$\sim$ 15.5$^{\prime\prime}$ away in the southern direction of K3-50C2, whereas
K3-50A appears as a stellar point-like source and is marked with a circle. The regions
K3-50B and K3-50C1 are optically invisible radio sources, implying that
the extinction is high enough to obscure the regions.
 By comparing the expected infrared Br$\alpha$ line fluxes as
predicted from radio continuum fluxes, to
the observed Br$\alpha$ line fluxes, Roelfsema et al. (1988) estimated visual extinction of 15, 26,
97 and 32 mag towards A (over 25$^{\prime\prime}$ aperture), B (over 5$^{\prime\prime}$ aperture), 
C1 and C2 (over 15$^{\prime\prime}$ aperture) respectively, while the extinction
towards D was found to be 2 mag (5$^{\prime\prime}$ aperture).
Observations at radio wavelengths suggest that these regions harbor at least 
one massive OB star (Harris 1975; Israel 1976; DePree et al. 1994). 
The  \uchii regions, astronomical masers, outflows, detection of 
high density tracer molecules and infrared (IR) sources with IR-excess are 
the trademarks of young star-forming regions.
The evidence of weak bipolar molecular outflow by Phillips \& Mampaso (1991) 
with CO $J$ = 2 - 1 transition, ionized outflow  by 
 DePree et al. (1994) with H76 alpha radio recombination line, 
detection of molecular core mapped in the \hco and CS (2-1) lines by 
Vogel \& Welch (1983) and Bronfman et al. (1996) suggest the 
youthfulness of the Sh 2-100 region. These observations along with the presence of 
astronomical masers (H$_2$O by Kurtz \& Hofner 2005; OH by Baudry \& Desmurs 2002) 
imply active star formation is going on  around Sh 2-100. The current star formation 
activity in the vicinity of Sh 2-100 region is probably the effect of feedback 
from the earlier generation stars of the W58 complex (Israel 1980).  

Though there are several studies in the radio and infrared  
domain, most of them are concentrated on K3-50A and K3-50C. To continue  
our  multiwavelength investigations of massive SFRs 
(cf. Ojha et al. 2004a; Ojha et al. 2004b; Ojha et al. 2004c; Tej et al.  2006; 
Samal et al. 2007; Pandey et al. 2008), we have studied an area of 8$^{\prime}$ \into 8$^{\prime}$ of
 Sh 2-100 region to understand the  star formation activity in this region. 
This paper presents new results from optical and 
infrared photometry, optical spectroscopy and low frequency radio continuum 
observations. Based on these observations, we have carried out a
 detailed study of the ionizing sources of individual \hii regions, physical 
characteristics and  the nature of associated lower mass population. With the following layout of the paper
we tried to interpret the possible star formation scenario of W58 cloud complex.  In Sect. 2, we
describe  our  observations  and  the  reduction  procedures.   In
Sect.  3, we  discuss other  available  datasets used  in the  present
study. Sect. 4 describes the general morphology of the region. 
Sect. 5 describes the 
infrared and radio components associated with the region.  In Sect. 6, we 
discuss the nature of individual regions. Sect. 7 is devoted to  
general discussion and 
star  formation scenario  in W58 complex. We present the  main conclusions 
of our results in Sect. 8.

\section{Observations and data reduction}

\subsection {Optical photometry}

$UBV$  CCD  photometric  observations  were  performed  for  the
Sh 2-100 region on 2006 October 24 and 25, using the 2K
$\times$ 2K  CCD system  at the f/13  Cassegrain focus  of the 104-cm
Sampurnanand telescope  (ST) of  ARIES, Nainital (India).   The 0.37
arcsec pixel$^{-1}$ plate scale  gives a field of view (FOV) of
$\sim$ 12$^{\prime}.6 \times 12^{\prime}.6$ on  the sky.  
The observing conditions were photometric and the average FWHM 
during the observing period was $\sim$ 1$^{\prime\prime}$.7-
2$^{\prime\prime}$.0.
To improve the signal-to-noise (S/N) ratio,  observations were
made in 2 $\times$ 2  pixel binning  mode. Several  bias frames  and
twilight  flat field exposures were taken during the observations.  We
observed the standard area SA101  (Landolt 1992) on 2006  October 25
several  times during the night to determine
the atmospheric extinction coefficients  and to 
calibrate  the  CCD systems. 
The initial processing of the
 CCD images was done using  the IRAF\footnote{IRAF is
distributed by the National Optical Astronomy Observatories, USA} data
reduction  package. Then, for  a given filter, frames of  same
exposure time were  co-added to improve the S/N.   
Photometric measurements of the stars
were performed  using DAOPHOT  II  
package  of MIDAS\footnote{MIDAS is developed and maintained by the European
Southern Observatory}.  The stellar images were  well
sampled and a variable point spread function  (PSF) was applied from several  uncontaminated
stars  present   in  the  frames.
The calibration from instrumental to standard system was done using the procedure 
outlined by Stetson (1987). The standardization residuals between the 
standard and transformed $V$ magnitude and colors are less than 0.02 mag. We used 
only those stars having photometric error in magnitudes $\leq 0.1 $ 
in the present analyses.

\subsection {Optical spectroscopy}
Using the Himalaya  Faint Object Spectrograph
Camera  (HFOSC) available with the 2.01 m  Himalayan {\it Chandra}
Telescope (HCT), operated  by the Indian Institute of Astrophysics,
Bangalore (India), we obtained optical spectra of possible ionizing sources of 
K3-50D, K3-50B and optical nebula associated with K3-50A on 2007 June 18 and 
2007 October 29.
The instrument is equipped with a  SITe 2K
$\times$ 4K pixel CCD. The central 2K $\times$ 2K region is used for
imaging which corresponds to  a  FOV  of   $\sim$   $10^{\prime}$
$\times$ $10^{\prime}$, with a pixel size of 0$^{\prime\prime}$.3.
The spectra  were obtained with a
grism (Gr 7) in the wavelength range of 
3500-7000 \AA~at a spectral dispersion of 1.45 \AA~pixel$^{-1}$.  The exposure time was 
900s for each source. The  spectroscopic observations were  carried out
under  good photometric  sky conditions.   
 The one-dimensional  spectra  were extracted  from  the bias-subtracted
and flat-field  corrected images  using  the optimal extraction
method (Horne 1986) of IRAF. Wavelength  calibration of  the spectra  was done 
using FeAr lamp source to identify emission lines of the lamp and 
we fitted a lower order polynomial to find a dispersion solution. For flux calibration of the nebular spectrum, we used
spectrophotometric standard  star BD 284211 (Massey et al. 1988) observed on the same night. 

\subsection{Near-Infrared Observations}

Imaging observations of the Sh 2-100 region centered on
$\alpha_{2000} = 20^{h}01^{m}44^{s}$,
$\delta_{2000} = +33^{\circ}33^{\prime}07^{\prime\prime}$,
 in  $J$ ($\lambda$ = 1.25  $\mu$m), $H$ ($\lambda$ = 1.63
$\mu$m), and \ks ($\lambda$ = 2.14  $\mu$m) bands were obtained on 2008
July 21 with the 1.4 m Infrared Survey Facility (IRSF) telescope and SIRIUS, a
three-color simultaneous camera equipped with three 1024 $\times$ 1024
HgCdTe arrays. The FOV in each band is $\sim$ 7$^{\prime}$.8
$\times$ 7$^{\prime}$.8, with a pixel scale of 0$^{\prime\prime}$.45 at 
the f/10 Cassegrain focus.  Further details of the
instrument are given in Nagashima et al. (1999) and Nagayama et
al. (2003).
A sky field was also observed centered on $\alpha_{2000} = 20^{h}00^{m}44^{s}$,
$\delta_{2000} = +33^{\circ}31^{\prime}36^{\prime\prime}$ and used for sky 
subtraction.
We obtained 120 dithered frames each of 10s, thus giving a total integration 
time of 1200s in each band.
The observing conditions were photometric and the average FWHM 
during the observing period was $\sim$ 
1$^{\prime\prime}$.8-2$^{\prime\prime}$.0. 
Data reduction was done using pipeline software based on
 IRAF package tasks. Dome flat-fielding and sky  subtraction
with a median sky frame were applied.  Photometry
of point sources was performed using 
the PSF algorithm ALLSTAR in the DAOPHOT package (Stetson 1987) within the IRAF environment.
The PSF was determined from 15 relatively bright and isolated stars of the field. 
We used an aperture radius of  1 FWHM,  with appropriate
aperture corrections per band for the final photometry.

For photometric calibration, we used  54 isolated 2MASS
Point Source Catalog (PSC) sources (see \S 3.1) with \ks mag range 12-14 and having error 
$<$ 0.05 mag.  The same 2MASS sources were used for absolute position
calibration and a position accuracy of better than $\pm$0$^{\prime\prime}$.1 was 
achieved.
The sources with \ks mag $<$ 11 were saturated in our image. 
For such bright sources, 2MASS PSC data were used.
For one source (i.e., a possible ionizing source of K3-50A), we have used the 
data (star id 68) from Alvarez et al. (2004), since the source was found to be
 \app 1.3 mag and 2.2 mag brighter than $K$-band and $H$-band, respectively,
than in Alvarez et al. (2004). 
Alvarez et al. (2004) observation covers an area of $\sim$ 80$^{\prime\prime}$ 
$\times$  80$^{\prime\prime}$ in the
vicinity of K3-50A with a resolution (FWHM) of 0$^{\prime\prime}$.22. 
A comparison of present photometry with that of 2MASS photometry
as a function of $J$ mag for the common sources  is shown in Fig. 2, where 
the 2MASS sources of photometric error $\leq$ 0.1 mag and `read-flag' 
values between 1 to 3 are selected to ensure good quality. 
We find that the photometric scatter in  magnitude increases at the
 fainter end, whereas the mean photometric uncertainties 
are $\le$ 0.1 in all the three bands.
To plot the data in color-color (CC) and
color-magnitude (CM) diagrams, the $JHK_{\rm s}$ data were transformed to
California Institute of Technology (CIT) system using the relation 
given at the 2MASS website\footnote{www.ipac.caltech.edu/2mass/releases/allsky/doc}. 

\subsection {Optical and NIR narrow-band imaging}

Optical CCD  narrow-band images of the nebula around Sh 2-100 region
were obtained in  H$\alpha$ ($\lambda$ =  6563 \AA,
$\Delta\lambda$ = 100 \AA)   and  [S II]
($\lambda$=  6724 \AA, $\Delta\lambda$  = 100 \AA) on 2007 June 18 
with HFOSC on the HCT. We used $R$-band as a nearby continuum for the
continuum subtraction.

Narrow-band NIR observations were  carried out in $Br\gamma$~
($\lambda$ = 2.16 $\mu$m, $\Delta\lambda$ = 0.019   $\mu$m)  and  
$K$ continuum  ($\lambda$ = 2.264  $\mu$m, $\Delta\lambda$  = 0.054
$\mu$m)  with  HCT on 2004 October 10, using the  NIR imager
\mbox{(NIRCAM)}, which  is a 0.8 -  2.5 $\mu$m camera with 512
$\times$ 512  HgCdTe array. For our observations, the NIRCAM was used
in the {\it Camera-B} mode which has a FOV of $\sim$ 3$^\prime$.6
$\times$  3$^\prime$.6. We obtained several dithered (by
$20^{\prime\prime}$) frames of the target centered on $\alpha_{2000} = 20^{h}01^{m}47^{s}$,
$\delta_{2000} = +33^{\circ}31^{\prime}38^{\prime\prime}$ in order 
to remove  bad pixels, cosmic rays and to eliminate  the   presence  
of  objects with   extended  emission  for construction of sky images.  
Typical integration times per frame were 90s, and 20s in the $Br\gamma$, 
and $K$ continuum bands, respectively. The images were co-added to 
obtain the final image in each band. We also obtained several dithered 
sky frames close to the target position in both the bands. All images 
were bias and dark-subtracted and flat-field corrected. The
continuum subtraction of the images was done after aligning and PSF matching. \\

Details of the log of observations are given in Table 1.

\subsection {Radio continuum observations}

In order to probe the ionized gas component, radio continuum
interferometric observations at 1280 MHz and 610 MHz   were carried out
on 2004 January 25 and 2004 April 07, using Giant Metrewave Radio Telescope (GMRT) array. The GMRT has a ``Y''-shaped hybrid 
configuration of 30 antennae, each 45 m in diameter. There are six 
antennae along each of the three arms (with arm length of $\sim$ 14 km). 
The rest of the twelve antennae are located in a random and 
compact $\rm 1\times1$ $\rm km^{2}$ arrangement near the center.
Details of  the GMRT antennae and their configurations can be found
in  Swarup et al.  (1991).  The largest angular  scales to which the  
GMRT is sensitive are 8$^{\prime}$  and 17$^{\prime}$ at 1280 and 610 MHz,
respectively.

The flux and phase calibrators were observed in order to derive the  
phase and amplitude gains of the antennae and  the details of the observations 
are  given in Table. 2.
 3C48 and 3C286 were used as the flux calibrators with assumed flux densities 
 of 17.23 Jy at 1280 MHz and 21.03 Jy at 610 MHz, respectively. The phase calibrators 
were 2052+365 at 1280 MHz and 1924+334 at 610 MHz observations, 
with  a bootstrapped flux density  of 9.59 Jy and  6.03 Jy, respectively.
The data analysis was done using AIPS. 
The estimated uncertainty of the flux calibration is within 
7\% at both frequencies. The image of the field was formed  by Fourier inversion and 
cleaning algorithm task IMAGR. The high resolution images ($\sim$ 3$^{\prime\prime}$.3 \into 
3$^{\prime\prime}$.0 at 1280 MHz and $\sim$ 5$^{\prime\prime}$.5 \into 4$^{\prime\prime}$.8
at 610 MHz) were made with a Briggs weighting function 
(robust factor = -1 to 0) halfway between uniform and natural weighting.
Few iterations of  self-calibration were carried out to remove
the residual effects of atmospheric and ionospheric phase corruptions
and to obtain improved maps. The system temperature correction was
done using sky temperature from the 408 MHz map of Haslam et
al. (1982). A correction factor equal to the ratio of the system
temperature toward the source and flux calibrator has been used to
scale the deconvolved images.

\section {Other Available Data Sets}

\subsection {Near-Infrared Data from $2MASS$}

For the IRSF photometric calibration, study of the bright end of the stellar population and 
astrometric calibration, we used 2MASS PSC (Cutri et al.  2003) in
the $J$ (1.25 $\mu$m), $H$ (1.65 $\mu$m) and $K{\rm_s}$ 
(2.17 $\mu$m) bands. The 2MASS data have positional accuracy  of better 
than $2^{\prime\prime}$.  Our  source selection  was based  on the  
`read-flag' values of ``1'', ``2'' or ``3'', which generally indicate 
the best quality detections, photometry and astrometry (http://www.ipac.caltech.edu/2mass/releases/allsky/doc). 

\subsection {Mid-infrared data from {\it Spitzer}}
The archived MIR data for the region were obtained with the 
Infrared Array Camera (IRAC; Fazio et al. 2004a) on board the {\it Spitzer} 
Space Telescope. 
The Basic Calibrated Data (BCD) images were downloaded from the {\it Spitzer} 
Space Telescope Archive using the Leopard package.  The IRAC observations of the region
were taken on  2005 October 22 (Program ID 20778, AOR key 15154944: A Young Stellar and Protostellar Census
of Galactic Ultracompact \hii Regions, PI: S. Carey).
Mosaics were built at 
the native instrument resolution of 1$^{\prime\prime}$.2 per pixel with the standard BCDs 
using the MOPEX (Mosaicker and Point Source Extractor) software provided by
{\it Spitzer} Science Center. 
To perform photometry with the DAOPHOT package provided within the IRAF environment,  we
first  converted the pixel values of IRAC images, which are in MJy/sr units to DN/s using 
 the conversion factor given in the ``IRAC Data Handbook'' version 3.0 (http://ssc.spitzer.caltech.edu/irac/dh/). 
We performed aperture photometry on the IRAC images with a source detection at the 5$\sigma$ level 
 above the average local background. Due to the crowded nature of the field, 
an aperture radius of 2 pixels  and a sky annulus extending from 2 to 6 pixels 
were used. The zero point magnitudes
and appropriate aperture corrections were taken from the the 
``IRAC Data Handbook'' version 3.0 for the final photometry.
The detected sources were examined visually in each band to clean the 
sample of non-stellar objects and false detections around bright stars. We 
accepted good detections as those sources which have photometric uncertainties less 
than 0.2 mag. 

\subsection {Mid-infrared data from $MSX$}

The {\it Midcourse Space Experiment}\footnote{This research made use
of data products from the Midcourse Space Experiment. Processing of
the data was funded by the Ballistic Missile Defense Organization with
additional support from NASA Office of Space Science. This research
also made use of the NASA/ IPAC Infrared Science Archive, which is
operated by the Jet Propulsion Laboratory, Caltech, under contract
with the NASA.}  (MSX) surveyed the entire Galactic plane within
$|$$b$$|$ $\le$ 5$^\circ$ in four MIR bands centered at
8.28 $\mu$m (A), 12.13 $\mu$m (C), 14.65 $\mu$m (D) and 21.34 $\mu$m
(E) at a spatial resolution of  18$^{\prime\prime}$.3 (Price et
al. 2001) and  a global absolute astrometric accuracy of  about
1$^{\prime\prime}$.9  (Egan et al. 2003).
Point sources have  been searched around Sh 2-100 region from MSX PSC 
Version 2.3 (Egan et  al. 2003).

\subsection {Mid- and far-infrared data from $IRAS-HIRES$}

The data from the {\it Infrared Astronomical Satellite} ($IRAS$)
survey in the two bands (25 and  60 $\mu$m ) for the region
around Sh 2-100 were {\it HIRES}-processed (Aumann  et al. 1990)  at the
Infrared Processing and Analysis Center (IPAC), Caltech, to obtain
high angular resolution maps.
We used $IRAS$-HIRES maps to generate the  intensity distribution map due to
thermal emission from the dust component 
 around Sh 2-100.

\section{General Morphology around Sh 2-100}

Fig. 3 represents a false $JHK_{\rm s}$ color-composite image
($J$, blue;  $H$, green; and $K_{\rm s}$, red) of Sh 2-100 region,
which reveals prominent nebulosities associated with the marked
\hii regions except K3-50C1, indicating that the extinction in the molecular
cloud towards C1 is high enough to attenuate the emission at 2.14 $\mu$m.
A dark lane is also observed in the direction of K3-50C2, bisecting its diffuse 
nebulosity into north and south components, which extend towards K3-50C1.
A  bridge between K3-50D and K3-50A is also seen in  Fig. 3,
which is not visible in POSS II image (Fig. 1), suggesting that the extinction
is high towards north-western side of K3-50D.
Fig. 3 also displays several red sources indicating the presence of young 
stellar sources still deeply embedded in the molecular cloud,
whereas bluest stars are likely to be evolved/foreground objects of the field.
 The background objects would also
appear red, but the locations of high column density regions should
minimize the  contamination of such objects.
Fig. 4 shows contours of the  610 MHz radio free-free  emission overlaid onto 
the IRSF \ksb image. In the \ksb image, the regions of recent 
star formation are traced by the bright nebulosity, whereas the \hii 
regions detected in centimeter surveys of the cloud trace high-mass 
star formation. It is well established that formation of high mass stars
in young complexes is accompanied with the simultaneous or 
subsequent formation of less massive sources,  which can be studied 
in NIR (Persi et al. 2000). 
 Our high angular resolution radio map at 1280 MHz is shown 
in Fig. 5, which was made with the robust parameter equal to -1, to see more
compact emission features.
In Fig. 5 one can notice a non-uniform 
distribution of free-free emission, indicating that Sh 2-100 is morphologically a complex region. 
 The low brightness components K3-50E and K3-50F are 
 resolved out in our high resolution map at 1280 MHz (Fig. 5).
 We therefore made a low resolution map ($\sim$ $\rm 44\arcsec \times 27\arcsec$) using {\it uv} range up to 10  $k\lambda$ with a robust parameter equal to  
0, to recover the weak diffuse emission associated to these sources as seen 
in 610 MHz map. The resulting map is discussed in \S 6.5. On the other hand, 
1280 MHz high resolution  map partially resolves  two \uchii regions of K3-50C, 
which are not resolved in our 610 MHz map.
 
\section {Infrared and Radio Properties of the Sh 2-100 star-forming region}
In  order to  examine the  nature of  the stellar  populations of Sh 2-100  
star-forming  region,  we have used NIR  CC and CM diagrams  
for all the sources detected in $JHK_{\rm s}$ bands, which are shown in  Fig. 6.

\subsection {NIR Color-Magnitude Diagram}

In $(H-K)$ versus $K$  CM diagram the nearly vertical solid
lines  represent zero-age-main-sequence (ZAMS) at a distance of 
8.7 kpc reddened by $A_V$ =  0,  10, 20, and  30  mag,  respectively.
The parallel slanting lines represent the reddening vectors to
the corresponding spectral type.  
Here, we would like to mention that uncertainty in spectral type
determination is mainly due to error in the distance determination as
well as the \ksb excess of the individual sources.
Fig. 6$a$ reveals a large number of ZAMS stars 
earlier than B2 which cannot be explained on the basis of integrated Lyman 
continuum (Lyc) photons from our low resolution radio map. Here, 
we have incorporated low resolution map, 
because the high resolution ($\sim$ 3$^{\prime\prime}$.3 \into 
3$^{\prime\prime}$.0) observations at 1280 MHz reveal only  compact sources
(see Fig. 5), which will underestimate the total radio flux from the entire ($\sim$ 
8$^{\prime}$ $\times$ 8$^{\prime}$) region.
Since Sh 2-100 is located in the Galactic plane ($l$ \app 70$^{\circ}$.2, 
$b$ \app 1$^{\circ}$.6) at a 
 distance of $\sim$ 8.7 kpc, it is natural that the region could be 
contaminated by Galactic field main-sequence (MS) and giant populations.  
Therefore in Fig. 6$a$, we overplotted the locus of giants (thick lines at \av = 0 and 3.3 mag) and 
supergiants (dashed line at \av = 0) 
using
intrinsic colors and absolute magnitudes from Cox (2000) for 
G0 to M5 and O9 to M5 stars, respectively.
From Fig. 6$a$, one can notice a  group of sources located within the reddening 
strips between \av = 0 and 3.3 mag for giants. Since the region is quite 
young (see \S 1), most of these sources must be field stars. 
However, as the region shows strong differential reddening 
(Roelfsema et al. 1988), it is worthwhile to mention that reddened early 
type MS stars may also fall within these reddening strips. The sample may also be 
contaminated by oxygen-rich  asymptotic giant branch (AGB) 
stars and carbon-rich giants, as both these classes of objects 
can display NIR colors indistinguishable from those of normal, 
reddened stars (Bessell \& Brett 1988). 
To estimate the probable  massive stellar  population,  we selected a 
relatively controlled region (marked as a rectangular box in Fig. 3) which does 
not show nebulosity in  $R$-band (see Fig. 1) as well as in \ksb (see Fig. 3).
The region is  also devoid of radio free-free emission above 5 mJy at 1415 MHz 
observation by Israel (1976). The 5 mJy flux at a distance 
of 8.7 kpc for an electron  temperature  of 10$^4$ $K$ corresponds to the number of  
Lyc photons $\sim$ 2.9 $\times$ 10$^{46}$ s$^{-1}$, 
which corresponds to a ZAMS spectral  type  close to B0 (Panagia 1973).
Therefore, if any young massive stars are present in the control region, their
ZAMS spectral types must be later than B0. 
The sources detected in the control region are shown in  Fig. 
6$b$ and their positions contradict the above fact, hence we 
believe that the stellar population in our studied region is significantly 
contaminated by field stars.  In comparison to the 
control region, a significant population in the Sh 2-100 
region shows red  $H-K$ colors,  possibly the young population of the region.   

\subsection {NIR Color-Color Diagram}
To further characterize the red sources, we use $J-H/H-K$ 
CC diagram as shown in  Fig. 6$c$. The solid thin  and thick dashed  
curves taken from Bessell \&  Brett (1988) represent the locus  of early MS and 
late giant  stars, respectively.
The parallel dashed lines are the reddening vectors for early MS and 
late giant stars (drawn from the base
and tip of the two branches). The dotted-line represents  the locus of 
Classical T Tauri stars (CTTS, Meyer et al. 1997). The reddening  vectors are plotted
using $A_J/A_V$ = 0.265, $A_H/A_V$ = 0.155, $A_K/A_V$  = 0.090 for CIT 
system (Cohen et al. 1981).
Since infrared excesses are interpreted as reflecting an 
early evolutionary stage for young stellar objects (YSOs), 
we classified the CC  diagram into three zones (F, T and P) 
to  study the nature of  sources (for details see Ojha et  al. 2004 b,c).
The sources in the ``F'' region are generally  considered to be 
either field stars,
Class III  objects or  Class II objects with small NIR  excess
(so-called weak-line T Tauri stars). ``T'' sources  
are considered  to be mostly
Class II objects with large NIR excess  (CTTS) or
reddened early type ZAMS stars, having excess emission in \ksb.
There  may  be an overlap in  NIR
colors  of the upper-end band  of Herbig Ae/Be  stars and the lower-end band
of T Tauri stars  in the ``T''  region (Hillenbrand et al. 1992).
The sources  in ``P'' region are most likely  Class I  objects (protostar-like  objects), 
having  circumstellar envelopes. A significant overlap between protostars 
and T Tauri stars is also seen in 
CC plane by Robitaille et al. (2006).  The Galactic evolved stars,
star-forming galaxies and active galactic nuclei (AGNs) display 
similar colors as those of YSOs. The thermal emissions from heated 
dust can also be identified as a point source  and may appear to the 
right of the ``F'' region like YSOs. Therefore,  NIR CC diagram alone is not 
an efficient tool to distinguish YSOs from other variety of objects, 
which requires spectroscopic information of individual objects. 
Due to non-availability of spectroscopic observations, we used a 
statistical approach to identify the probable YSOs in the region by comparing the 
distribution of sources in NIR CC diagram of the Sh 2-100 region 
with that of the control region as shown in Fig 6$d$. 
Comparison reveals that sources lying in the ``T'' region and 
redward of ``T'' region" with (J-H) $\geq$ 0.8 in Fig 6$c$ are 
likely to be the young stellar population of the Sh 2-100 region. 
We do not discard the possibility of a few sources 
with colors below CTTS locus having been 
missed out in this selection criteria. In this case, we may be underestimating
the number of YSO candidates.

\subsection {IRAC Color-Color Diagram}
The circumstellar disk emission dominates at longer wavelengths, where the
spectral energy distribution (SED) significantly  
deviates from the pure photospheric emission. Hence, the IRAC bands can 
be used to identify  IR-excess emission sources (cf. Allen et al. 2004, 
Megeath et al. 2004).  However, contamination from 
extragalactic sources such as polycyclic aromatic hydrocarbons (PAH)-rich star-forming galaxies
and AGNs, limit the use of this method as both types of sources display colors similar to that of YSOs.
Various authors have investigated the contamination of AGNs and galaxies in the IRAC CC
diagrams. We followed the approach of Gutermuth et al. (2008) and Stern et al. (2005) 
to identify galaxies and AGNs in IRAC CC diagrams. The zones identified by 
Gutermuth et al. (2008) in [3.6]-[5.8]/[4.5]-[5.8] and [4.5]-[5.8]/[5.8]-[8.0] CC diagrams 
for PAH-rich star-forming galaxies are marked by solid lines in Figs. 7$a$ and 7$b$, where 
small dots represent the sources identified in all the four IRAC bands 
within a $\sim$ 8\arcm $\times$ 8\arcm~region, covering the FOV of our NIR observations.  
The sources marked as crosses in Figs. 7$a$ and 7$b$ are star-forming galaxies. 
Stern et al. (2005) found broad-line AGNs  in a specific zone in the 
[3.6]-[4.5]/[5.8]-[8.0] CC diagram using optical spectroscopic observations, 
however this zone is also occupied by protostars (Allen et al. 2004). 
Similarly, in a nearby star-forming region, Fazio et al. (2004b) found that 50\% of the sources with 
[3.6] $>$ 14.5 mag are extragalactic in nature.  Since our star-forming region is at a farther distance, 
some of the YSOs in our sample may appear fainter than 14.5 mag in the IRAC 3.6 $\mu$m band. 
Our search of AGNs by combining both the methods as mentioned above,
results no contamination of AGNs in our sample of YSOs. 
After identifying the extragalactic contamination
we used IRAC [5.8]-[8.0]/[3.6]-[4.5] CC diagram, shown in Fig. 7$c$ to identify 
the IR-excess sources (i.e., YSOs). In the CC diagram, the regions occupied by Class II, 
Class I/II and Class 0/I objects are indicated following the classification scheme 
by Allen et al. (2004). However, as noted by Whitney et al. (2003b), the SEDs of highly inclined 
Class II sources closely resemble the SEDs, and therefore,
the colors of Class I sources (Whitney et al. 2003b; Allen et al. 2004). 
These facts  make the boundary between Class I and Class II sources somewhat
indistinct and also Class II objects hidden behind large quantity of dust can also
be shifted to Class I zone. Here, all the  sources with 
[5.8]-[8.0] $\ge$ 0.4 and [3.6]-[4.5] $\ge$ 0.1 are 
assumed to be probable YSOs after eliminating star-forming galaxies.  
On the basis of the above criteria, the [5.8]-[8.0]/[3.6]-[4.5] CC 
diagram yields 27 YSOs, which are marked as squares in Fig. 7$c$.
The search of YSOs on the basis of [5.8]-[8.0]/[3.6]-[4.5] CC diagram could 
be affected by the lower sensitivity of IRAC 5.8 and 8.0 $\mu$m bands to detect
faint sources.  As  4.5 $\mu$m  emission traces hot circumstellar dust and is less affected by
PAHs, we used a combination of $H$, $K$, and IRAC 4.5 $\mu$m bands to detect IR-excess sources.
Fig. 7$d$ shows the $H-K$/$K-[4.5]$ CC diagram. 
The solid curve in 
the diagram represents the locus of MS late dwarfs (Patten et al. 2006) 
and the dashed line denotes the reddening vector from the tip of M6 dwarf, 
using the average extinction law  by Flaherty et al. (2007). The stars with 
IR-excess are located to the right of the reddening vector and are marked 
as circles. The squares represent  the YSOs selected from the IRAC [5.8]-[8.0]/[3.6]-[4.5] 
CC diagram.  Since most of the field stars in the $K$/$H-K$ CM diagram
(see Fig. 6$b$) have colors $H-K$ $\leq$ 0.35, we therefore adopted a conservative 
approach to select IR-excess sources in Fig. 7$d$ by considering sources having 
$H-K$ $\geq$ 0.35 and which are also lie to the right of the reddening vector.
As can be seen in Fig. 7$d$ the majority of the YSOs selected on the basis of Fig. 7$c$
fall above these color cutoffs. 

Combining IRAC photometry with NIR observations, we find a 
total of 150 likely YSO candidates with circumstellar disks, that include 
Class I and Class II sources, which indicate a site of active star 
formation. The spatial distributions of these YSO candidates are used to 
study the star formation scenario in this complex, along with their correlation 
to the associated gas and dust (see \S 7.2).

\subsection {Ionized gas distribution}
 There are  several radio interferometric studies, mainly towards the
regions A, B, C and D in the frequency range of 5-15 GHz (Harris 1975;
 Kurtz et al. 1994; DePree et al. 1994) with resolutions better than 
3$^{\prime\prime}$.5, available in the literature.
Our radio continuum maps at 610 MHz \& 1280 MHz are shown in Figs. 4 and
5, with a resolution of 5$^{\prime\prime}$.5 \into 4$^{\prime\prime}$.8 and
3$^{\prime\prime}$.3 \into 3$^{\prime\prime}$.0, respectively. To our knowledge these
observations are the highest angular resolution images at low frequency to date.
The 1280 MHz map shows the radio emission from the five 
 components A, B, C1, C2 and D. The components C1 and C2 are embedded 
in an envelope of more diffuse emission, whereas the component A  shows
 extended emission in northwest and southwest directions. The extended emission
 in the southwest appears to connect to the low brightness emission seen to the north of K3-50D.
  The measured peak position, integrated flux density and
 angular diameter, deconvolved from the beam for roughly spherical compact sources A, 
 C2 and C1, were determined by fitting a two-dimensional 
elliptical Gaussian using the task JMFIT in AIPS.
For irregular, diffuse  extended sources  D and B, these observed parameters were obtained
using the task TVSTAT in AIPS. 
 The flux densities and dimensions for these extended sources were 
obtained by integrating over the approximate angular size of the contour map
above 3$\sigma$ (where $\sigma$ is the {\it rms} noise in the map). 
These observed estimations are given in Table 3.
An error of 10-15 $\%$ is expected to be associated with these values due to the
combination of the {\it rms} noise in the map ($\sigma$), 
the flux calibration  error (within 7\%) associated with source flux (S)
and the uncertainty in flux extraction of the source 
($\sigma_{\rm{source}}$). The combined error is estimated as;
$\sigma_{\rm{S}}$ = $\sqrt{{\sigma}^2 + (0.07S)^2 + \sigma^2_{\rm{source}}}$.  
Here, we expected a negligible amount of the missing flux from the compact 
sources, however in the case of extended sources, the uncertainty in the flux estimation 
could be high, hence the quoted error for such sources should be considered as a 
conservative lower limit. The integrated flux and size of the cores of 
C1 and C2 should be considered as an upper limit, since the cores are embedded 
in an envelope of diffuse emission and the present angular resolution is 
insufficient to resolve C1 and C2. 
 It is not known whether the extended emissions of the envelopes 
are either ionized by the sources in the cores and/or (partly) by the nearby external 
source. The radial velocity of the extended emissions and the compact cores
are required to know their physical association.
The images show that all the sources have complicated morphology. The 
determination of important physical parameters is hampered due to the 
complicated morphology of the sources because these parameters, such as 
density and surface brightness along the line of sight, depend 
on the source structure. In the present study, we calculate the 
physical parameters assuming the simplest geometry of uniform spherical  model.  
The physical parameters are such as: $n_e$, the rms electron density; EM, the emission measure;
 $\tau$$_c$, the optical depth;
$N_{Lyc}$, the number of Lyman continuum photons for a
spherically symmetric \hii region; and  the spectral type of
the ionizing source, assuming a single ZAMS star from the $N_{Lyc}$. 
These parameters were derived from the measured
angular diameter and integrated flux density using the formulae given
by Panagia \& Walmsley (1978) and Mezger \& Henderson (1967) for spherical, homogeneous
nebulae. We have assumed an electron temperature of 10$^{4}$ K and source size as the 
geometrical average of the major and minor axes, as listed in Table 3, to derive the physical 
parameters. 
The derived physical parameters are listed in Table 4. However, the uncertainties are
associated with the values given in Table 4 because of our  limited knowledge of the 
source structures, dust  absorption of Lyc photons (Garay et al. 1993) at
low frequency for \chii and \uchii regions (e.g., K3-50A \& K3-50C) and missing flux from the
low density components (e.g., K3-50D \& K3-50B).
As the gas seems to be clumpy in nature, the rms electron density of the \hii regions 
is expected to be lower than the density at the peak positions. 
The sources K3-50C1 and K3-50A are optically thick at 1280 MHz, therefore
the spectral type, electron density  derived by this method should be 
considered as the lower limit. 
The  spectral types of the ionizing sources for the components K3-50B and K3-50D
may be earlier than the present estimation possibly due to missing flux 
from the low density components. 
 The spectral types suggest that all the sources are ionized by massive 
late O-type ZAMS stars and their physical parameters point to a different 
evolutionary class of \hii regions (Kurtz et al. 2000).

\subsection{Candidates for the ionizing sources of the \hii regions}
The OB type stars associated with K3-50 region are significantly contaminated by 
the luminous field stars (see \S 5.1). Therefore, in order 
to identify the probable ionizing candidates of the individual \hii regions, we 
used our $JHK_{\rm s}$ catalogue. We  searched for ionizing sources 
in a circular area of radius $\sim$ 0.5 pc from their radio-emission peak, except for 
the source K3-50D which shows core-halo morphology (see \S 6.1). We identified the ionizing 
source of K3-50D using optical spectroscopy and discussed in section 6.1.
The  probable sources are 
shown in $(J-H)$ versus $(H-K)$ CC diagram (Fig. 8$a$) as well as in $(J-H)$ versus $J$ CM 
diagram (Fig. 8$b$). Fig. 9 shows \ksb image with the sources located
within the radius $\sim$ 0.5 pc of each \hii region, along with the ionizing source 
of K3-50D. The sources are labeled with numbers 
corresponding to the \hii regions.  In Fig. 8$a$, sources (A1, A2, C2 and F2) show NIR excess. 
Therefore, the spectral type estimation on the basis of $(J-H)$ versus $J$ CM diagram 
should be considered as an upper limit for these sources.
In order to distinguish the Galactic field stars 
from the OB stars, we further constrain our analysis by using a 
reddening-free quantity Q = $(J-H)$ - 1.70 $(H-\ks)$, which is $\sim$ 0 for early 
type stars with no NIR excess and $\sim$ 0.4-0.5 for Galactic field stars and 
negative for NIR excess sources (Comer\'{o}n \& Pasquali 2005). As the 
sources (A1, A2, C2 and F2) show considerable NIR excess and may be the candidates for
ionizing sources, we selected the sources with Q value $<$ 0.3 as possible ionizing sources. 
 It is worthwhile to point out that the sample identified using 
the criterian Q $<$ 0.3 may also be contaminated by objects such as;  
AGB stars, carbon stars, foreground F-type stars and background A-type giants
(Comer\'{o}n \& Pasquali 2005).
Since we are looking for early type stars within the radio emitting 
region in a circular area of radius $\sim$ 0.5 pc around the radio emission peak,
we presume the probability of contamination is less, however  this can not be ruled out. 
 Moreover, in $(J-H)$ versus $J$ CM diagram, we
selected luminous sources 
that have spectral type earlier than B3, which further reduces the possibility of contamination.
The estimated photometric spectral types for the probable ionizing sources 
(details are discussed in their individual sections)
 with Q $<$ 0.3 along with the 
spectral type estimated from radio observations at frequencies in which 
these \hii regions are optically thin, are  mentioned in Table 5. Here, it is noted that
the ZAMS spectral types given in Table 5 are taken from Panagia (1973), corresponding to the
 $N_{Lyc}$ photons s$^{-1}$ reported by DePree et al. (1994) from 14.7 GHz observations.
It is evident from Table 5 that all the \hii regions are excited by  early-B to late-O 
type of stars, which confirms the estimation by  radio observations, except the source A 
(see \S 6.2 for further discussion on source A).

\section{ Nature of Individual Sources}

\subsection{K3-50D}
The morphological features of the nebula 
at different optical and NIR bands are shown in Fig. 10.  The
ionized gas is  seen in H$\alpha$ with an extended ring,
whereas \sii  image shows a limb brightened shell towards the western side.
The Br$\gamma$ image traces a semi-circular shell of ionized  gas  with the shell  
opening towards the north-western side. The same morphology is also seen in IRSF \ksb 
image. 
In the IRSF \ksb image, two bright sources are visible within the ring of
ionized gas. The star situated at the center of nebula is the main ionizing
source of ZAMS spectral type $\sim$ O5 (\S 5.5), whereas the colors ($J-H$ $\app$ 0.32, 
$H-K$ $\app$ 0.20) of the second source situated at 12$^{\prime\prime}$ towards the south-western direction
from the central star suggest a reddened  giant, which is supported by its $Q$ value of
 $\sim$ 0.38. A plausible model for this type of morphology could be the displacement and
ionization of the gas due to the 
stellar wind produced by the massive star at the center of 
the nebula. 
The one-dimensional spectrum of the star in  the  range  3950-4750 \AA~is shown
in   Fig. 11$a$.
We classified the spectrum by comparing the observed  spectrum with those given by Walborn \&
Fitzpatrick (1990). Besides hydrogen Balmer lines (H$_\gamma$,  H$_\delta$,
H$_\epsilon$, etc), the spectrum displays prominent \heii line at
$\lambda$ 4200 \AA, and strong \heii line at $\lambda$ 4686 \AA.
These features indicate that the star should be an O-type of luminosity class V.
In the case of O-type stars the classification of different subclasses are based on
the strengths of the \heii lines  at  $\lambda$ 4541 \AA~ and $\lambda$ 4471 \AA.
The equal strength of the lines
implies a spectral type of O7, whereas the greater the
strength of $\lambda$ 4541 \AA~ \heii
line as compared to $\lambda$ 4471 \AA~ line implies a spectral type
earlier than O7.
If  the strength of  \hei + \heii $\lambda$ 4026 \AA~
$\le$ $\lambda$ 4200 \AA~ \heii, it favors classification earlier than
$\sim$ O5.5.  Comparing the strengths of these lines we
adopt the classification as $\sim$  O4V,
which is in agreement with the estimation by Georgelin  (1975).
 Roelfsema et al. (1988) estimated a visual extinction of 2 mag 
towards K3-50D from the ratio of expected
Br$\alpha$ emission predicted from radio emissions to the observed Br$\alpha$ emission, whereas 
Persson \& Frogel (1974) estimated a visual extinction of $\sim$ 4.4 mag using the spectrophotometric 
measurements of hydrogen lines of the nebula associated with the K3-50D.
To estimate the  extinction  towards  the region, we use the $(U-B)/(B-V)$  CC  diagram.
 Morphologically K3-50D shows a bright star at the center and a partial ring of ionized gas of diameter
$\sim$ 40$^{\prime\prime}$, therefore we restrict our optical analysis within a circular area 
of diameter $\sim$ 50$^{\prime\prime}$ to avoid field star contamination. 
The $(U-B)$ versus $(B-V)$ CC diagram of the region is shown in
 Fig. 11$b$.
 To match the observations, the ZAMS  by Schmidt-Kaler (1982) is shifted
along the reddening   vector  having  a  normal  slope  of
$\frac{E(U-B)}{E(B-V)} = 0.72$.  The  CC   diagram  indicates a
minimum reddening of  $E(B-V)_{min}$ = 1.36 mag  for the  stars 
within $\sim$ 50$^{\prime\prime}$  region. Using the relation \av = 3.1 $\times$ $E(B-V)$,
the estimated value of \av comes out to be $\app$ 4.2 mag. We also estimated the average visual
extinction for the region, using stars with spectral type earlier than A0 by  `Q' method 
(Johnson \& Morgan 1953). The intrinsic colors of these stars
 were obtained from the relation $(B-V)_0$ = 0.332 $\times$ Q
(Hillenbrand et al. 1993). To check the bias of the contamination
of other field sources to this value, we estimated the visual extinction from the ionizing source of K3-50D.
The O-type MS star shows a narrow range in their intrinsic colors. Adopting an intrinsic color $(B-V)_0$ = -0.33 
(Cox 2000) for the ionizing star of K3-50D and using the measured photometric color $(B-V)$ = 1.04, we derived an 
excess $E(B-V)$ = 1.37, which in turn gives \av \app 4.2 mag, which agrees well with the estimation based on the CC diagram.
Since the distance estimates to the region are uncertain and vary from 4.5 to 12.5 kpc 
(Israel 1976), the present spectral classification of the ionizing star and reliable
estimation of extinction together with the present apparent magnitude of the ionizing source enables us 
to derive its distance.
Using the present spectral classification (O4V) along with \av \app 4.2 mag, V = 13.31 
and \mv-spectral type calibration table of Vacca et al. (1996), we estimated
a distance of $\sim$  8.5 kpc for the K3-50D region.
However, the spectroscopic distance determination 
is strongly dependent on the assumed value of \mv and  the values of \mv in case of 
O4V star vary from -5.4 (Schmidt-Kaler 1982) to -5.9 (Conti 1988) in the 
literature, which in turn can vary distance from 7.9 kpc to 10 kpc. 
However, recent investigation by Martins et al. (2005) found
considerable shift in calibration of effective temperature in
early type stars. The \mv value of O4V star (\mv = -5.5) by Martins et al. (2005) 
results in a distance of 8.6 kpc to the region. 
These estimations are consistent with the average kinematic distance ($\sim$ 8.7 kpc) 
towards K3-50 and supports its association with the W58 molecular cloud.

\subsection {K3-50A}
K3-50A is a  \chii region, ionized by a single O5.5V star and its angular
size is $\sim$ 4$^{\prime\prime}$.1  \into  3$^{\prime\prime}$.4 at 14.7 GHz 
(Kurtz et al. 1994).  The radio continuum peak is coincident with  10 $\mu$m peak, 
whereas the associated optical emission (see \S 1) is located to the south-west 
($\sim$ 2.5$^{\prime\prime}$) of the radio 
peak (Wynn-Williams et al. 1977) and the region is possibly ionized 
by multiple sources (Okamoto et al. 2003).
We detected two NIR sources in all the three bands 
within $\sim$ 0.5 pc radius of the radio peak. 
The source A1 shows a large IR-excess ($H-K$ $\app$ 3.35) 
and is also more luminous ($K$ $\app$ 9.11) than any reddened O-type stars of 
the complex. The second bright source A2 is located outside  the 
radio size ($\sim$ 4$^{\prime\prime}$) of K3-50A, hence the probability of this source 
as one of the  ionizing candidates is rather low. Therefore, A1 
is the most probable ionizing source of K3-50A.
We speculate that all the point sources close to the nebula  (e.g., A2) may be strongly 
contaminated by the intense \ksb nebulosity of K3-50A.
The crowding in the region has already been noticed by Okamoto et al. (2003), 
therefore, the spectral classification on the basis of NIR photometry may 
be affected by source contamination along with bright nebulosity and 
will yield an earlier spectral type.

\subsubsection {Spectrum of the Nebula}
Fig. 10$a$  shows the continuum-subtracted H$\alpha$ + \nii~image of the
optical nebula associated with K3-50A.
The spectrum of the nebula is presented in Fig. 12. The nebula shows
forbidden emission lines as seen in the case of \hii regions. We 
obtained the emission line fluxes by Gaussian fitting
of the line profiles and by direct integration of the fluxes  over the
local continuum with SPLOT routine of the IRAF package. 
Uncertainties in these line intensities depend upon the strength of 
the lines and blending with adjacent lines. The measurement errors 
for the isolated bright lines (fluxes larger than 
$\sim$ 10$^{-14}$ ergs cm$^{-2}$ s$^{-1}$) are expected
 to be less than 10\%, whereas for the weaker and
blended lines like H$\alpha$ + \nii~and [S~{\sc ii}], the
measurement uncertainties can be as large as 20\%.  Our measurements
are consistent with that of Persson \& Frogel (1974) for
the common lines.

The physical parameters of the nebula can be derived from the diagnostic
of line ratios with certain assumptions.
Since our spectrum contains emission lines of hydrogen, their relative intensities can
be used to estimate the reddening for the region. 
We obtained the  reddening coefficient  C($H_\beta$) \app 3.16,
by comparing the observed ratio of
H$\alpha$/H$_\beta$ with the theoretical ones calculated by
Osterbrock (1989), for the temperature 10$^{4}$ K and density 10$^{4}$ cm$^{-3}$, using 
Case-B recombination theory and extinction law of Cardelli,
Clayton \& Mathis (1989, hereafter CCM 89).
Here, we would like to mention that the presence of extinction increases 
with the observed ratio, 
because extinction affects the H$_\beta$ wavelength more than the H$\alpha$ wavelength 
and also the ratio
depends on the  geometry of the dust and gas cloud. 
However,  Persson \& Frogel (1974) 
found variable extinction within the region by comparing the radio, infrared 
and optical emission fluxes. 
Wynn-Williams et al. (1977) found that
these emissions are not  coincident and the location of peak intensity
moves from the radio center towards the southwest direction as the wavelength 
decreases. The position of the optically visible nebula is approximately
2$^{\prime\prime}$.5 away from the peak of radio position. 
 On the basis of a deep 9.7 $\mu$m absorption feature in the MIR spectra, 
it is interpreted that K3-50A is partly obscured by a dense dust cloud 
(Thronson \& Harper 1979 and references therein), which
supports similar conclusion of  Wynn-Williams et al. (1977). 
To explain the strong extinction gradient across the region, 
Wynn-Williams et al. (1977) proposed a possible geometry that the K3-50A is a 
partially ionization bounded \chii region formed near the edge of an obscuring cloud and 
is now eating its way to the less dense region at the edge, like the Orion Nebula \hii region. 

Assuming that the optical lines  represent the physical properties of local environment, 
we retain our analysis with  \av  $\sim$ 6.79 mag, which has been derived using 
the extinction law \av $\sim$ 2.145 \into C($H_\beta$) (CCM 89).  All  the line 
fluxes were  dereddened using  \av $\sim$ 6.79 mag and extinction law of 
CCM 89.  Table 6 lists the  observed and reddening
corrected emission line intensities relative to H$_\beta$.  We
estimate the electron density $\sim$ 0.67 \into 10$^4$ cm$^{-3}$ 
 using 5-level atom program by Shaw \&  Dufour  (1994) from the \sii ($\lambda$ 6717 / $\lambda$
6731) line ratios, assuming  $T_e$ = 10$^{4}$ K. This assumption is not
critical because the density calculated from \sii lines are nearly
independent of $T_e$. The \sii line ratio ($\sim$ 0.54) lies close
to  the high density limit, probably due to  the
collisional quenching of forbidden lines, hence the estimated electron density
should be considered as a conservative lower limit though it might be higher.
It is noted that the density derived from the \sii lines is  
representative of a nebula $\sim$ 2$^{\prime\prime}$.5 away from
the high-density core, traced by high frequency high resolution radio observations. Therefore, 
the electron density estimated with high frequency radio observations is expected to 
be higher than density from  \sii lines. We estimate the electron temperature  $T_e$ (\nii) 
from the temperature sensitive intrinsic \nii (6548 \AA + 6583 \AA/ 5755 \AA)  line intensity ratio,
which turns out to be $\sim$ 10496 K. However, it should be noted that  
5755 \AA~ line is  quite faint as seen in the spectrum.
A roughly linear, increasing trend in [O III] (4959 \AA + 5007 \AA)/$H_\beta$
with stellar temperature is expected, because the nebular volume containing 
doubly ionized oxygen increases with the higher ionizing flux  emitted by 
early spectral type stars. 
By comparing the [O III] (4959 \AA + 5007 \AA)/$H_\beta$ value $\sim$ 1.79 of K3-50A, with 
that of values in the range 1.71 to 1.86 for the Galactic \hii regions ionized by
a single star (Kennicutt et al. 2000), a rough estimation of stellar temperature of
the ionizing star hotter than 35900 K and cooler than 42700 K can be 
obtained. Thus, from the stellar temperature, a MS spectral  type between O9 to O6.5 
(Vacca et al. 1996) is expected as  the probable exciting source of the nebula.  

\subsubsection  {Sub-millimeter emission from cold dust}
As star-forming regions are  optically thin for wavelengths
longer than $\sim$ 200 $\mu$m,  thermal dust emission can be
used to determine the physical parameters such as mass and
density of star-forming regions. Continuum sub-millimeter fluxes
at 850 $\mu$m and 450 $\mu$m for the K3-50A were obtained from
Thompson et al. (2006). The integrated flux densities at 450 $\mu$m
and 850 $\mu$m are 256 $\pm$ 79.3 Jy and 37 $\pm$ 4.3 Jy,  respectively.
The peak flux densities are 53.6 $\pm$ 16.1  Jy/beam
and 13.6 $\pm$ 1.4 Jy/beam at 450 and 850 $\mu $m, respectively.
The 850 $\mu$m map is shown in Fig. 13. The bright
continuum emission is concentrated towards the central region of
K3-50A, showing a core peaking at $\alpha_{2000} = 20^{h}01^{m}45^{s}$,
$\delta_{2000} = +33^{\circ}32^{\prime}44^{\prime\prime}$,  with
extended emission along the northeast and southwest directions.
For optically thin emission, the dust mass can be estimated from the
standard method of Hildebrand (1983) and the total mass
(dust plus gas) of the cloud is given by
\begin{equation}
M = \frac{d^{2}F_{\nu}C_{\nu}}{B_{\nu}(T_{\rm d})},
\label{eqn:dustmass}
\end{equation}
Where $F_{\nu}$ is the flux density at frequency $\nu$,
$B_{\nu}(T_{\rm d})$ is the Planck function evaluated at frequency $\nu$ and
dust temperature $T_{\rm d}$. The parameter $C_{\nu}$ is a mass conversion
factor combining both the dust-to-gas ratio and the frequency-dependent dust
 opacity $\kappa_{\nu}$ and $d$ is the distance to the source in kpc.
Various values for $C_{\nu}$ are quoted in the literature; here we
have adopted a value for $C_{\nu}$ of 50 g\, cm$^{-2}$ at 850 $\mu$m,
following the method of Kerton et al. (2001), which is appropriate for
dense molecular cores.  The integrated flux over the region at 
850 $\mu$m gives the mass of the cloud as $\sim$ 7800 $M_\odot$ for 
an assumed average dust temperature of 30 K for the entire cloud.
The main error in this estimate is likely due  to the uncertainty in the
dust temperature. Because of non-linear dependence of the
Planck function on temperature, a small change in temperature leads to a large change 
in the calculated dust mass (e.g., a 5 K error in temperature leads to an 
error of $\sim$ 40\% in mass estimation). Hence we consider the present 
mass estimate from the dust emission to be accurate only within an order of magnitude.

\subsubsection {Spectral Energy Distribution}
The star A1 shows large infrared excess (Fig. 8$a$) and is located close to the
peak of the molecular core having weak bipolar outflow (Phillips \& Mampaso 1991), 
a phenomenon that is thought to be closely related to the accretion process. 
 Molecular outflows are common phenomenon in the early stage of massive 
star formation prior to formation of \uchii region (e.g., Beuther et al. 2002) and 
also during the \uchii phase (e.g., Shepherd \& Churchwell 1996). 
To investigate the nature of A1, we constructed SED using the recently available 
grid of models and fitting tools of Robitaille et al. (2006, 2007). The models 
are computed using a Monte-Carlo based
radiation transfer code (Whitney et al. 2003a, b), using several combinations 
of central star, disk, in-falling envelope and bipolar cavity, for a 
reasonably large parameter space.
To produce the  observed  SED, we have used the data from the following: 1.25-2.17 $\mu$m  (Alvarez et al. 2004),  
3.5 and 10.1 $\mu$m (Neugebaure \& Garmire 1970); 8-21 $\mu$m, MSX PSC (Egan
et al. 2003); 
450 and 850 $\mu$m (Thompson et al. 2006). 
 The SED is shown in Fig. 14. Here, we would like to mention that 
the angular resolutions for most of the data are not
sufficient to derive precise  parameters for a single YSO.
We set MSX and sub-mm fluxes as upper limits to account for possible
contribution from other sources that might be included within the large beam of
the observations. We further set 20 to 10$\%$ error in NIR and MIR 
flux estimates due to possible uncertainties in the calibration, the extinction 
and intrinsic object variability.
A distance range of 7 to 10 kpc and the visual 
extinction in the range from 15 
to 30 mag are used as input parameters to the model to construct the SEDs. The parameters obtained from the 
weighted mean and standard deviation of all the models  with ${\chi}^2 - {\chi_{best}}^2 < 3$
(per data point), weighted by inverse square of ${\chi}^2$ of each model, 
suggest the associated source is a massive YSO with mass
$M_\star$ = 42 $\pm$ 4.4 $M_\odot$ and  of total luminosity Log($L_\star$) = 5.6 $\pm$ 0.2 $L_\odot$. 
The models also predict that the source is deeply embedded 
behind  17.5 $\pm$ 1.8 mag of visual extinction and has an envelope accretion 
rate of Log($\dot{M}_{env}$) =  -3.8$\pm$ 0.2 $M_\odot$ yr$^{-1}$ with no 
disk. The SEDs of high mass protostars are dominated by the envelope flux, where the disk flux
is almost embedded within the envelope flux. Therefore, the models with envelopes alone and the models with disks 
and envelopes successfully fit the same SED (Grave \& Kumar 2009). It is to be noted that these models are 
valid for isolated objects. Therefore, considering the presence of multiple point sources claimed by Okamoto et al. (2003), 
the above parameters should be taken with caution. Nevertheless these models 
provide quantitative information about the nature of the object.
\subsection {K3-50B}
The \ksb image of  K3-50B shows unusual distribution of dust within 
the \hii region with an elongated hole from east to west (see Fig. 3). 
The morphology of 610 MHz map (Fig. 4) suggests a blister type 
\hii region, where a sharp ionization bounded edge can be
seen  towards  the western direction. Towards the east, the gradient
of flux density indicates a density-bounded side of the blister morphology. 
The high resolution image at 1280 MHz (see Fig. 5) also shows 
non-uniform clumpy distribution of ionized gas. The complex structure of radio 
emission suggests that it may have been produced through excitation by multiple stars 
(e.g., Ojha et al. 2004a; Garay et al. 1993) or presence of density 
inhomogeneities within the \hii region excited by a single luminous star.
We detect four sources (marked in Fig. 9) within a radius of $\sim$ 0.5 pc 
from  the center of the nebula. The source B4  is a  bright infrared source 
($K$ $\app$ 11.27, $H-K$ $\app$ 0.88) with  $Q$ $\app$ 0.06, whereas the source 
B3  ($K$ $\app$ 12.47, $H-K$ = 0.31) has $Q$  $\app$ 0.31. 
The source B3 is bright in the optical wavelengths and being situated at the central region of the nebula gives an impression 
of it being one of the ionizing candidates. 
The  optical spectrum of the source B3 in the range 3950-4750 \AA~ 
is shown in Fig. 15. Due to low S/N ratio,  the 
exact spectral classification of the source is rather difficult, but the 
absence of absorption lines (\heii at $\lambda$ 4200 \& $\lambda$ 4686 \AA, \hei at $\lambda$ 4026 \& 
$\lambda$ 4471 \AA, \siii at $\lambda$ 4128 \& $\lambda$ 4130 \AA~ and \mgii at $\lambda$ 4481 \AA), 
suggest that source B3 is not an OB-type star. The source B1 has $Q$ $\app$  
0.49, along with $H-K$ $\app$ 0.61, which suggests a reddened giant rather than a
ZAMS star. The positions of the sources B1 and B3 on the CC diagram (Fig. 8$a$)
fall close to the giant locus, confirming the above argument. The source B2 with
$Q$ $\sim$ 0.12, indicates that it could be one of the ionizing candidates 
of ZAMS spectral type of B2-B3, but its low visual extinction (\av $\app$ 3.8 mag) 
in comparison  to  \av \app 26 mag (Roelfsema et al. 1988) towards K3-50B suggests 
a field star. The source B2 seems to have spectral type later than B2.
As the star of spectral type later than B2V  emits $<$ 2 $\%$ of its total 
output energy in terms Lyc photons per second, therefore the source B2 
cannot be one of the main ionizing candidates of K3-50B.
 The position of the source B4 on NIR CM diagram (Fig. 8$b$) shows the 
characteristics of ZAMS  spectral type $\sim$ O5, which is consistent 
with spectral type ($\sim$ O6-O5.5) derived from Lyc fluxes from radio 
observations, therefore B4 can be the most probable ionizing candidate for K3-50B.
The spectral type estimation of B4 is not biased by excess emission as it does 
not possess NIR excess, therefore the complex structures of the K3-50B 
region are probably due to density inhomogeneities 
rather than due to multiple  early-type sources in the region.

\subsection {K3-50C}
K3-50C is located $\sim$ 2$^{\prime}$.5 northeast from K3-50A and 
consists of two radio continuum sources C1 and C2. These sources are
separated by $\sim$ 15$^{\prime\prime}$ and C1 is  stronger than 
C2  at 8.4 GHz (Kurtz et al. 1994) and 15 GHz (DePree et al. 1994) 
radio wavelengths. 
However, below 20 $\mu$m  C2 is 
much stronger than C1 and C1 is barely detected at 20 $\mu$m and 
is not seen  at shorter wavelengths (Wynn-Williams et al. 1977), indicating
that C1 is more deeply embedded in molecular cloud of high visual 
extinction as compared to C2. C2 is located at the center of a 
prominent \ksb nebulosity in K3-50C region, where we detect a bright NIR 
point source close to the radio peak. The positions of the source C2
on CC and CM diagrams (Figs. 8$a$ \& 8$b$) indicate a reddened ZAMS 
spectral type between B0 - O9.5 with NIR excess. 
The NIR spectral type  estimation agrees well with the 
$\sim$ O9 ZAMS spectral type derived from radio continuum observations. 
The position of C1 falls on the dark  lane that bisects \ksb 
diffuse emission into north and south components (Fig. 3).
A large variation of extinction is found in the direction of C1. At the peak 
of radio continuum emission no Br$\alpha$ was detected, indicating a visual extinction 
of  $\geq$ 190 mag (Roelfsema et al. 1988), which is the possible cause for the 
non-detection of ionizing source(s) in our infrared survey around the radio peak.
The N$_{Lyc}$ photons s$^{-1}$ from the region suggest
that C1 is ionized by a single O8-O9 ZAMS star.
 The spectral type estimation on the basis of radio continuum photons is an
indirect approach in the absence of estimation based on direct stellar radiation from the star(s).
Therefore, here the possibility that more than one star might be responsible
for the ionization of C1 cannot be excluded, since massive star formation usually proceeds in cluster mode. 
In this case, the spectral type of the most luminous star must be later type than that derived from the 
single-star hypothesis.

\subsection {K3-50E and K3-50F}
K3-50E and K3-50F are two faint optical nebulosities separated by
2\arcmin.2 (Fig. 1).  Our low resolution ($\sim$ 44$^{\prime\prime}$ $\times$ 27$^{\prime\prime}$) 
map at 1280 MHz (see Fig. 16) shows continuum emission from ionized gas around both  
the regions, with K3-50F being relatively brighter than K3-50E.
In Fig. 4,  bright \ksb sources can be seen  within the radio contours, which
are also visible  in the optical (K3-50E: $V$ = 17.16, $B-V$ = 1.543; K3-50F: 
$V$ = 14.95, $B-V$ = 1.282). These sources are  located within the radius of $\sim$ 0.5 pc from
the radio peak, and so are most likely the ionizing sources of the nebulae. 
Their visual extinctions estimated using the `Q' method (\S 6.1) suggest that 
they are extincted by visual extinction of 5.6 and 4.8 mag, respectively.

 There are no definite estimations of distance for K3-50E and K3-50F 
in the literature but they are assumed as $\sim$ 8.7 kpc (Israel 1976).
Hence, to estimate approximate distance we use the following relation
given by Comer\'{o}n \& Torra  (2001) for an ionization 
bounded optically thin \hii region with an electron
temperature of $\sim$ 10$^4$ K :
\begin{equation}
D(kpc) \approx  5.35 \times 10^{-3} [\frac{\nu^{0.1}(GHz)
S_{\nu}(J_y)}{\Sigma^{N}_{i=1}10^{-1.26 H_{oi}}}]^{0.233} 
\end{equation}
where S$_\nu$ is the flux density in Jy at frequency $\nu$ (in GHz). 
$H_{oi}$ is the dereddened $H$-band magnitude of `i' th ionizing star.
  Due to the small sizes of both the regions and our selection method to identify possible
ionizing sources described in section 5.5, we believe that the two optically bright sources
are most likely the ionizing sources of the \hii regions.
Using the above approximate relation, we obtain distances of $\sim$ 7.6 kpc and 
$\sim$ 8.4 kpc for K3-50E and K3-50F, respectively. Here, we assume that 
both the regions are ionization bounded and ionized by a single star.
In Eq. 2 we have used  radio fluxes ($\app$ 0.04 Jy for K3-50E and $\app$ 0.13 Jy 
for K3-50F) estimated from the low resolution map (Fig. 16), the 
$H$-band magnitudes ($\app$ 12.75 mag for K3-50E and $\app$  12.25 mag for K3-50F)  
of the ionizing stars and extinction at $H$-band ($\app$ 0.87 mag for K3-50E and
 $\app$ 0.74 mag for K3-50F). 
 Although the above approximate distance estimates agree with the kinematic distance (7.9 to 9.3 kpc)
reported towards the direction of K3-50 SFR, the validity of the
method is constrained by the number of assumptions concerning membership, dereddening, 
dust absorption of Lyc photons, leakage of ionizing photons from the 
\hii regions and uncertainty in the radio flux estimations. The more precise  
distance can be estimated using observed photometric magnitudes and accurate spectral type of 
the ionizing sources as described in section 6.1 for K3-50D. We estimated Lyc photons 
per second from the integrated fluxes at an assumed distance of 8.7 kpc and that 
come out to be $\sim$ 1.5 $\times$ 10$^{47}$ s$^{-1}$ and $\sim$ 4.8 $\times$ 10$^{47}$ s$^{-1}$ 
for the K3-50E and K3-50F, respectively. These fluxes
correspond to ZAMS spectral types of B0.5-B0 and B0-O9.5, respectively, 
consistent with the ZAMS spectral 
types estimated from $J-H$ versus $J$ CM diagram (see Fig. 8$b$ and Table 5).

\section {General discussion}

\subsection {Extinction towards individual regions}
The massive stars (M $>$ 8$M_\odot$) do not have pre-main-sequence (PMS) 
phase and the object reaches 
ZAMS while still embedded in the molecular cloud (Palla \& Stahler 1990). We used
probable ionizing sources (see Table 5) of the individual \hii regions to 
derive their  extinction values. 
We traced the ionizing  stars back to their intrinsic ZAMS position in $(J-H)$ versus $J$ CM diagram 
(Fig. 8$b$) along the  reddening vector. 
However, the  uncertainty  in estimation of extinction 
cannot be ruled out due to excess emission in $J$ and $H$ bands.
This could be the case for the ionizing star A1 of K3-50A region, 
which shows maximum NIR excess among all the YSOs.  The visual extinction toward regions
A, B, C2, D, E and F is estimated to be 31.2, 16.1, 21.2, 4.4, 5.9 and 5.5 mag, respectively. 
Here it is to be noted that the extinction estimates for A1 and C2 must be
towards the higher side as these stars have significant NIR excess. For example,  
in the case of A1, the extinction value derived from the SED fitting (\S 6.2.3)
comes out to be 17.5 $\pm$ 1.8 mag.   
The above discussion along with the average extinction found towards C1 (see \S 1) 
indicate that  the K3-50 SFR shows variable extinction in the range of 4.4 - 97 mag.
A comparison of extinction found
towards K3-50 to the  $^{12}$CO(1-0) map by Israel  (1980) for W58 complex
at a resolution of 2$^{\prime}$.3 reveals the following facts: the position 
 of K3-50C coincides with the $^{12}$CO(1-0) peak, whereas the regions K3-50D, 
K3-50E and K3-50F are  located towards the edge, and therefore suffer less  
amount of extinction (see Fig. 17). 
The relatively better resolution ($\sim$ 1$^{\prime}$) 1 mm continuum map of 
Wynn-Williams et al. (1977) shows mainly two peaks (towards K3-50C and K3-50A) 
of dust emission, indicating that K3-50B is less obscured in comparison with  both these sources. 
This agrees well with its low extinction value.

\subsection{Spatial distribution of YSOs}
The  IR-excess in the case of YSOs can be due to circumstellar disk/envelope 
around the YSOs or weaker contribution from reflected stellar radiation 
of the dust emission. In any case IR-excess represents the association of young sources 
in a star-forming region. The spatial distribution of probable YSOs viz. 
Class II YSOs (asterisks),  Class  I  YSOs (triangles), possible YSOs 
detected only in  $H$ and $K_{\rm s}$  bands with  $H-K$ $\geq$ 1.8 
(filled circles), and  selected from $H-K/K-[4.5]$ CC diagram (open circles)  
is shown in Fig. 17.  Most of the Class I sources and sources with 
$H-K$ $\geq$ 1.8  are mainly concentrated towards the regions 
A, B and C2. The detection of sources with $H-K$ $\geq$ 1.8 towards A, B and C2
suggests that the extinction is so large that many YSOs may be too heavily 
extincted to be detected in $J$-band.
Since most of these sources are also associated with \ksb nebulosity, 
they possibly have intrinsic NIR excess as well as local extinction. 
However, the  IR-excess sources identified from $H-K/K-[4.5]$ CC diagram 
are spreaded over the entire region with less concentration towards 
the south-eastern direction. 
The spatial distribution of these IR-excess sources
is well correlated with the diffuse emission seen in 8 $\mu$m band of 
{\it Spitzer} as well as with the intensity of CO emission.
 An inspection of the MSX PSC yields six sources within the investigated region
 with positive flux values in A (8.3 $\mu$m), C (12.1 $\mu$m), D (14.7 $\mu$m) and E (21.3 $\mu$m) 
of MSX bands. The positions of MSX point sources are shown in Fig. 17 as crosses.
 The spatial distribution of four sources coincides with the \ksb nebulosity associated to the 
\hii regions. The position of one source (G070.2990+01.5762) falls on the
north-eastern side of the K3-50A and the other source (G070.3164+01.6493) coincides with the 
intense compact \ksb nebulosity seen between K3-50E and K3-50F (see Fig. 3).
An IRAS point source (IRAS 19597+3327A) is also detected in the proximity of 
G070.3164+01.6493 (hereafter IRAS-B, see details in appendix). 
To know the nature of these sources, we use CC plot as suggested by
Lumsden et al. (2002) who  derived criteria
for identifying various Galactic plane sources using the MSX bands data.
Fig. 18 shows the CC plot for the MSX point sources. The demarcations shown in  
Fig. 18 were derived from known sample of objects by Lumsden et al. (2002), where
a significant overlap between massive YSOs and \chii regions is seen in their sample
having colors $F_{\rm 21}$/$F_{\rm 8}$ $\ge$ 8 and $F_{\rm 14}$/$F_{\rm 12}$ $\ge$ 1. 
Figure 18  shows that most of our sources  fall in the overlap region 
of massive YSOs and \chii regions.
To study the nature of these sources we constructed SEDs with the 
help of their NIR and IRAC counterparts. For the construction of SEDs, 
IRAC sources with photometric error less than 0.25 have been used.
Fig. 19 displays the SEDs for
the sources  between 1.25 $\mu$m to  8 $\mu$m.
 Protostars are classified on the basis of their spectral slopes from their SEDs.
The general consensus is that rising and flat spectrum sources are, 
on an average,  younger and are in a more embedded phase of star formation 
(Allen et al. 2004). The SEDs of massive stars associated to  D, E and F 
show almost flatter spectrum  or a minor increasing trend, whereas
those of IRAS-B and  C2 show rising trend
at longer wavelengths.
From SEDs it appears that the source C2 is relatively younger 
than the other sources, neglecting the effect of inclination, 
stellar temperature, contribution from the scattered light and extinction, 
which can cause diversity in the spectral slopes.

\subsection {Age of the massive stars}
The ages of young clusters are typically derived from the post-main-sequence
evolutionary tracks, for the earliest members, if significant evolution has occurred. 
The presence of early O-type MS star (e.g., O4V in the case of K3-50D) 
in a star-forming region would indicate that the age of the region should 
be $<$ 3 Myr,  as lifetime of an O4
star is of the order of 3 Myr (Meynet et al. 1994). Massive OB stars 
do not have optically visible PMS phase. As Sh 2-100 region   
contains a few optically detected early-B stars and the  contraction time scale for 
such stars (B3V; 9$\msun$ to B0V; 18$\msun$) 
to reach ZAMS is approximately few $\times$ 10$^{5}$ yr (Bernasconi \& Maeder 1996), 
hence the expected ages of massive stars associated to the Sh 2-100 region must be of the order 
of few Myr. To estimate the ages of ionizing 
sources of K3-50D, K3-50E and K3-50F,  we
used V/B-V CM diagram. Fig. 20 shows intrinsic V/B-V CM diagram,
 where the stars are dereddened individually using the $Q$ method as discussed in 
section 6.1.  We visually fit  solar metallicity theoretical 
isochrone of 1 Myr by Girardi et al. (2002), assuming a distance of 
8.7 kpc for the region. However, due to the distance uncertainty 
associated with the individual sources, uncertainty in the estimation of extinction based on Q 
method and  the resolution of the 
isochrones to distinguish the ages between 1-3 Myr, the derived age ($\sim$ 1 Myr)
should be considered as  an approximate estimation. Probable 
ionizing stars of the remaining regions are embedded  and  not detected 
in the optical band. Consequently, their age estimation is rather difficult. 
Most of these stars show NIR excess, 
which is a signature of the presence of an inner disk. As the time scale for 
destruction of disks ranges from less than 10$^{5}$ yr for an early O star
 to $\sim$ 5 \into 10$^{5}$ yr for B0V star (Bik et al. 2006 and references 
therein),  the ages of the massive stars 
associated with K3-50A, K3-50B and K3-50C2 should be less than 1 Myr.  

\subsection {Mass Spectrum of Young Stellar Objects}
 Considering the presence of \chii \& \uchii regions, astronomical masers 
and intense gas and dust, similar to the W3 Main region studied by 
Ojha et al. (2004b), 
we assumed PMS age of 0.5 Myr to 1 Myr for the YSOs to estimate their masses.
To  minimize the 
effect of NIR excess emission on mass estimation we preferred $J$-band luminosity 
rather than $H$ or $K$ band.  Fig. 21 represents
 $(J-H)$ versus $J$ CM diagram for the YSOs. The solid and  dotted curve lines  
in  Fig. 21 denote the loci of 0.5 Myr and 1 Myr PMS isochrones
by Siess et al. (2000). The solid and dotted slanting lines are the
reddening vectors for  0.3 and 4 $\msun$ stars for the 0.5 Myr 
and 1 Myr isochrones, respectively. The different symbols represent
the YSOs selected from NIR and IRAC CC diagram and are the same as in Fig. 17.
The reddening vectors for the assumed
age suggest that the majority of the YSOs have  masses  in the range of
0.3$\msun$ - 4$\msun$ indicating that stellar population in
Sh 2-100 is mainly composed of low-mass YSOs.
 Although, the assumed age for the region is the subject
of uncertainty and may not be accurate as compared to the ages estimated
with the spectroscopic information of the sources in star clusters,
the distibution of YSOs with wide colors (Fig. 21) probably indicates
the combined effect of variable extinction, weak contribution of excess emission 
in $J$ and $H$ bands and/or sources in different evolutionary stages.
Therefore, it would be very useful to get radial velocity information of 
the YSOs to confirm their association and ages with spectroscopic information.

\subsection {A possible evolutionary status of the \hii regions}
The regions K3-50E and K3-50F contain massive B0-B0.5V MS stars. These early-type stars
usually form in the dense core of a molecular cloud and can be seen in millimeter, mid-, far-infrared 
and/or as compact radio emissions. The presence of weak extended radio emission (Fig. 4) and
lack of dust continuum emission at 25 $\mu$m and 60 $\mu$m (see Fig. 22) indicate that the 
associated \hii regions are probably  in an advanced stage of evolution. 
The presence of relatively intense radio emission 
at 1280 MHz and dust emission at  25 $\mu$m and 60 $\mu$m, respectively, indicate that 
K3-50D is younger or of age comparable with K3-50E \& K3-50F. 
 All the ionizing sources of these \hii regions do not show infrared excess and their detection
in the optical band suggests their evolved nature. The optical V/V-I diagram suggests that 
the age of the ionizing sources is of the order of 1 - 2 Myr.  However, non-detection of 
ionzing source(s) in optical, relatively intense \ks-band emission as well as 
radio emission as compared to K3-50D suggest that K3-50B should be younger than K3-50D.
The regions K3-50A and K3-50C (consisting of two sources C1 and C2) are possibly younger than K3-50B, due to
their association with intense $^{12}$CO(1-0) emission, compact radio sizes and 
the ionzing sources possess infrared excess (e.g., C2 and A). The detection of OH masers (Baudry \& Desmurs 2002) and 
ionized outflows (DePree et al. 1994;  Balser et al. 2001) in both the cores
indicate that they are not dynamically evolved objects and suggest that the 
star formation activity around K3-50C and K3-50A  should not be more
than $\sim$ few $\times$ 10$^5$ yr. The component C1 is hidden behind a visual extinction $\geq$ 190 mag, 
therefore invisible in wavelength shorter than 20 $\mu$m, whereas the ionizing source of
K3-50A is detected in NIR and embedded in molecular cloud having visual extinction $\sim$ 17.5 mag. 
K3-50C1 has lower dust temperature than K3-50A, possibly indicating that K3-50C1 is more deeply embedded 
in the molecular cloud as compared to K3-50A (Hunter et al. 1997). The above discussion 
reveals that K3-50C1 is most likely the youngest source of the region.

\subsection  {Star formation scenario in W58}
W58 is a molecular cloud complex of an angular extent of about 60 arcmin. 
The $^{12}$CO distribution 
mainly reveals three distinct components (see Fig.3 of Israel 1980) and the 
associated SFRs to these components are K3-50, Sh 2-99 (S99) and W58G.
 Fig. 23 shows 8 $\mu$m  emission from MSX A-band along with the distribution of
\hi emission around the SFR complex (thick solid line) 
taken from Fig. 6 of Israel (1980). The positions of different sources found towards the direction 
of W58 cloud complex are shown as crosses. 
The most intense northern 
CO component associated with K3-50 SFR peaks close to K3-50C, where most of the star 
formation activity is currently going on. The western 
CO component is associated with an optically visible \hii region 
(Sh 2-99/G70.15+1.73) situated at a distance of $\sim$ 11$^{\prime}$ from 
K3-50A towards the south-western direction. The weaker southern CO 
component is associated with an \hii region (W58G/G69.92+1.52)  which is $\sim$ 23$^{\prime}$ 
away from K3-50A. In the case of \hii region (W58G), a radial velocity of -65 km s$^{-1}$ has been 
measured, which differs significantly from the radial velocity - 22 to -25 km 
s$^{-1}$  found in the direction of W58. This indicates that W58G most likely is not related to 
W58 cloud complex (Israel 1980 and references therein), whereas the radial velocity  of 
-22.9 km s$^{-1}$ for 
Sh 2-99 (Blair et al. 1975) suggests its association with W58. 

With the help of \hi emission observed by  Read (1981) with a resolution of 2$^{\prime}$.5 $\times$ 4$^{\prime}$.6,
Israel (1980) pointed out an \hi/\hii expanding shell of diameter $\sim$ 75 pc and suggested that the 
current star formation activity in Sh 2-99 and K3-50 region might have been 
induced by the interaction  of this shell produced by stellar winds from the older generation stars of W58 cloud complex.
To recognize such a shell, a potential way is to use the PAH emissions at 8 $\mu$m 
arising from the photodissociation region, which occurs at the 
interface between the molecular and neutral gas and thought to be excited by
sub-Lyman UV photons from the massive star(s).
Consequently, it can be used as a tracer of the neutral shell observed
in the \hi 21-cm line emission  at the edge of expanding \hii region 
(Deharveng et al. 2005). Fig. 23 shows the 8 $\mu$m  PAH emission surrounding the  
bubble noticed by Israel (1980) and roughly coincides with the \hi emission 
in terms of discrete patches. The absence of  8 $\mu$m  emission in the interior of the \hi  shell 
can be interpreted as the destruction of PAH molecules by intense UV radiation from the associated massive star(s).
The shape and large dimension of \hi shell suggest that the stars producing stellar wind should be 
located inside the shell. A search of {\it Simbad} database results in a WR star (WR 131) of WN7h 
(hydrogen rich) subtype (Figer et al. 1997), whose position is marked in Fig. 23. The hydrogen rich WN7 stars are 
descendants of massive stars with initial masses above 50-60 $\msun$ 
(e.g., Crowther et al. 1995). Indeed, progenitors of such stars are 
believed to be core-hydrogen burning Of stars with strong stellar winds (Crowther et al. 1995).
The  presence of WR 131 in the \hi shell and its approximate distance of $\sim$ 9.1 kpc 
(Georgelin et al. 1988) comparable  with the distance estimates of W58 complex 
($\sim$ 7.9 to 9.3 kpc), favors a probable association between the WR 131 and W58 complex. 
The catalog of Reed (1998) yields a few B-type stars which
are projected  inside the bubble, but their distances are not known so as to confirm
their association. If indeed these sources are associated then the presence of 
\hi shell can be due to the combined energy input from these sources.

WR 131 is not located at the center of the \hi shell but at an off-center position, which
could be due to the combination of an inhomogeneous medium, proper motion of 
the star, anisotropic winds and projection effect along the 
line of sight. If we assume WR 131 is the source to produce
\hi shell due to its strong stellar winds and since WR phase is  short lived ($\sim$ 10$^{5}$ yr) and  
the presence of WR star limits the maximal age $\sim$ 4.5 - 5 Myr of a cluster (Meynet \& Maeder 2005), 
an upper limit of age $\sim$ 4 - 4.5 Myr can be assigned to WR 131, considering the estimated 
dynamical age  ($\sim$ $10^{7}$ yr) of the bubble by Israel (1980). 
If the age of WR 131 is $\sim$ 4 - 4.5 Myr 
then one would expect the ages of the  second  generation stars, (if any), 
at the periphery  of the bubble, should be less than that of a WR star. 
The presence of  star formation  activity, which might have started  
$\leq$ 2 Myr ago is evident at the north-eastern edge of the shell, with the existence of Class I 
sources, astronomical masers, \uchii and \chii regions 
containing young OB stars of ages 1-2 Myr. The presence of seven \hii regions
at different spatial locations and possibly at different evolutionary stages 
suggests star formation activity has proceeded at multiple sites in the cloud.  The presence of
WR 131 star of age $\sim$ 4 - 4.5 Myr and  young massive stars still in accretion phase (e.g., C1 and A),
associated with the same complex, indicates distribution of stars of different ages in the region. 
Israel  (1980) suggested that at present the shell is colliding with a larger \hi cloud to the northeast of 
W58 (i.e., east of K3-50C). In this conjuncture, one would expect Sh 2-99 and K3-50D to be relatively 
evolved regions and K3-50C to be the least evolved source as we move from southwest to northeast. 
On a large scale, it seems to be consistent with the scenario of sequential star formation
because the probable age of K3-50D is $\sim$ 1 - 2 Myr, whereas the expected age of K3-50C1 is 
of  $\sim$ few $\times$ 10$^{5}$ yr. This is further supported by the presence of more intense 
molecular material observed towards K3-50C than Sh 2-99 (Israel  1980), along with the average extinction 
gradient of 4.2 to 97 mag from K3-50D to K3-50C1. However, 
 it is not known whether the different condensations  of Sh 2-100 region were in different stages of 
evolution before the passage of ionization front generated by the stellar winds of WR 131. It 
is also difficult to anticipate the projection/ orientation  effect on the location of the WR star with respect to the Sh 2-100 region. The initial mass of the cloud associated with the individual region is also an important parameter as it governs evolution  of the cloud in presence of external radiation pressure. Numerical models predict that if the pressure due to external ionized gas is too small to compress a massive  molecular gas, the radiation field will not have any dynamic effect on the cloud (Bertoldi \& McKee 1990; Lefloch  \& Lazareff 1994).
At present the sequential star formation around Sh 2-100 region due 
to the interaction of \hi bubble is suggestive. A more complete picture can be revealed with the
 help of high resolution, high sensitivity \hi and CO observations covering the entire 
W58 complex. A positional and chronological study of the YSOs associated with the complex 
(e.g., Lee et al. 2005; Ogura 2006)  can support the validity of sequential star formation.
Since the region studied here is deeply embedded and shows a large variation
in extinction, we would like to determine the intrinsic  properties of the low mass PMS stars 
identified from the NIR and IRAC CC diagrams with follow up deep optical/NIR spectroscopic observations 
in order to determine their relative ages by constructing
intrinsic Hertzsprung-Russell diagrams. 
These observations will allow us to have a better understanding  of 
star formation history across the entire region and a much clearer picture
of the triggered star formation across the W58 complex.

\section{Conclusions}
In this paper we have presented a multi-wavelength study of
the stellar contents and physical environment of star-forming region (K3-50)
in the proximity of diffuse \hii region  Sh 2-100.
K3-50 consists of seven \hii 
regions (A, B, C1, C2, D, E and F) of different classes and are thought to be a part of 
W58 molecular cloud complex located at a distance of $\sim$ 8.7 kpc.
 Our main conclusions are as  follows:

1. We have identified the ionizing star of the optical visible nebula K3-50D. 
Optical spectroscopy  suggests that an O4V star is 
responsible for the ionization. We have estimated a distance of $\sim$
8.6 kpc for the K3-50D region.  The approximate distances for   
the K3-50E and K3-50F regions have been estimated from the NIR photometry 
and radio continuum flux densities following the methodology by 
Comer\'{o}n \& Torra (2001). The distance estimation suggests that 
both the regions are likely the part of the W58 complex. 

2. We identified the probable exciting sources for six \hii regions using 
the $J$ vs. $J-H$ CM diagram. The photometric spectral types of the ionizing
sources agree well within a subclass with that derived from radio observations 
based on the number of Lyc photons. 

3. We derived physical parameters for all the five \hii regions using 1280 MHz
radio continuum  observation. The \hii regions cover a wide range  
in size ($\sim$ 0.2 to 1.6 pc), optical depth ($\sim$ 0.18 to 1.92),
rms electron density ($\sim$ 740 to 6390 cm$^{-3}$) and rate of 
Lyc photons ($\sim$ 5.5 $\times$ 10$^{47}$ s$^{-1}$
to $\sim$ 1.1 $\times$ 10$^{49}$ s$^{-1}$). These physical parameters
reflect different evolutionary stages of \hii regions.

4. We estimated the total mass of the cloud associated with one of the \hii
regions (K3-50A) which was found to be $\sim$ 7800 $\msun$.
At the center of the cloud core, we detected a possible  massive YSO
with $M_{\star}$ $\sim$ 42 $\pm$ 4.4 $M_\odot$, accreting with an effective
envelope accretion rate of Log($\dot{M}_{env}$) $\sim$  -3.8$\pm$ 0.2 $M_\odot$ yr$^{-1}$.

5. Combining IRAC photometry with ground-based NIR observations within an area of $\sim$ 8$^{\prime}$ $\times$
8$^{\prime}$, we found a total of 150 objects with circumstellar disks (i.e., YSOs), suggesting 
a site of active star formation. The distribution of these YSOs (Class II, Class I and $H-K$ $>$ 1.8 sources) 
in Sh 2-100 SFR correlates well with the association of gas and dust within the region. 

6. The distribution of YSOs at the periphery of \hi shell on a large scale supports the 
speculation by Israel (1980)  that the star formation activity 
around K3-50 region might have been induced by 
an expanding spherical bubble which was created by strong  stellar winds from 
older generation OB stars of the complex in the south-eastern direction.  

The Sh 2-100 SFR region represents a broad sample of different stages of massive star formation 
all in the same cloud, from the cold, dense core at source C1 to the diffuse H II region at source F. 
The conclusions drawn  in this paper on this
cloud should be relevant and applicable to further deep study of the region with high 
resolution and high sensitivity instruments.

\section{Acknowledgments}
 We thank the anonymous referee for a critical reading of the paper and
several useful comments and suggestions, which greatly improved the 
scientific content of the paper.
 The authors would like to thank the staff of HCT operated by Indian Institute of Astrophysics, Bangalore, IRSF at South Africa in joint partnership
between S.A.A.O and Nagoya University of Japan and GMRT managed by 
National Center for Radio Astrophysics of the Tata Institute of Fundamental Research (TIFR) for their assistance and support
during observations. MRS would like to thank
the TIFR for the kind hospitality during his visits to the 
institute, where a part of the work reported was carried out. MRS also
thanks T. P. Prabhu and A. Tej for helpful discussions,
as well as N. Chauhan and J. Jose for their helpful discussions on the {\it Spitzer}
data reduction. 

\vskip 1cm
\begin{center}
\end{center}
{\centerline { \large Appendix A: Selected interesting region}} 
IRAS-B is  situated at $\sim$ 3$^{\prime}$.3  northwest of K3-50A. 
An analysis of the IRAS fluxes reveals that this source has 
color characteristics
(e.g., log (F25 )/log (F12 ) = 2.02 and log (F60 )/log (F12) =
5.06) consistent with being an \uchii region (Wood \& Churchwell 1989).
A prominent \ksb nebulosity seen in the direction of IRAS-B is 
offset  by $\sim$ 15$^{\prime\prime}$ in the western direction from the position of  IRAS-B (see Fig. 24).
Our NIR images reveal that the \ksb nebulosity contains few point sources and 
the position of the MSX point source (G070.3164+01.6493) falls on the \ksb 
nebulosity. The offset of IRAS-B may be due to the large IRAS beam. 
Therefore, the more 
accurate position of the IR source associated with IRAS-B can be inferred 
from the nearby MSX point source (G070.3164+01.6493), since the MSX 
survey has global absolute astrometric accuracy of 1$^{\prime\prime}$.9. 
The  MSX CC plot (see Fig. 18) for the source suggests IRAS-B to 
be a massive YSO.
Since there is no study available in the literature for IRAS-B, the 
distance to the source is unknown, which  puts a limit to support in favor of a 
physical connection of this source with Sh 2-100, rather than a chance of 
alignment along the line of sight. However, if we assume a distance 8.7 kpc, 
our analysis leads to many useful conclusions.
We acknowledge that this assumption is highly speculative, hence the results should be taken
with caution. Analysis of NIR CM and CC diagrams suggest that the most luminous member associated
with IRAS-B  is a ZAMS star of spectral type $\sim$ O7-O8 with NIR excess.  Since the source
shows NIR excess, the spectral class assigned on the basis of CM diagram 
could have been estimated as an early type rather than its real spectral type, but this still disagrees with 
the ZAMS spectral type (earlier than O3) derived from  IRAS point source fluxes using the 
equation of Casoli et al. (1986). 
This discrepancy could be due to large IRAS beam and/or presence of low mass stars, which  can
contribute significantly to the radiation emitted in the FIR. If we assume that an O8 star is 
present in the region, the absence of radio emission (up to $\sim$ 5 mJy) at 1280 MHz 
indicates that (i) either the ionizing star is heavily enshrouded by the dust so
that all its ionizing radiation is absorbed  or (ii) the \hii region is
suppressed by the accretion of matter (see, e.g., Yorke 1986) or 
(iii) the source must be a ZAMS star of spectral type later than B0. 
The spectral type B0 has been  estimated from the Lyc photons derived 
using  $\sim$ 5 mJy flux (3$\sigma$ flux of the map at 1280 MHz) and  the calibration table of 
Panagia (1973), for an electron temperature of 10$^{4}$ K. 
In order to get a qualitative estimation of the
evolutionary stage of the source, we have fitted the models 
by Robitaille et al. (2006),  which are already described in section 6.2.3. 
The SED is shown in Fig. 24.
 Though, the number of data points used for the SED fitting are quite less,
nevertheless as stated by Robitaille et al (2007), some of the parameters can be
constrained more narrowly than the others, depending upon the available fluxes.
The parameters computed from the models as described in  section 6.2.3,
imply a source of mass  $M_\star$ = 10.7 $\pm$ 1.8 $M_\odot$ and total luminosity 
 Log$L_\star$ = 3.4 $\pm$ 0.2 $L_\odot$ lying behind 6.1 $\pm$ 0.2 mag of
visual extinction. In the absence of sub-mm and mm data points, the resulting 
uncertainties in other parameters may be large, and therefore not quoted here.
The mass of the object suggests a star of spectral type B2-B3V, but the lack of radio emission around 
the star at such a low visual extinction requires follow up high sensitivity, high frequency  
radio observations, along with \ksb spectroscopy to obtain the accurate nature of the massive star 
and its distance.

\begin{table}
\caption{Log of observational data.}
\begin{tabular}{lllll}
\hline
\hline
Date (UT)   &  Filter     & No of & Exposure (sec) & Telescope \\
& &\multicolumn{1}{c}{frames} & \multicolumn{1}{c}{per frame} \\
\hline
2006 Oct 24 &\hspace{2mm} U & \hspace{2mm}    3         &\hspace{4mm} 1000 & ST\\
2006 Oct 25 &\hspace{2mm} U & \hspace{2mm}    3         &\hspace{4mm} 300& ST\\
2006 Oct 24 &\hspace{2mm} B & \hspace{2mm}    3         &\hspace{4mm} 900& ST\\
2006 Oct 25 &\hspace{2mm} B & \hspace{2mm}    3         &\hspace{4mm} 300& ST\\
2006 Oct 24 &\hspace{2mm} V & \hspace{2mm}    3         &\hspace{4mm} 600& ST\\
2006 Oct 25 &\hspace{2mm} V & \hspace{2mm}    3         &\hspace{4mm} 100& ST\\
2007 Oct 29 &\hspace{2mm} Gr 7 & \hspace{2mm}  1         &\hspace{4mm} 900& HCT\\
2007 June 18 &\hspace{2mm} Gr 7 & \hspace{2mm}  2         &\hspace{4mm} 900& HCT\\
2008 July 21 &\hspace{2mm} J & \hspace{2mm}    120         &\hspace{4mm} 10& IRSF\\
2008 July 21 &\hspace{2mm} H & \hspace{2mm}    120       &\hspace{4mm} 10& IRSF\\
2008 July 21 &\hspace{2mm} K$_{\rm s}$ & \hspace{2mm}    120        &\hspace{4mm} 10& IRSF\\
2007 June 18 &\hspace{2mm} H$\alpha$ & \hspace{2mm} 1      &\hspace{4mm} 300& HCT\\
2007 June 18 &\hspace{2mm} R &  \hspace{2mm}    1         &\hspace{4mm} 300& HCT\\
2007 June 18 &\hspace{2mm} [SII] &  \hspace{2mm}    1         &\hspace{4mm} 300& HCT\\
2004 Oct 10 &\hspace{2mm} Br$\gamma$ & \hspace{2mm}  21         &\hspace{4mm} 90& HCT\\
2004 Oct 10 &\hspace{2mm} K$_{cont}$ & \hspace{2mm}   21         &\hspace{4mm} 20& HCT\\
\hline
\end{tabular}
\end{table}
\begin{table}
\caption{Details of the GMRT radio continuum observations.}
\label{radio.tab}
\begin{tabular}{l|c|c}
\hline\hline
 Rest freq. (MHz)         & 1280  & 610 \\
\hline
Date of Obs.      & 25 Jan 2004 & 07 Apr 2004 \\
Phase center      & $\alpha_{2000}=20^{h}01^{m}45.50^{s}$ & $\alpha_{2000}=20^{h}01^{m}45.50^{s}$   \\
                  & $\delta_{2000}=33^{h}32^{m}43.02^{s}$ & $\delta_{2000}=33^{h}32^{m}43.02^{s}$   \\
Flux calibrator &  3C48    & 3C286 \\
Phase calibrator &  2052+365    & 1924+334 \\
On source int. time &  1 (hr)    &  3 (hr) \\
Cont. Bandwidth  & 16  (MHz)       & 16 (MHz)  \\
Synth. beam   &  3\arcsec.3 $\times$ 3\arcsec.0 &
5\arcsec.5 $\times$ 4\arcsec.8 \\
rms noise  &  2.2 mJy/beam           & 0.9 mJy/beam \\
\hline
\end{tabular}
\end{table}
\begin{table}
\caption{Observed integrated flux and angular size for associated \hii regions}
\begin{tabular}{cccccc}
\hline\hline
Frequency &\multicolumn{2}{c}{} &\multicolumn{1}{c}{1280 MHz} &\multicolumn{1}{c}{1280 MHz} \\ 
\hline
Source& \multicolumn{2}{c}{Peak Position} & Int. Flux  & Angular Size\\
  & \multicolumn{1}{c}{$\alpha$(J2000)}& \multicolumn{1}{c}{$\delta$(J2000)} & \multicolumn{1}{c}{Jy} & \multicolumn{1}{c}{} \\
\hline
A  &20:01:45.57 &+33:32:42.86  &0.387& 7\arcs.1 $\times$ 4\arcs.6 \\
B  &20:01:48.01 &+33:33:03.21  &1.680&36\arcs.4 $\times$ 33\arcs.3 \\
C1 &20:01:53.87 &+33:34:15.52  &0.156&8\arcs.2 $\times$ 5\arcs.3  \\
C2 &20:01:55.08 &+33:34:16.44  &0.093& 5\arcs.8 $\times$ 4\arcs.2 \\
D  &20:01:47.67 &+33:31:50.61  &1.880&42\arcs.9 $\times$ 37\arcs.8\\
\hline
\end{tabular}
 \end{table}
\begin{table}
\caption{Derived parameters for the \hii regions from radio observations}
\begin{tabular}{cccccccc}
\hline\hline
Source &\multicolumn{1}{c}{$\tau_c$} &\multicolumn{1}{c}{rms $n_{e}$} &\multicolumn{1}{c}{EM} & \multicolumn{1}{c}{log N$_{Lyc}$} & ZAMS$\dag$ \\
  & \multicolumn{1}{c}{}
          & \multicolumn{1}{c}{(10$^{3}$ cm$^{-3}$)}
          & \multicolumn{1}{c}{(10$^{6}$ cm$^{-6}$ pc)} & \multicolumn{1}{c}{(ph. s$^{-1}$)} &
          \multicolumn{1}{c}{Spectral type}\\ \hline
\hline
A&1.92&6.39&9.84&48.36 &O8\\
B &0.22&0.88&1.14&49.00&O6\\

C1 &1.38&5.04&7.07&48.34 &O8 \\

C2 &0.84&4.89&4.28&47.74 &O9.5 \\

D &0.18&0.74&0.95&49.04&O6.5 \\
\hline
\end{tabular}
\flushleft\small{ $\dag$ From Panagia (1973)}
 \end{table}

\begin{table} \centering
\caption{Spectral types of the ionizing sources of the \hii regions}
\begin{tabular}{cccccccc}
\hline\hline
   &  \hspace{4mm}  Spectral Type    & \\
\hline
Source & \multicolumn{1}{c}{NIR estimation $\dag$}
          & \multicolumn{1}{c}{Radio estimation (Ref.)} \\ \hline
A& \hspace{2mm} $>$ O5  & \hspace{2mm} O5.5 (1) \\
B & \hspace{2mm} O5  & \hspace{2mm} O5.5 (1)\\
C1 & \hspace{2mm} --- & \hspace{2mm} O9 (1)\\
C2 & \hspace{2mm} B0-O9.5 & \hspace{2mm} O9.5 (1)\\
D &\hspace{2mm} O5 & \hspace{2mm} O6 (1)\\
E &\hspace{2mm} B0.5& \hspace{2mm} O8 (2) \\
F &\hspace{2mm} B0 & \hspace{2mm} O9.5 (2) \\
\hline
\end{tabular}
\flushleft\small{ $\dag$ From this work}
\flushleft\small{ (1) De Pree et al. 1994; (2) Israel (1976)}
 \end{table}

\begin{table}
\caption{Observed and dereddened line intensities derived from the optical spectrum of K3-50A nebula}
\begin{tabular}{llll}
 \hline
 \hline
  $\lambda$ (\AA)   & Lines &  Obs. Flux $\dag$   & Dered. Flux $\ddag$  \\
\hline
3726, 9    &[O~{\sc ii}]         &0.299             &3.121    \\
4340      &H$\gamma$            &0.189             &0.591   \\
4861      &H$\beta$             &1.000             &1.000    \\
4959      &[O~{\sc iii}]        &0.502             &0.416   \\
5007      &[O~{\sc iii}]         &1.814            &1.378   \\
5755      &[N~{\sc ii}]         &0.059             &0.016   \\
5876      &He~{\sc i}           &0.412             &0.094    \\
6548      &[N~{\sc ii}]         &1.796             &0.208   \\
6563      &H$\alpha$            &24.79             &2.839    \\
6583      &[N~{\sc ii}]          &6.74             &0.758   \\
6678      &He~{\sc i}           &0.281             &0.029    \\
6717      &[S~{\sc ii}]         &0.502             &0.050   \\
6731      &[S~{\sc ii}]         &0.932             &0.091    \\
7065      &He~{\sc i}           &0.457             &0.032   \\
7136      &[Ar~{\sc iii}]       &0.986             &0.065    \\
7319, 20   &[O~{\sc ii}]         &0.350             &0.019    \\
7330, 31   &[O~{\sc ii}]         &0.733             &0.040    \\
\hline
\end{tabular}
\flushleft\small{$\dag$ in the unit of H$_\beta$ = 2.210 \into 10$^{-14}$ ergs
cm$^{-2}$ s$^{-1}$}
\flushleft\small{ $\ddag$ in the unit of  H$_\beta$ = 3.203 \into
10$^{-11}$ ergs cm$^{-2}$ s$^{-1}$}
\end{table}

\begin{figure*}
\centering
\includegraphics[width=13cm]{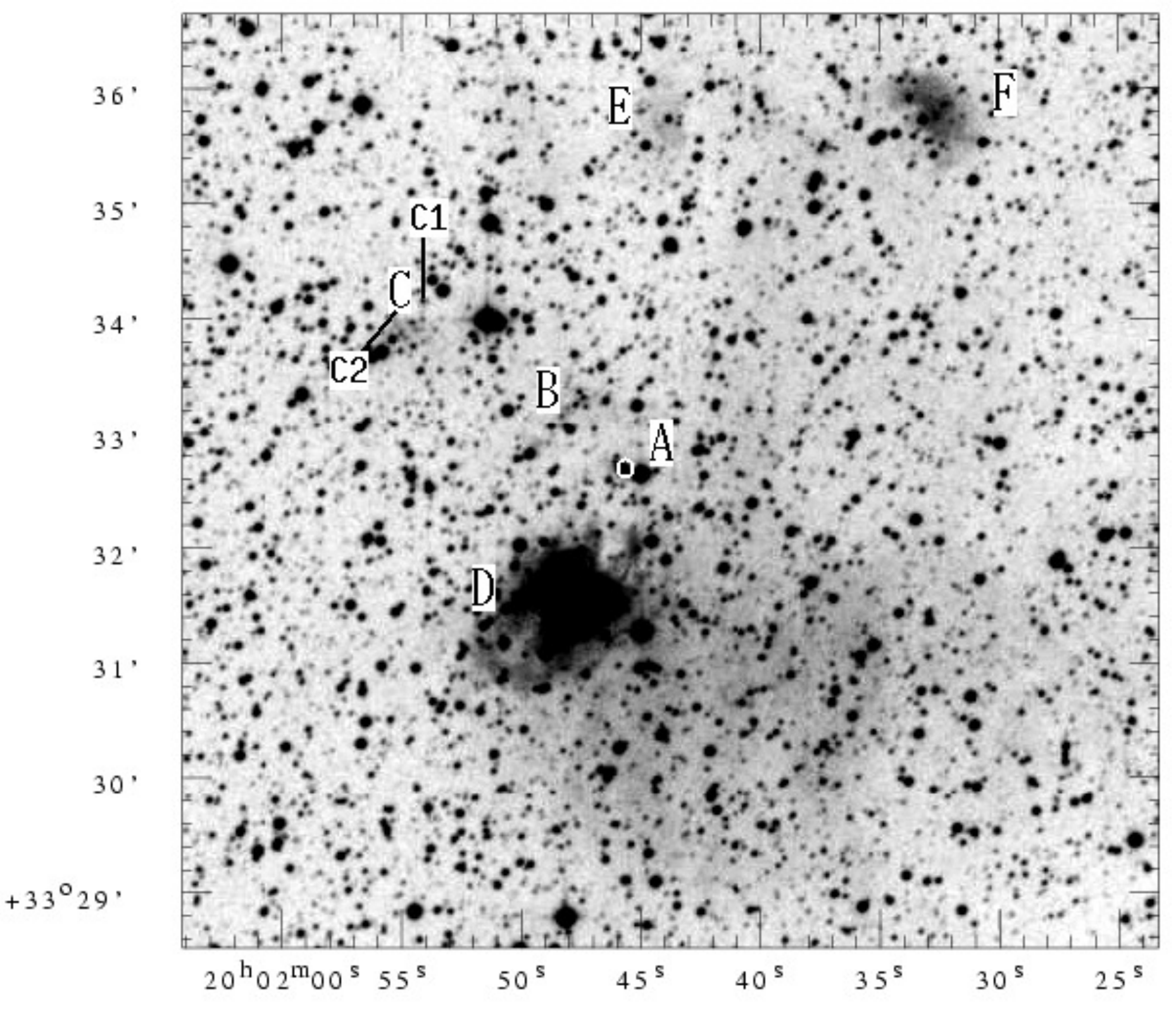}
\caption{POSS II $R$-band image of the region around Sh 2-100. 
The names (A, B, C1, C2, D, E and F) represent  the different
associated \hii regions namely K3-50A, K3-50B, K3-50C1, K3-50C2, K3-50D,
 K3-50E and K3-50F, respectively, following the nomenclature by  Israel (1976).
The RA and DEC coordinates are in J2000.0 epoch.}
\label{fig1}
\end{figure*}

\begin{figure}
\centering
\includegraphics[width=7cm, height=6cm]{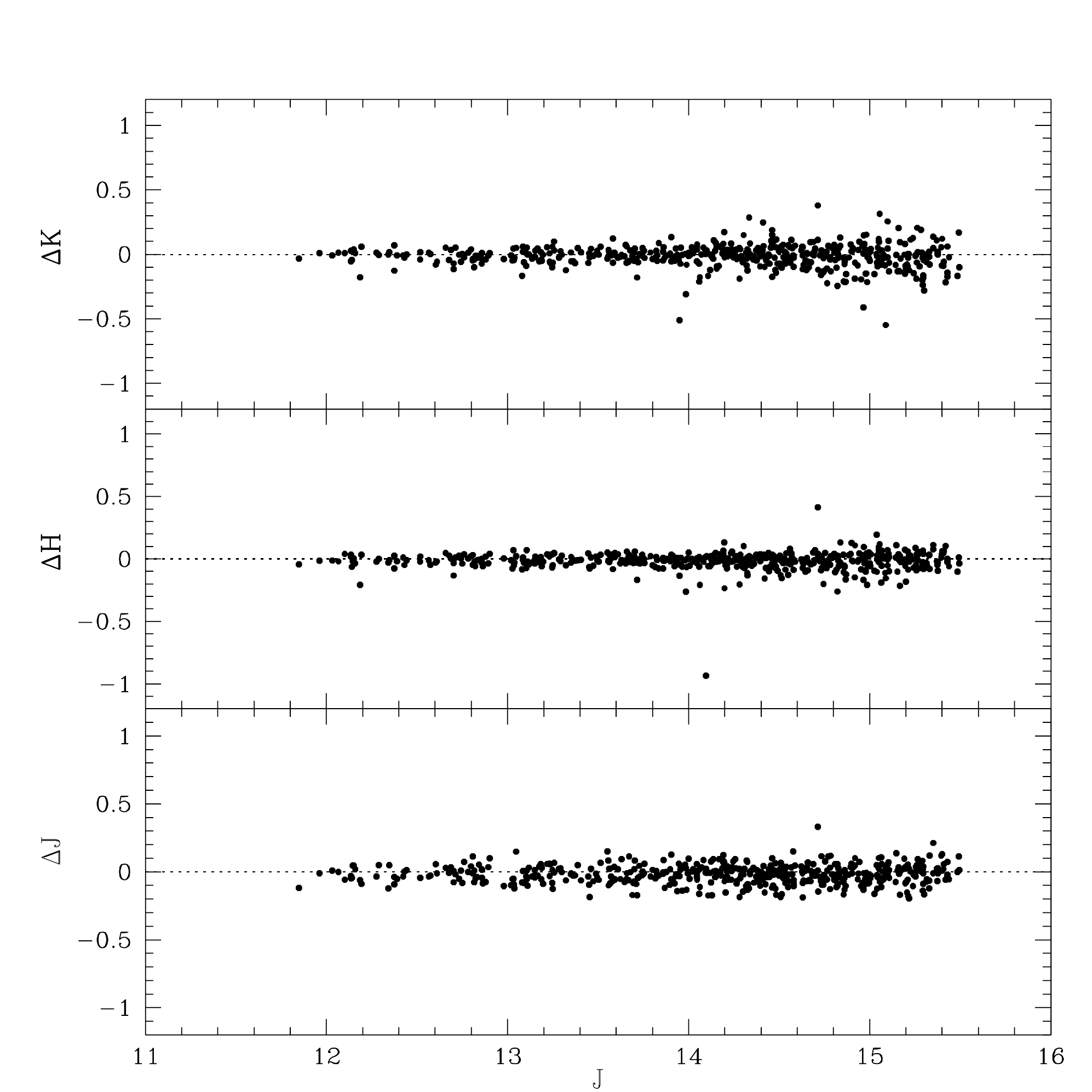}
\caption{Comparison of IRSF/SIRIUS $JHK_{\rm s}$ photometry with the
2MASS data for the common sources. The
difference $\Delta$ (IRSF - 2MASS) in mag is plotted as a function
of $J$ mag.}
\label{fig2}
\end{figure}

\begin{figure*}
\centering
\includegraphics[width= 12 cm ]{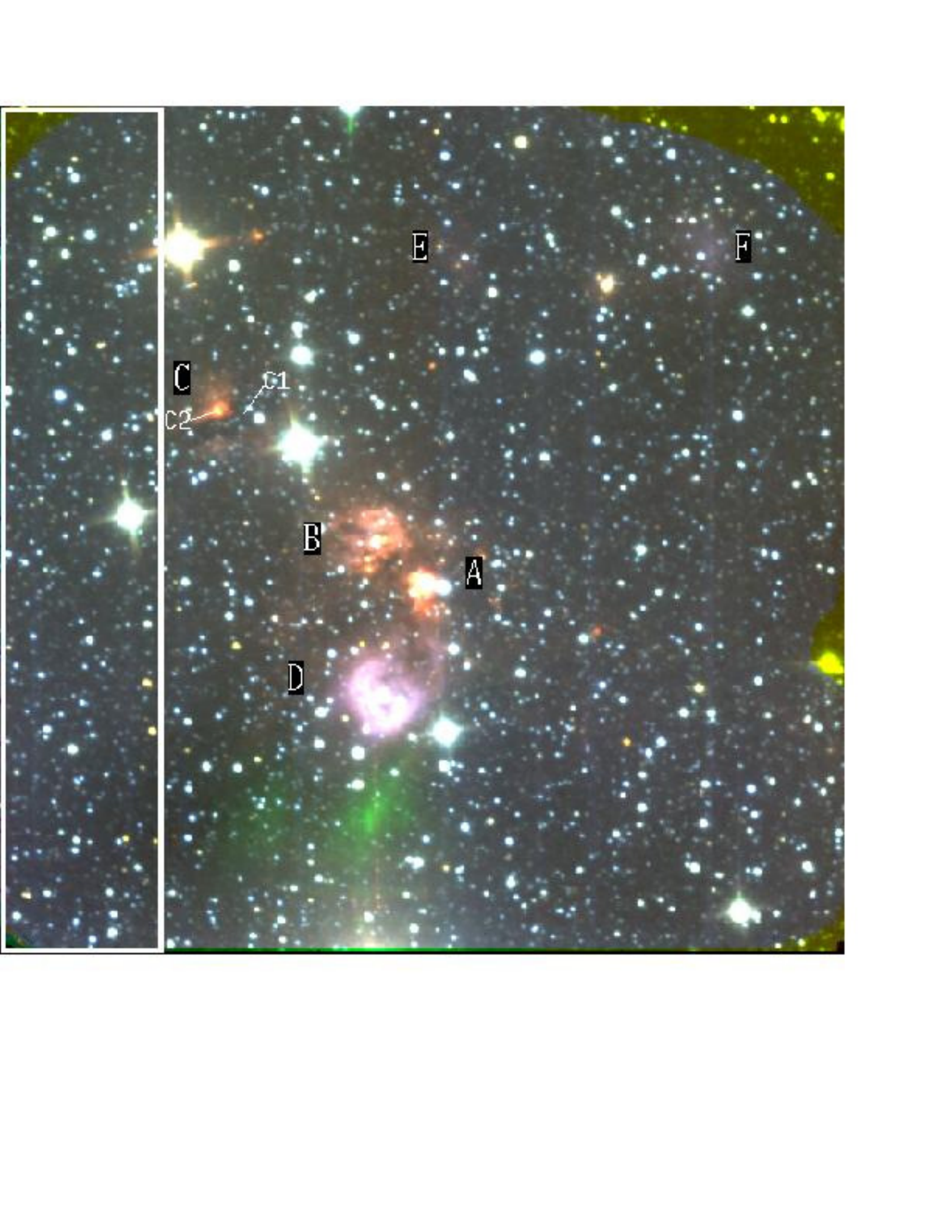}
\caption{\jhk color-composite image of the Sh 2-100 star-forming region ($J$, blue; $H$, green; 
and \ks, red). Locations of the different regions are also marked. The
rectangular box shown by white lines to the left represents the region selected as  relative
control field (see \S 5.1). The enhanced green color seen below the region D is the artifact of the
$H$-band due to a bright star. North is up and east is to the left.}
\label{fig3}
\end{figure*}

\begin{figure*}
\centering
\includegraphics[width= 12 cm]{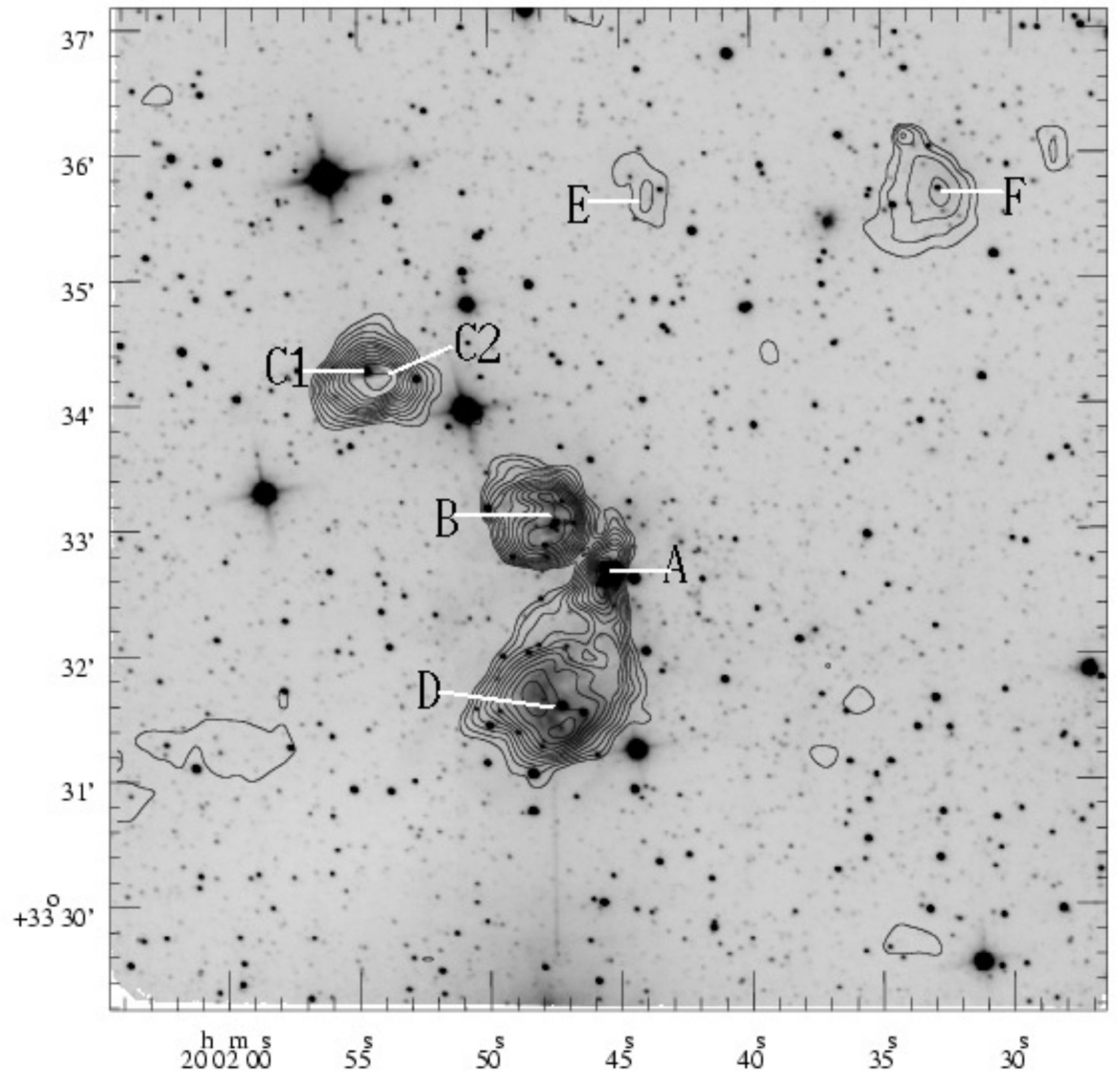}
\caption{IRSF $K_{\rm s}$-band image in logarithmic scale overlaid by
the GMRT radio continuum contours at 610 MHz. The contour levels are at
0.88 \into (3, 5, 8, 12, 17, 22, 28, 35, 43, 51, 70, 81, 93, 102) mJy/beam,
where $\sim$ 0.88 mJy/beam is the rms noise in the map at the resolution
of $\sim$ $5\arcsec.5 \times 4\arcsec.8 $.
Individual \hii regions are also marked in the figure.
The labelled axes are in J2000 coordinates.}
\label{fig4}
\end{figure*}

\begin{figure*}
\centering
\includegraphics[width= 12 cm, height=12 cm]{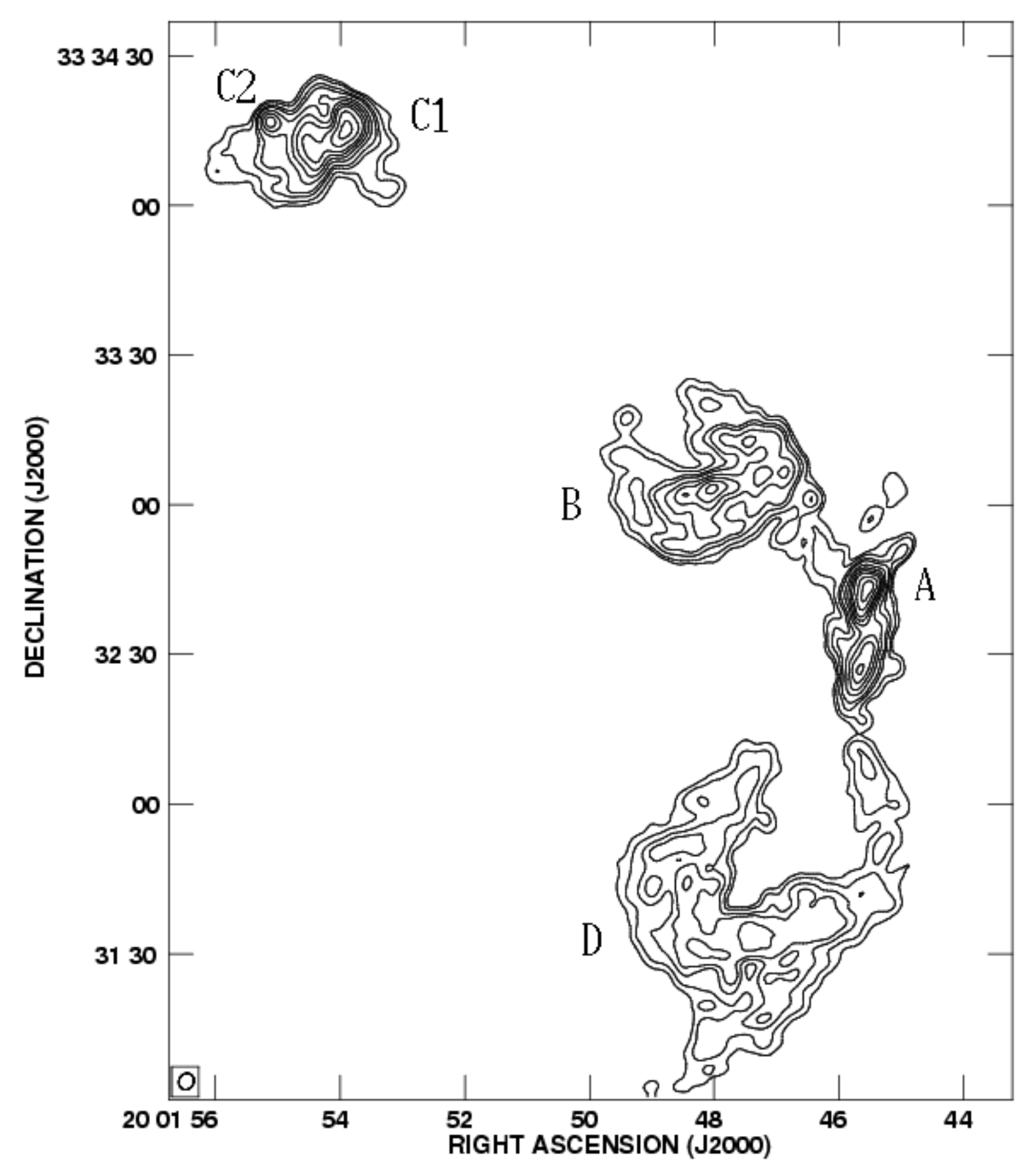}
\caption{High resolution radio continuum map at 1280 MHz
for the region around Sh 2-100 with a resolution
of $\rm 3\arcsec.3 \times 3\arcsec.0$. The contour
levels are at 2.2 \into (3, 4, 6, 8, 12, 16, 20, 24, 30, 36) mJy/beam,
where $\sim$ 2.2 mJy/beam is the rms noise in the map. The labelled
axes are in J2000 coordinates. Individual \hii regions are also marked.}
\label{fig5}
\end{figure*}

\begin{figure*}
\centering{
\includegraphics[angle=00, scale = 0.37, trim = 00 00 00 00, clip]{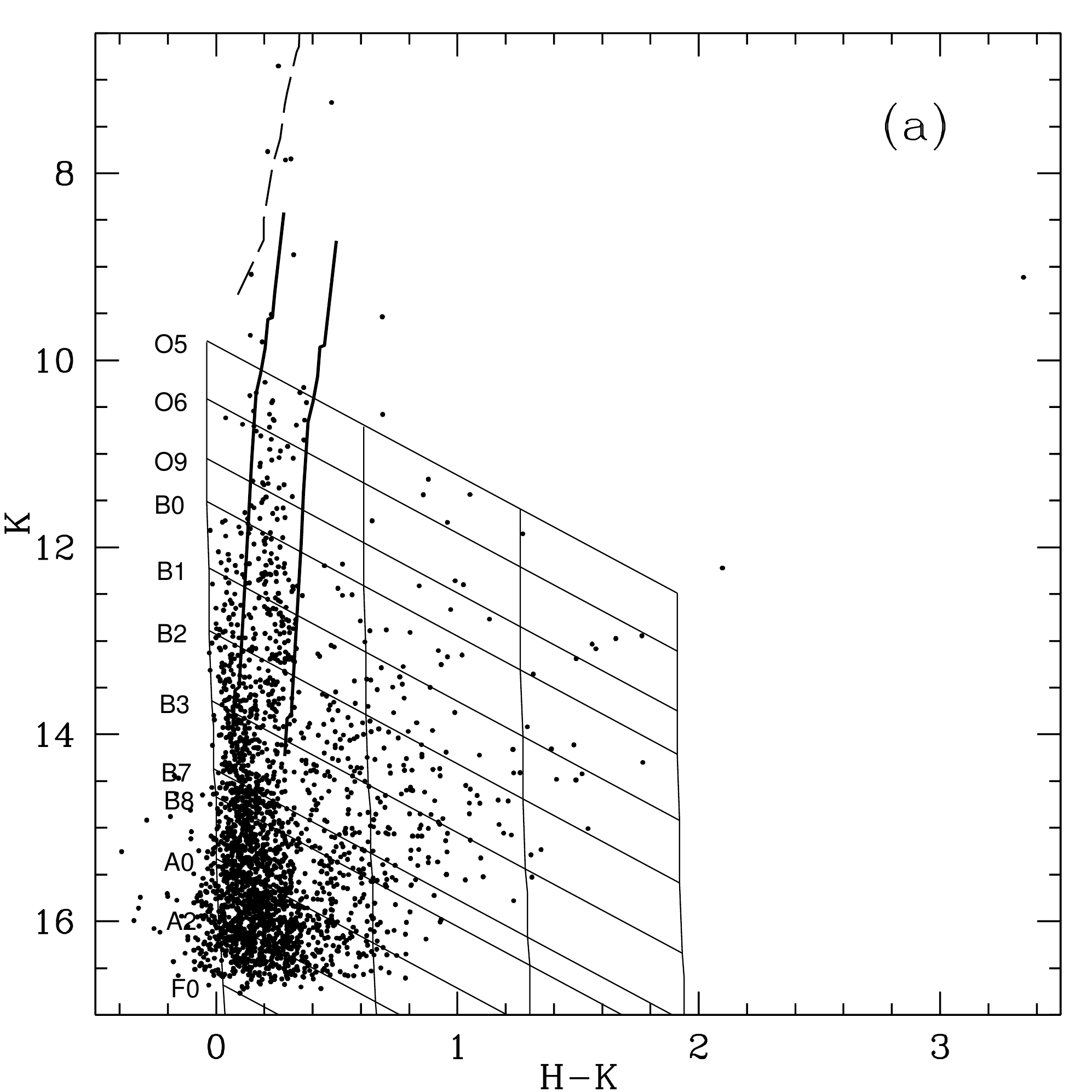}
\includegraphics[angle=00, scale = 0.37, trim = 00 00 00 00, clip]{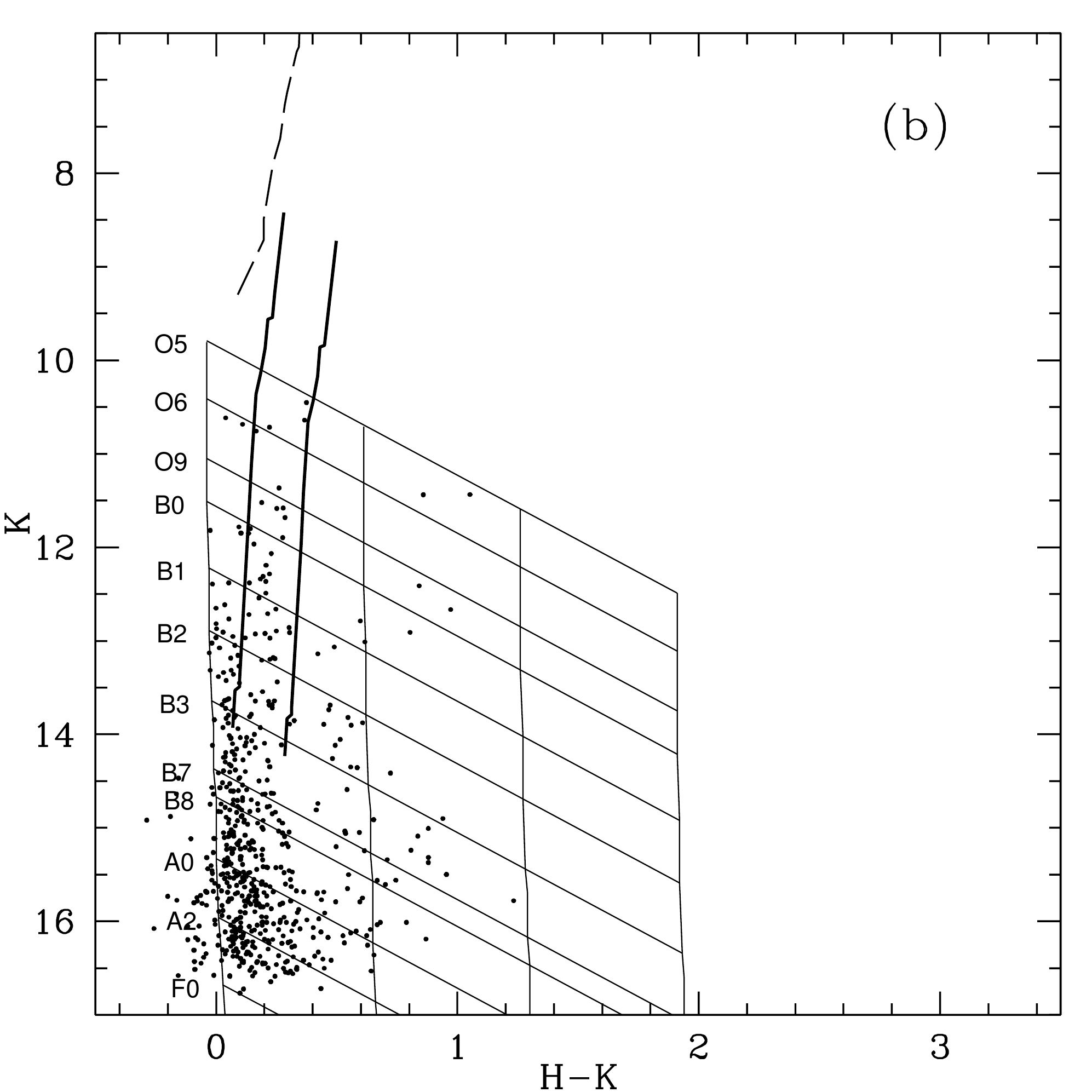}
\includegraphics[angle=00, scale = 0.37, trim = 00 00 00 00, clip]{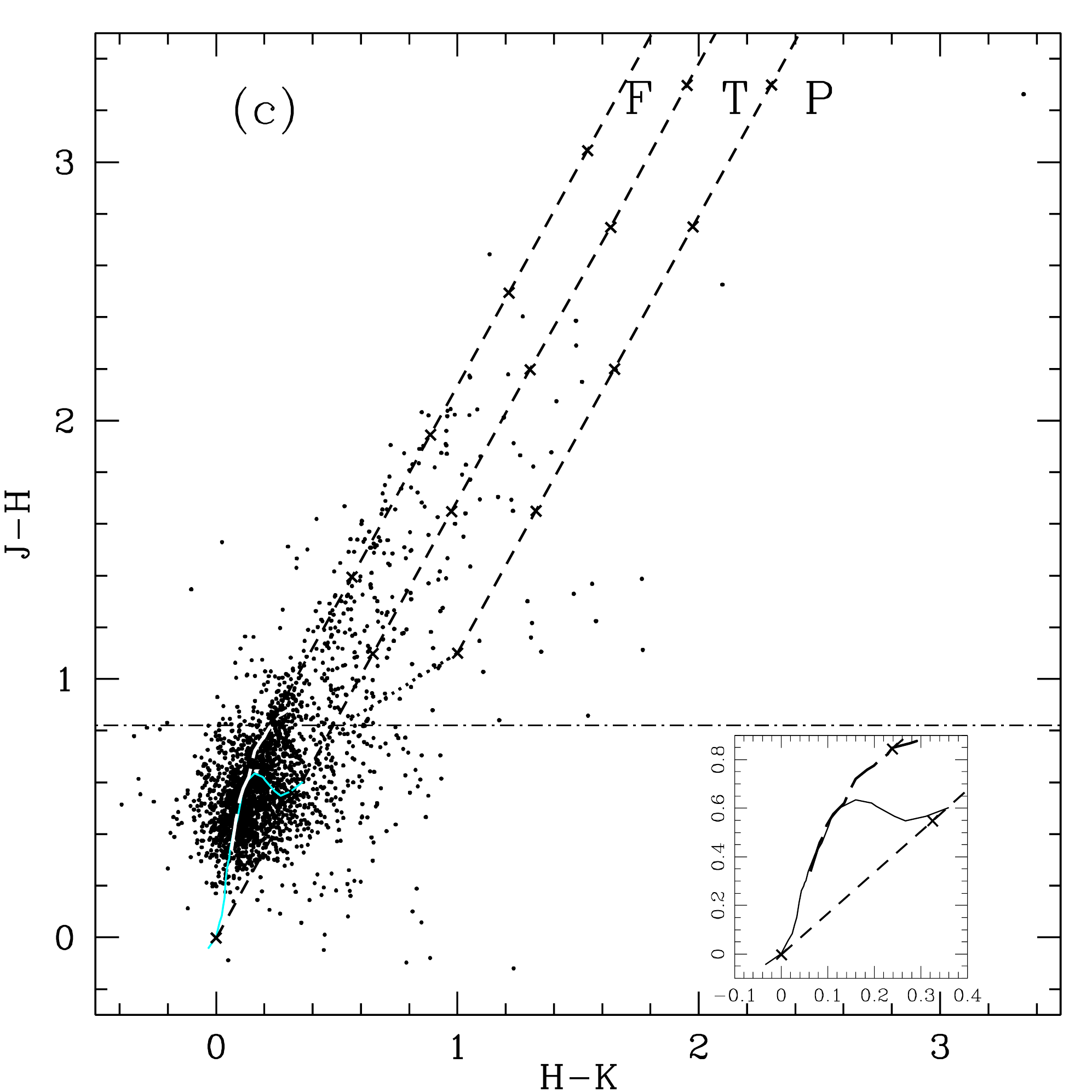}
\includegraphics[angle=00, scale = 0.37, trim = 00 00 00 00, clip]{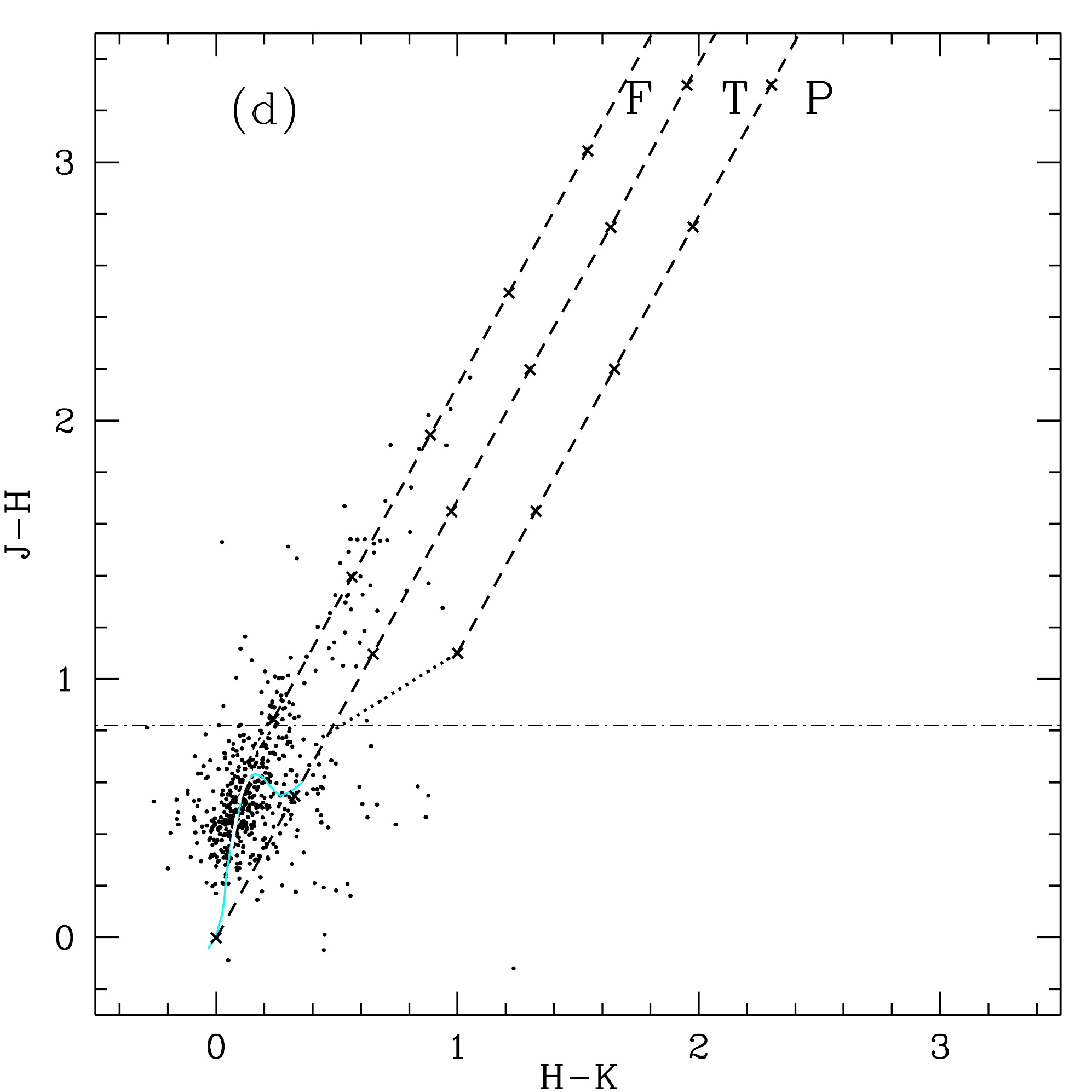}}
\caption{(a) NIR CM diagram of the Sh 2-100 star-forming region, showing
the ZAMS locus (thin vertical solid
line) reddened by \av = 0, 10, 20, and 30 mag, respectively, and supergiant
locus (thick dashed line) at zero reddening. The
 thick solid lines represent the loci of giants reddened by
\av = 0 and \av  = 3.3  mag, respectively. All the loci are drawn at the
distance of 8.7 kpc. Slanting horizontal lines represent the standard
reddening vector with a length of \av = 30 mag, drawn from the MS locus
corresponding to different spectral types. (b) NIR CM diagram for the relative control region. (c) NIR CC diagram of the 
Sh 2-100 region for the sources
detected in the $JHK_{\rm s}$ bands. The continuous
line marks the locus of  MS and the thick dashed line is the locus of giant
stars (see inset box). The three parallel dashed lines represent the reddening vectors with
crosses representing a visual extinction of \av = 5 mag. The locus of
CTTS is also shown by a dotted line.
The horizontal dashed-dotted line
represents  the $J-H$  color cut-off to distinguish NIR-excess stars
from probable other kind of sources (see the text). 
(d) NIR CC diagram for the relative control region.}
\label{fig6}
\end{figure*}

\begin{figure*}
\centering
\includegraphics[width=15 cm]{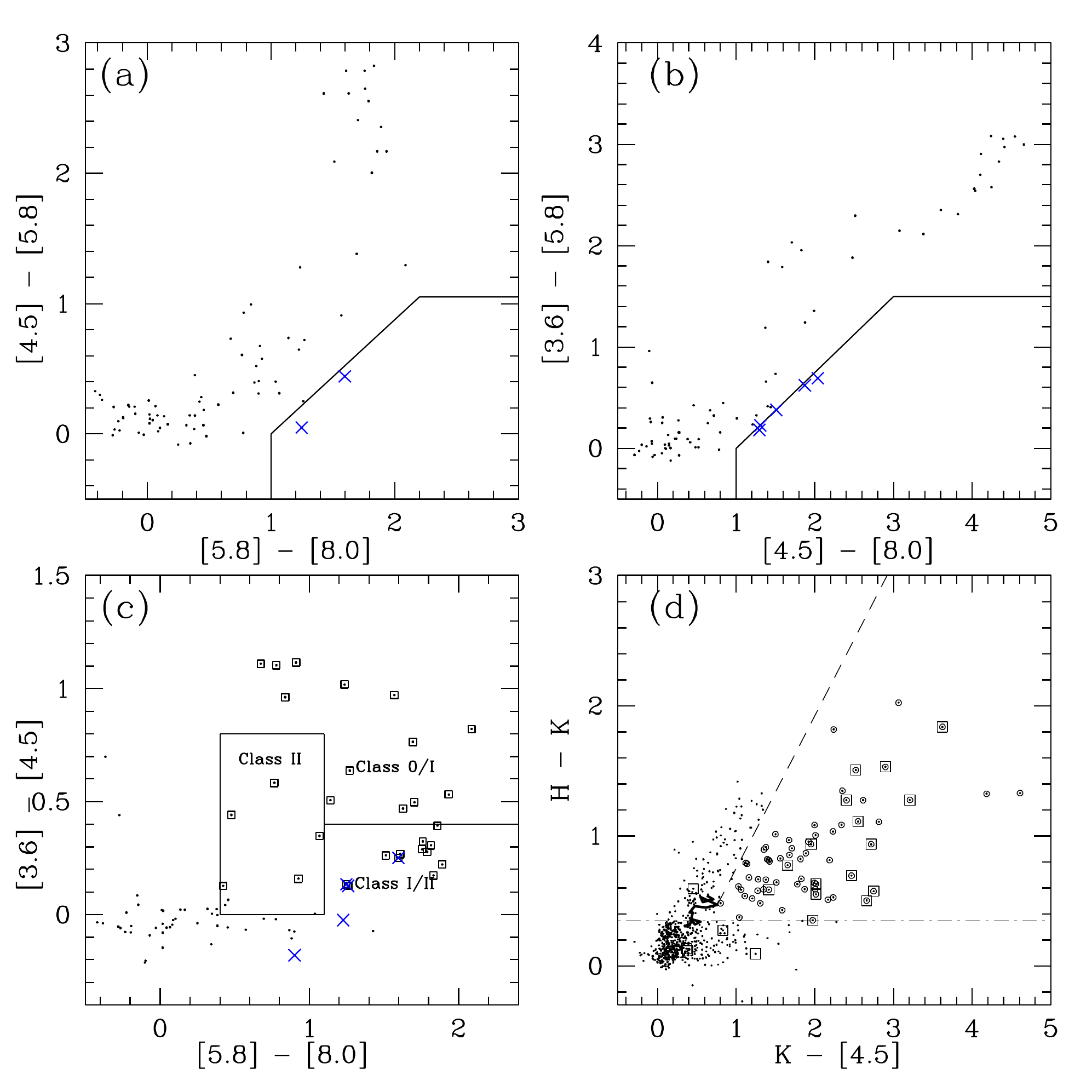}
\caption{ (a) \& (b).  IRAC CC diagrams to identify star-forming 
galaxies, using the method described in Gutermuth et al. (2008). The zones
where these extragalactic sources are located in the CC diagrams are 
marked with solid lines. 
(c). IRAC CC diagram to identify YSOs. The different zones indicate 
the locations of Class II, Class I and Class 0 sources from Allen et al. (2004). 
(d). The $H-K$/$K-[4.5]$ CC diagram. 
The curved line is the MS locus of  late M type dwarfs 
(Patten et al. 2006). The long-dashed line corresponds to the 
reddening vector (from Flaherty et al. 2007). The dashed-dotted horizontal 
line represents the $H-K$ color cut off for the selection of YSOs 
(see the text).}
\label{fig7}
\end{figure*}

\begin{figure*}
\centering
\qquad{
\includegraphics[width=7.5 cm]{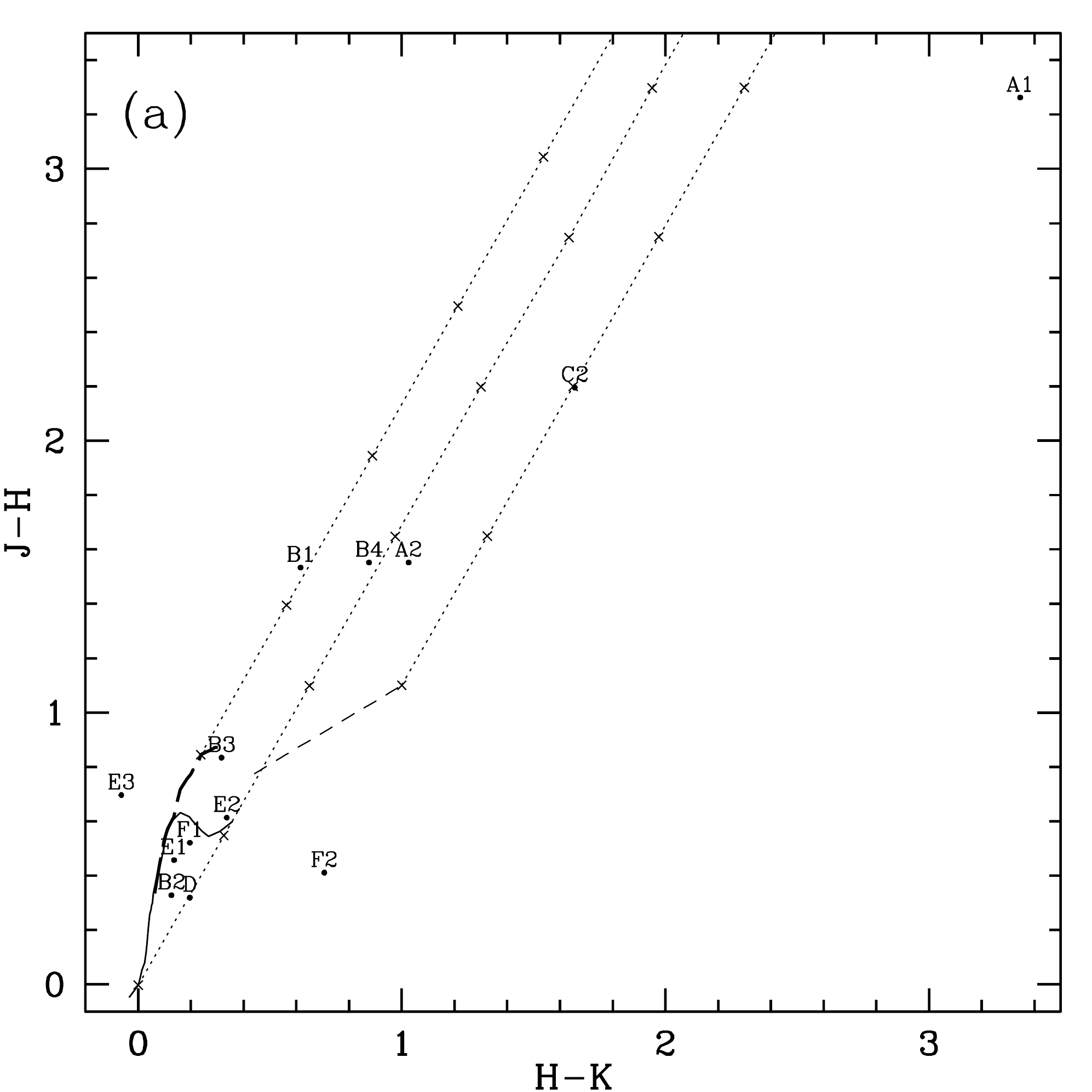}
\includegraphics[width=7.5 cm]{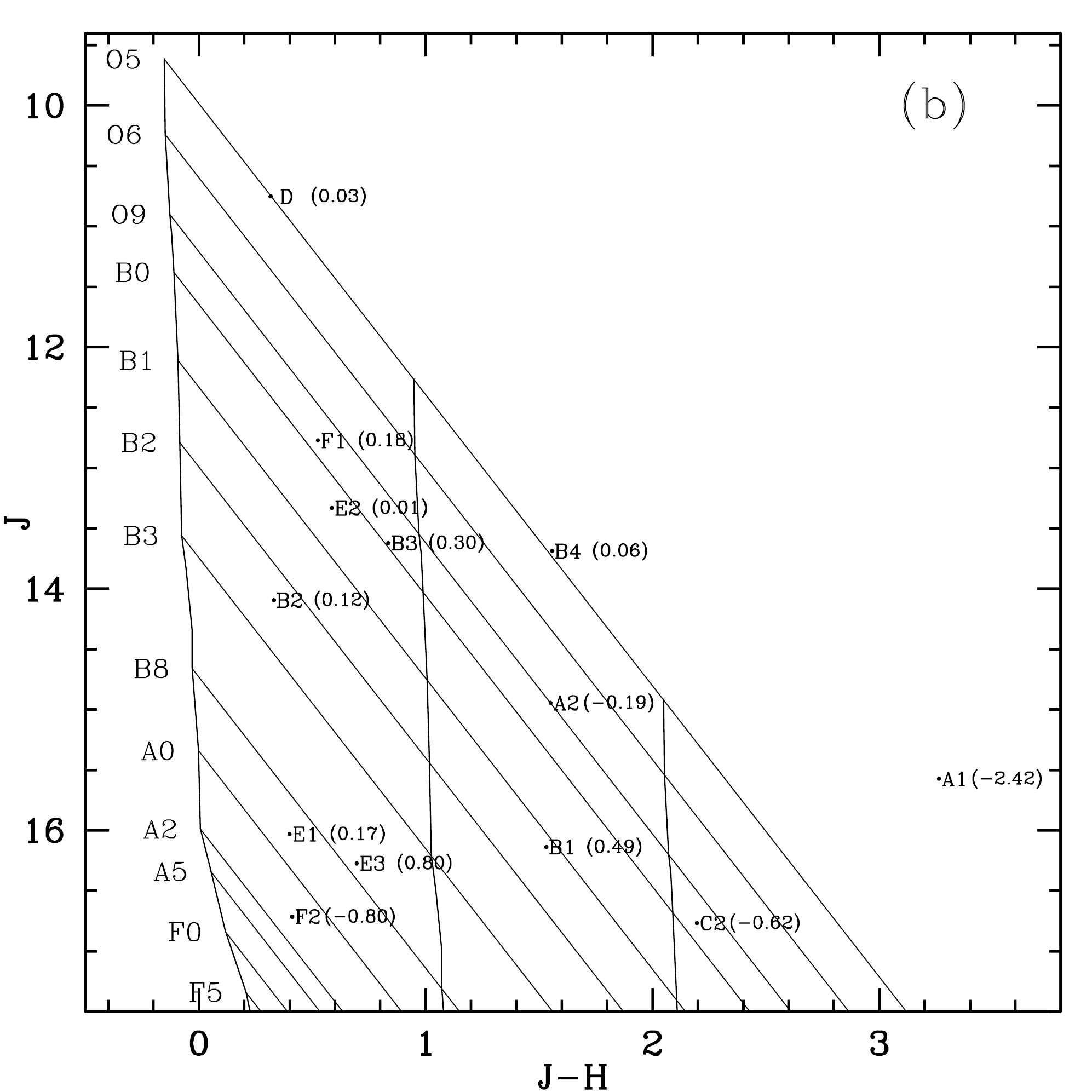}}
\caption{(a) CC diagram  for the sources within 0.5 pc radius of the \hii regions. 
The giant, dwarf, CTT loci and the reddening vectors are plotted in a similar 
manner as in Fig. 6$c$.  (b) $J-H$ vs. $J$ CM diagram for the same sources with 
their $Q$ values. The nearly vertical solid lines represent the ZAMS loci at a distance of
8.7 kpc reddened by \av = 0, 10  and 20 mag. The slanting lines show the
reddening vectors corresponding to  ZAMS spectral type.}
\label{fig8}
\end{figure*}

\begin{figure*}
   \centering
   \includegraphics[width=12 cm]{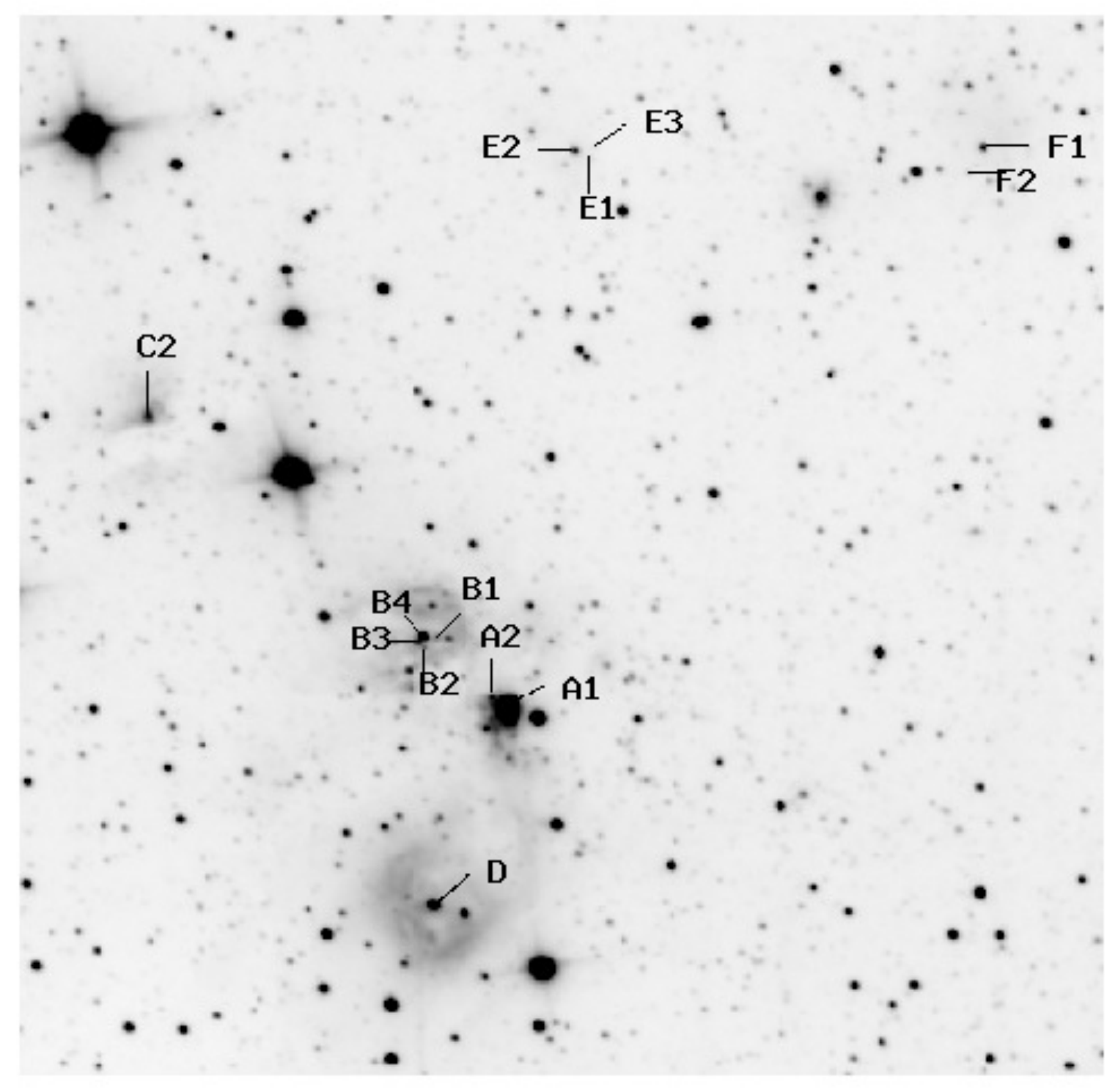}
      \caption{Spatial distribution of the sources found within 0.5 pc radius 
of each \hii region superposed on the \ksb image. The sources are labeled 
with numbers corresponding to the \hii regions.
North is up and east is to the left.}
\label{fig9}
\end{figure*}

\clearpage
\begin{figure*}
   \centering
   \includegraphics[width=10 cm]{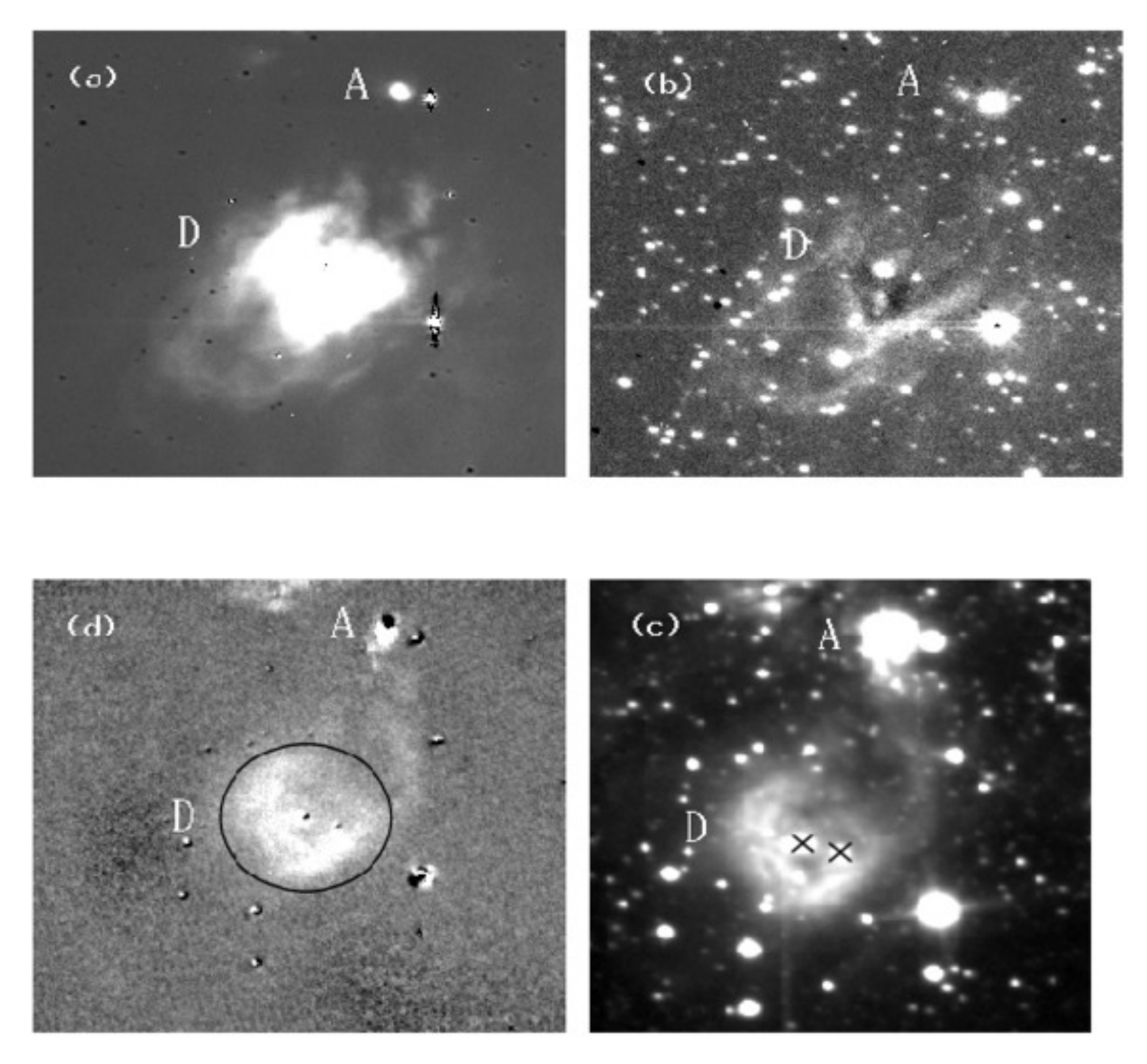}
      \caption{ Images of K3-50D and K3-50A (clockwise from the top left panel) :
 (a) The continuum-subtracted image of  H$\alpha$ (6563 \AA) + \nii (6548 \AA~ and 6584 \AA), 
(b) Bias and flat-field corrected image of \sii at 6717\AA + 6731 \AA~
(c) IRSF $K_{\rm s}$-band image with two bright sources  (marked with crosses)
 inside the semi-circular shell 
(d) Continuum-subtracted Br$\gamma$ image, marked
with a circle of $\sim$ 50$^{\prime\prime}$ radius around central
bright star.}
\label{fig10}
\end{figure*}

\begin{figure*}
   \centering
   \includegraphics[width=8 cm, height=8 cm, trim = 40 00 40 00]{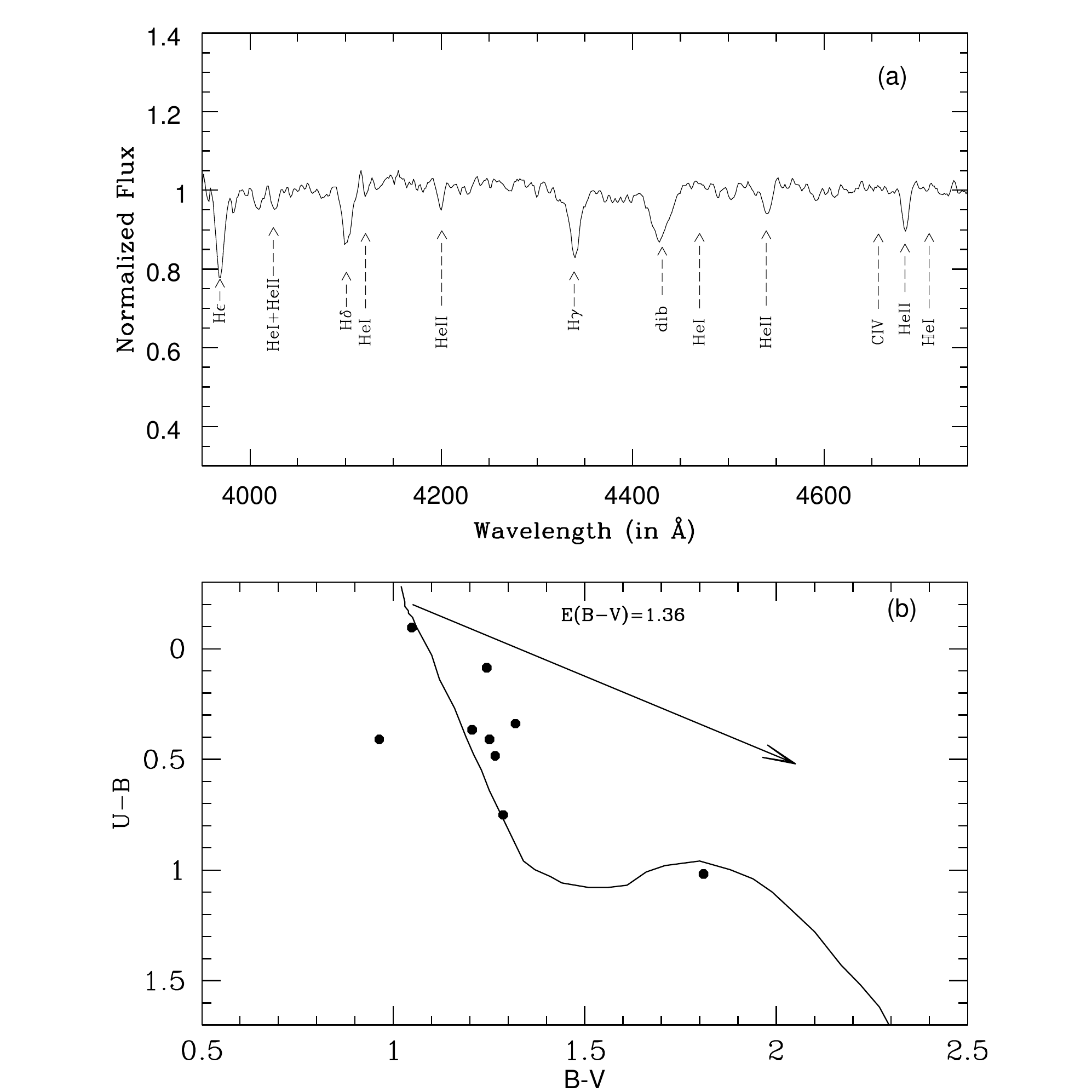}
      \caption{(a) The blue part of the spectrum of the most luminous  star 
(V $\sim$ 13.31 mag) of K3-50D, displaying characteristics similar to 
$\sim$ O4  MS member. (b) The $(U-B)/(B-V)$ CC diagram for the stars
lying within 50$^{\prime\prime}$ of K3-50D region. The continuous curve represents
intrinsic ZAMS by Schmidt-Kaler (1982), shifted along the reddening vector
of 0.72 slope for $E(B-V)_{min}$ = 1.36 mag.}
         \label{fig11}
   \end{figure*}

\begin{figure*}
   \centering
   \includegraphics[width=16 cm, height=6.5 cm]{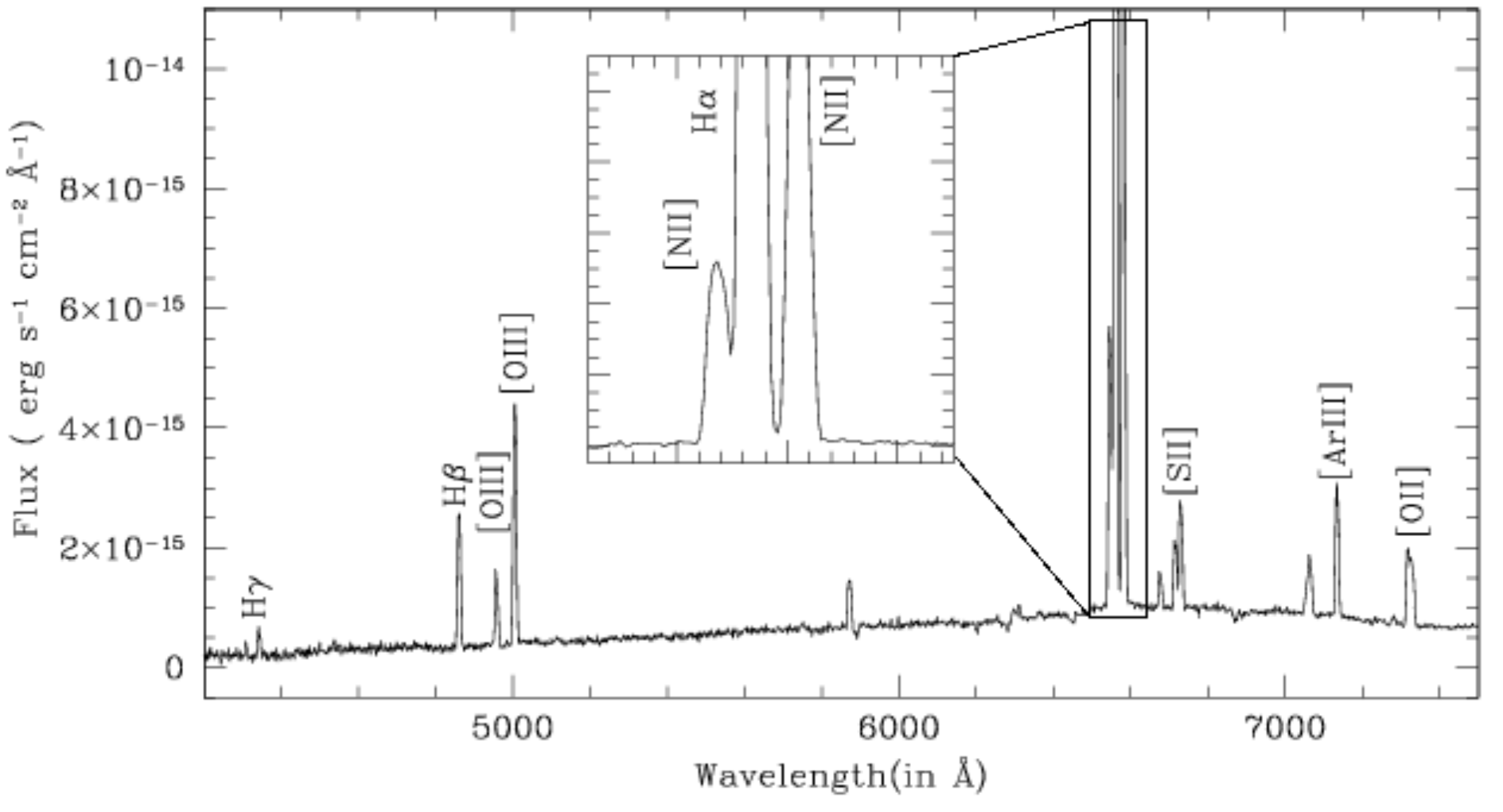}
      \caption{The observed flux-calibrated optical spectrum of the compact nebula associated with K3-50A.}
         \label{fig12}
   \end{figure*}

\begin{figure}
\centering
\includegraphics[width=7.5 cm, height=7cm]{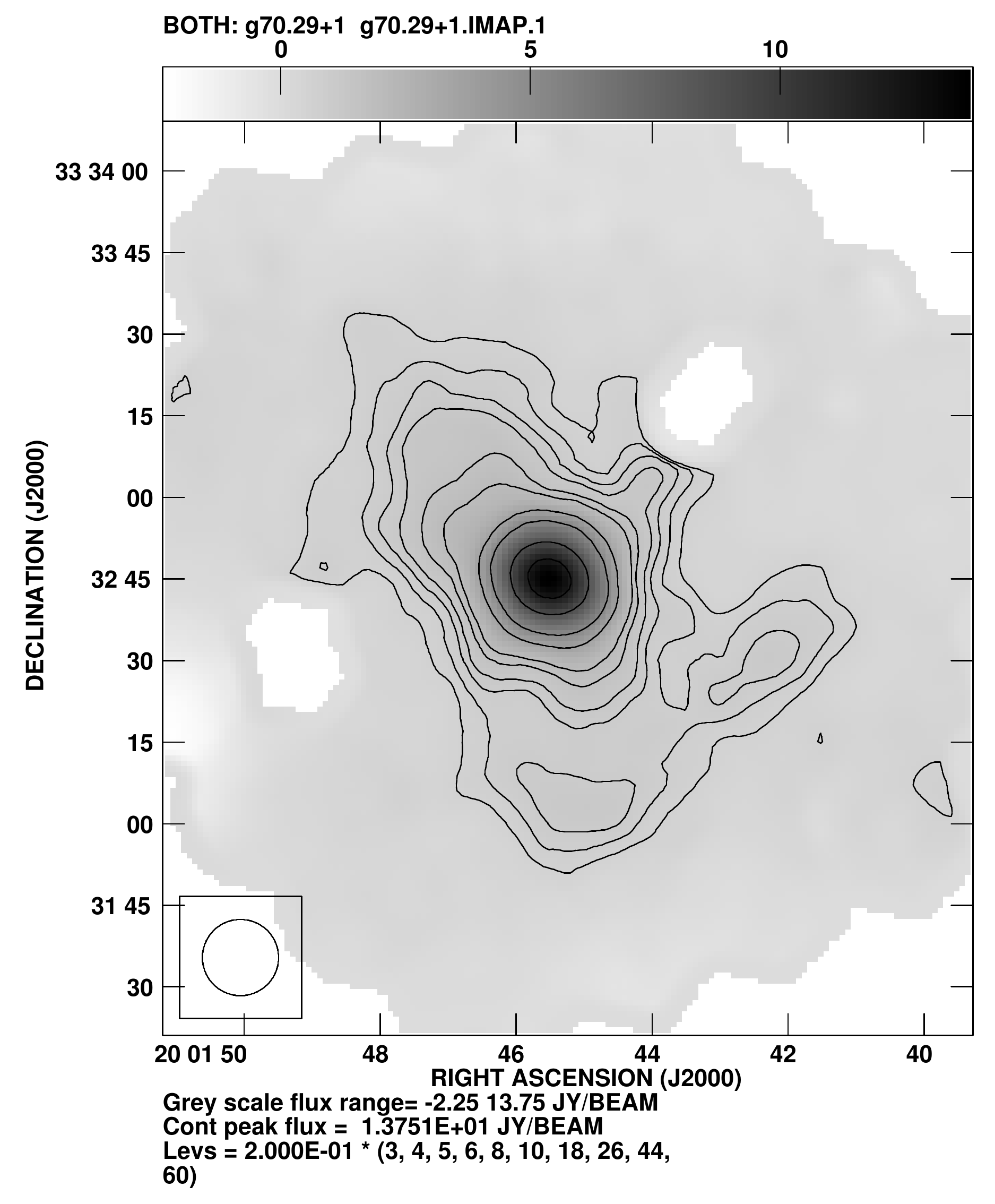}
\caption{Contour map of  spatial distribution of cold dust emission at
 850 $\mu$m around K3-50A region. The contour levels are drawn above 3$\sigma$,
  where $\sigma$ is the rms noise ($\sim$ 2 mJy) in the map and  the  FWHM of
  the symmetric 2-D Gaussian beam is $\rm 14\arcsec.0$. The labelled axes are in J2000 coordinates.}
\label{fig13}
\end{figure}

\begin{figure}
\centering
\includegraphics[width=8 cm]{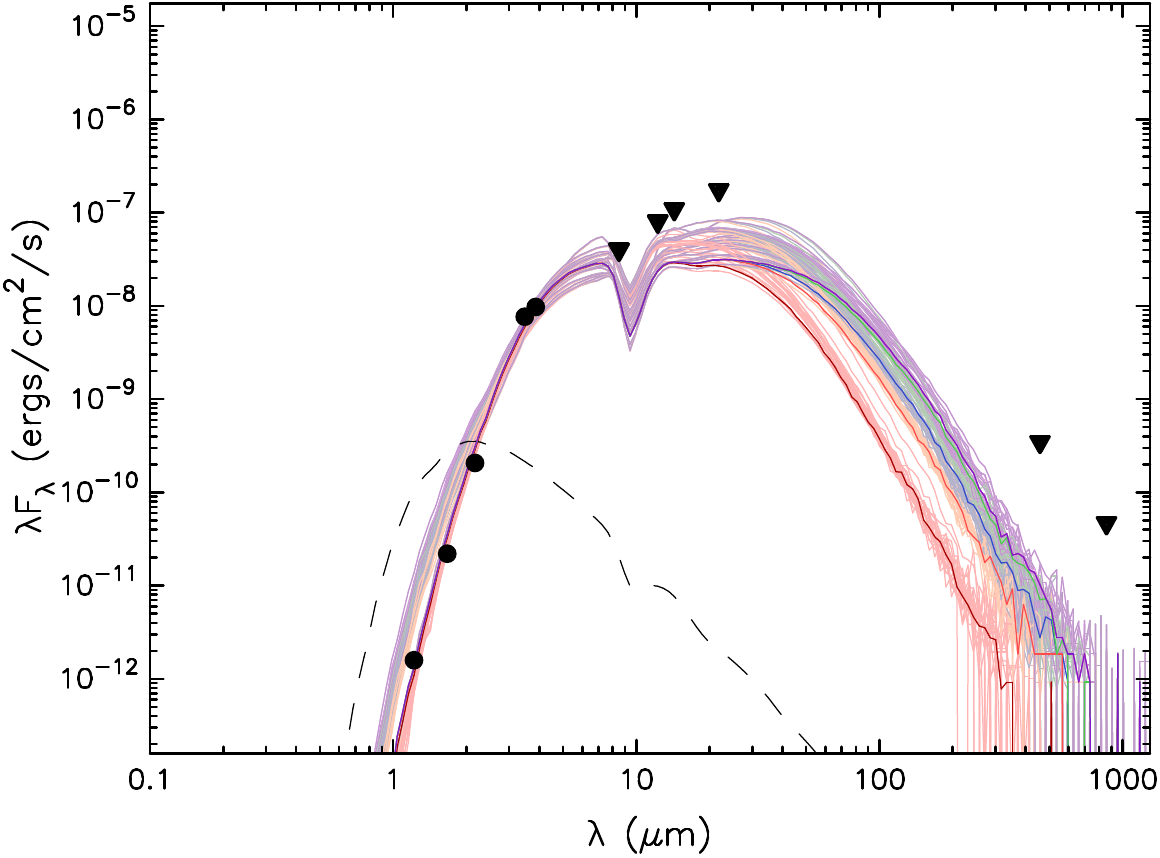}
\caption{SED for the luminous source (A1) at the center of K3-50A region. The filled circles represent
flux values while the triangles denote upper limits of the flux values. The solid lines show
all the models that fit the data reasonably with ${\chi}^2 - {\chi_{best}}^2 < 3$. The
dashed line shows the SED of the stellar photosphere in the best-fitting model. }
\label{fig14}
\end{figure}

\begin{figure}
   \centering
   \includegraphics[width=8 cm, height = 6 cm, trim = 40 50 40 130]{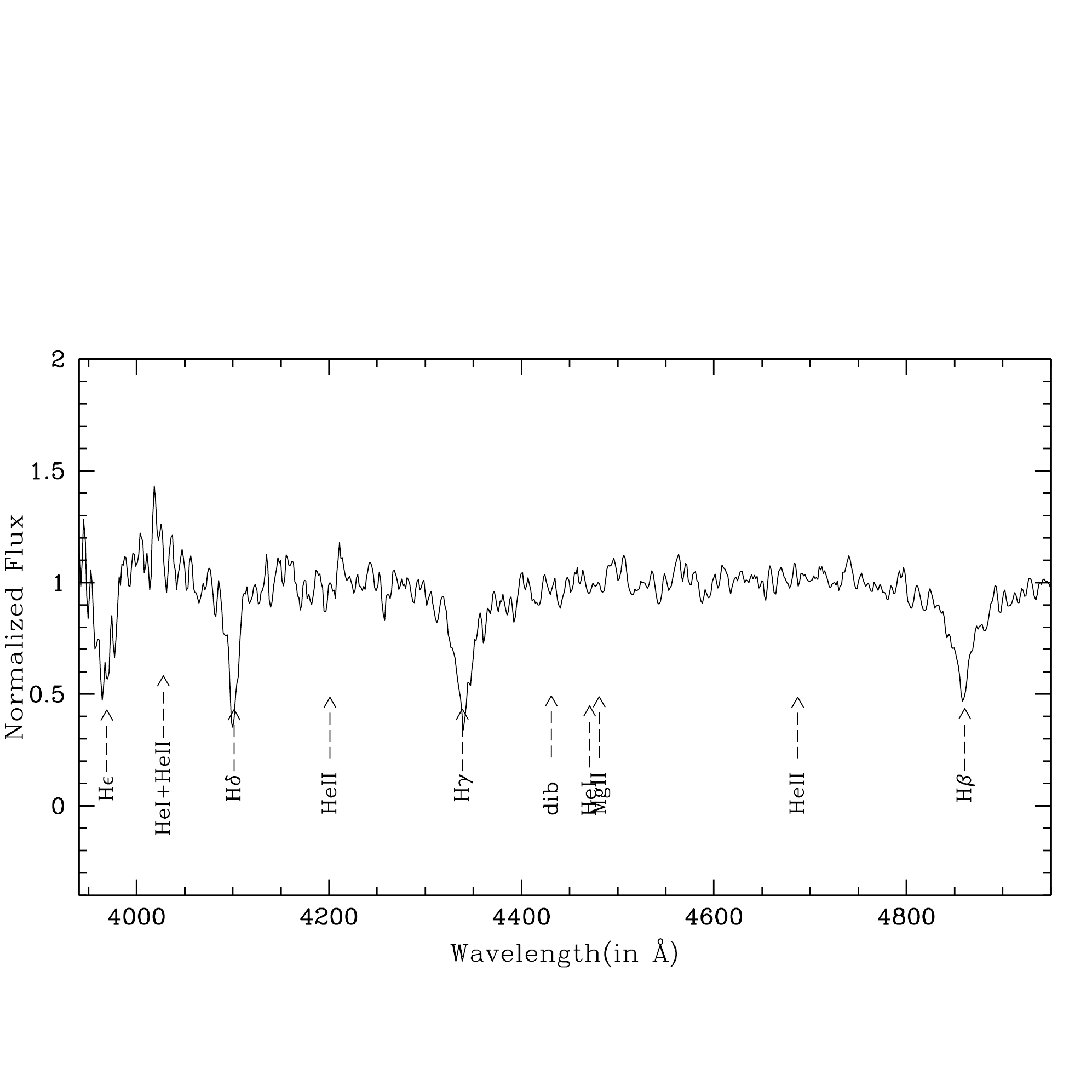}
      \caption{The blue part of the optical spectrum of the star B3 of K3-50B region.}
         \label{fig15}
   \end{figure}

\begin{figure}
\centering
\includegraphics[width=8 cm]{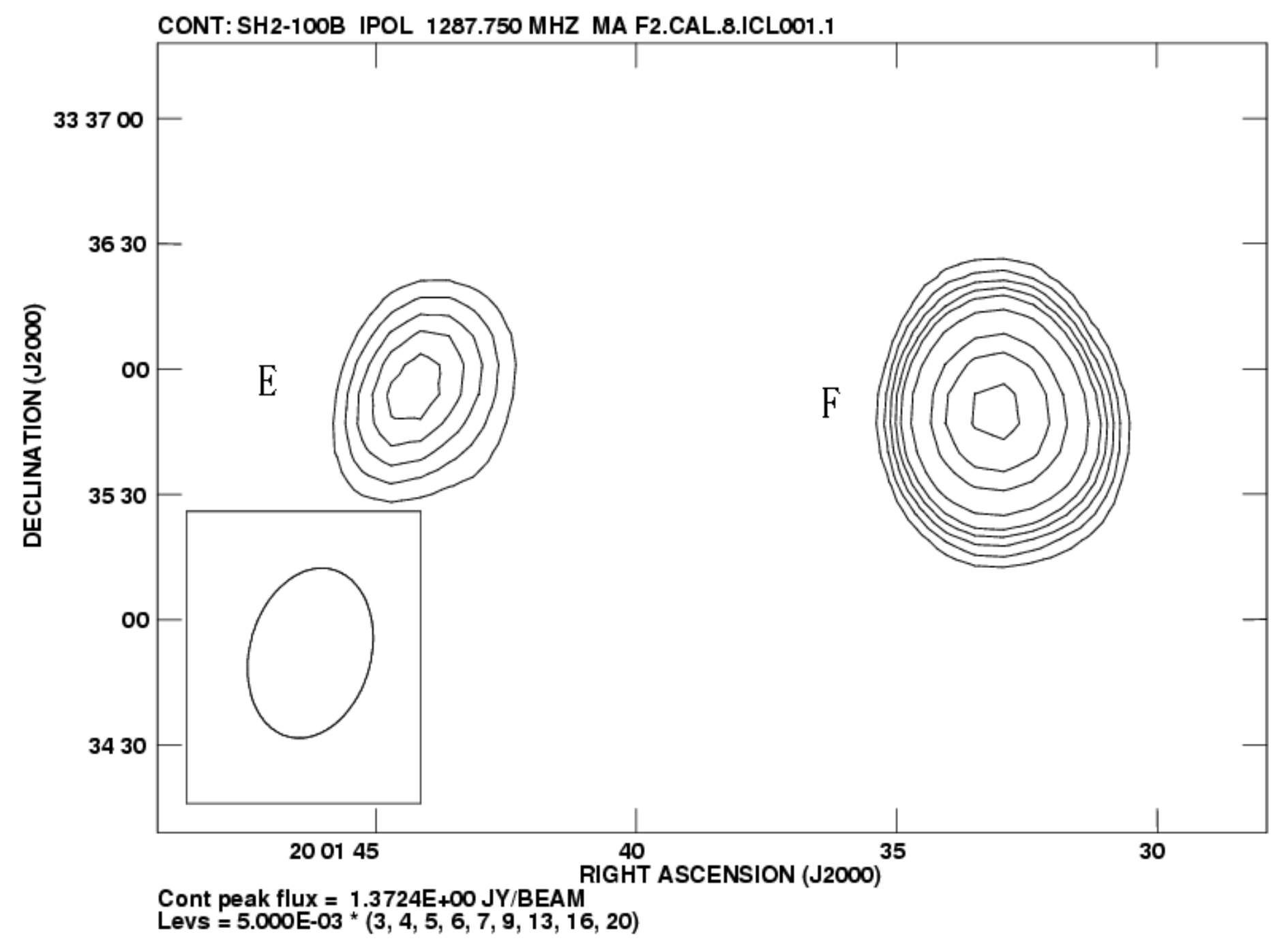}
\caption{Low resolution ($\sim$ $\rm 44\arcsec \times 27\arcsec$) radio continuum map at 1280 MHz
for the regions K3-50E and K3-50F. The contour
levels are at 5.0 \into (3, 4, 5, 6, 7, 9, 13, 16, 20)  mJy/beam,
where $\sim$ 5.0 mJy/beam is the rms noise in the map. The labelled axes are in J2000 coordinates.}
\label{fig16}
\end{figure}

\begin{figure*}
   \centering
   \includegraphics[width=15cm, height= 18 cm]{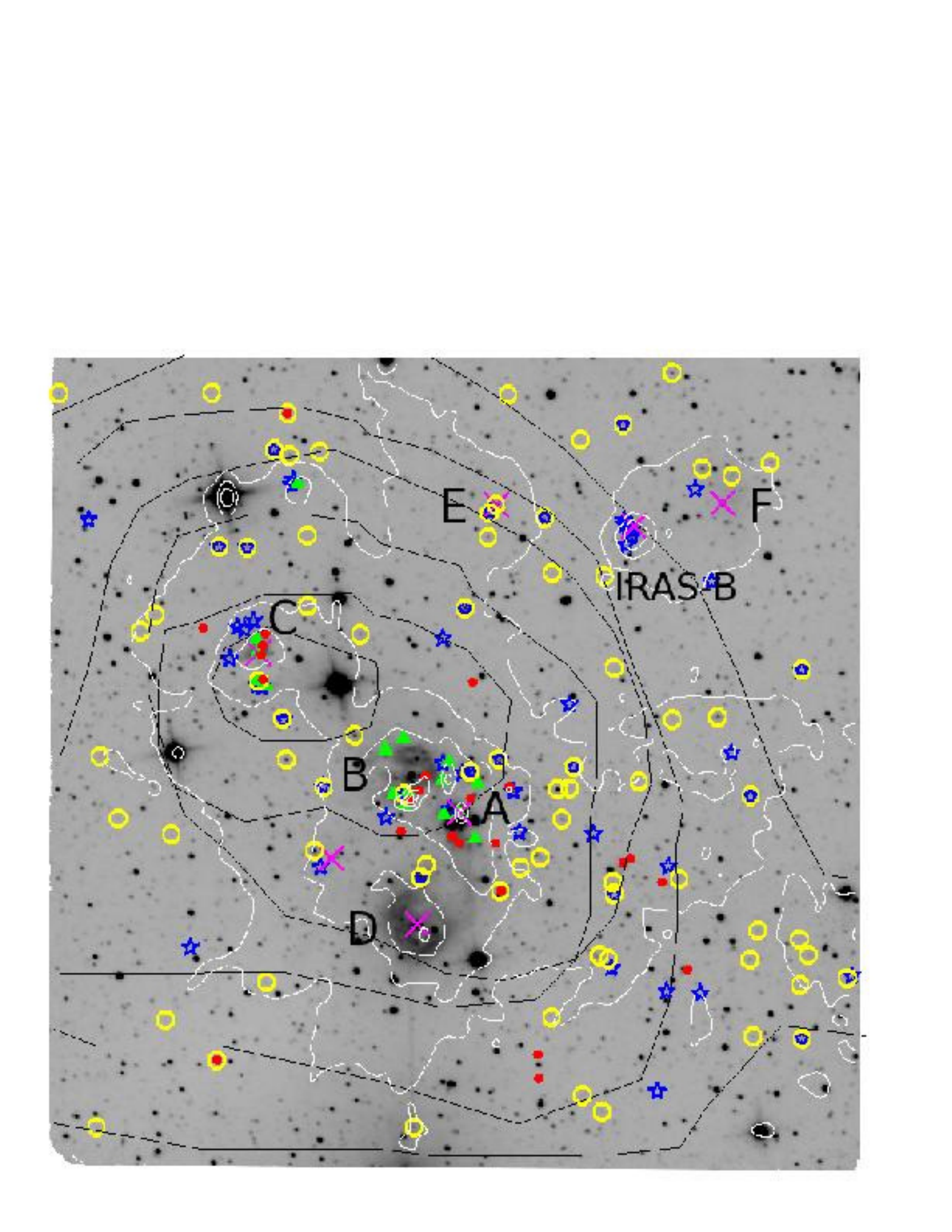}
      \caption{
             The  spatial distributions of Class II (asterisks), Class I
         (filled triangles), sources  detected only in $H$ and $K_s$ bands
         with $H-K$ $\geq$ 1.8  (filled circles), YSOs detected from
         $H-K/K-[4.5]$ CC diagram (open circles) and MSX point sources (crosses), overlaid on \ksb image
               (for color plot see online electronic version).
         The thin contours represent  $^{12}$CO emission from Israel (1980)
          at a resolution of 2$^{\prime}$.3 integrated over a velocity range of
          13 km s$^{-1}$. The thin white contours represent 8 $\mu$m emission
          from {\it Spitzer}. The contours start at 54 MJy sr$^{-1}$
         ($\sim$ 3$\sigma$, where $\sigma$ is the rms noise of the map)          and the innermost contour corresponds to a value of 450 MJy sr$^{-1}$.
         North is up and east is to the left.}
         \label{fig17}
   \end{figure*}

\begin{figure}
   \centering
   \includegraphics[width=8 cm, height=7cm]{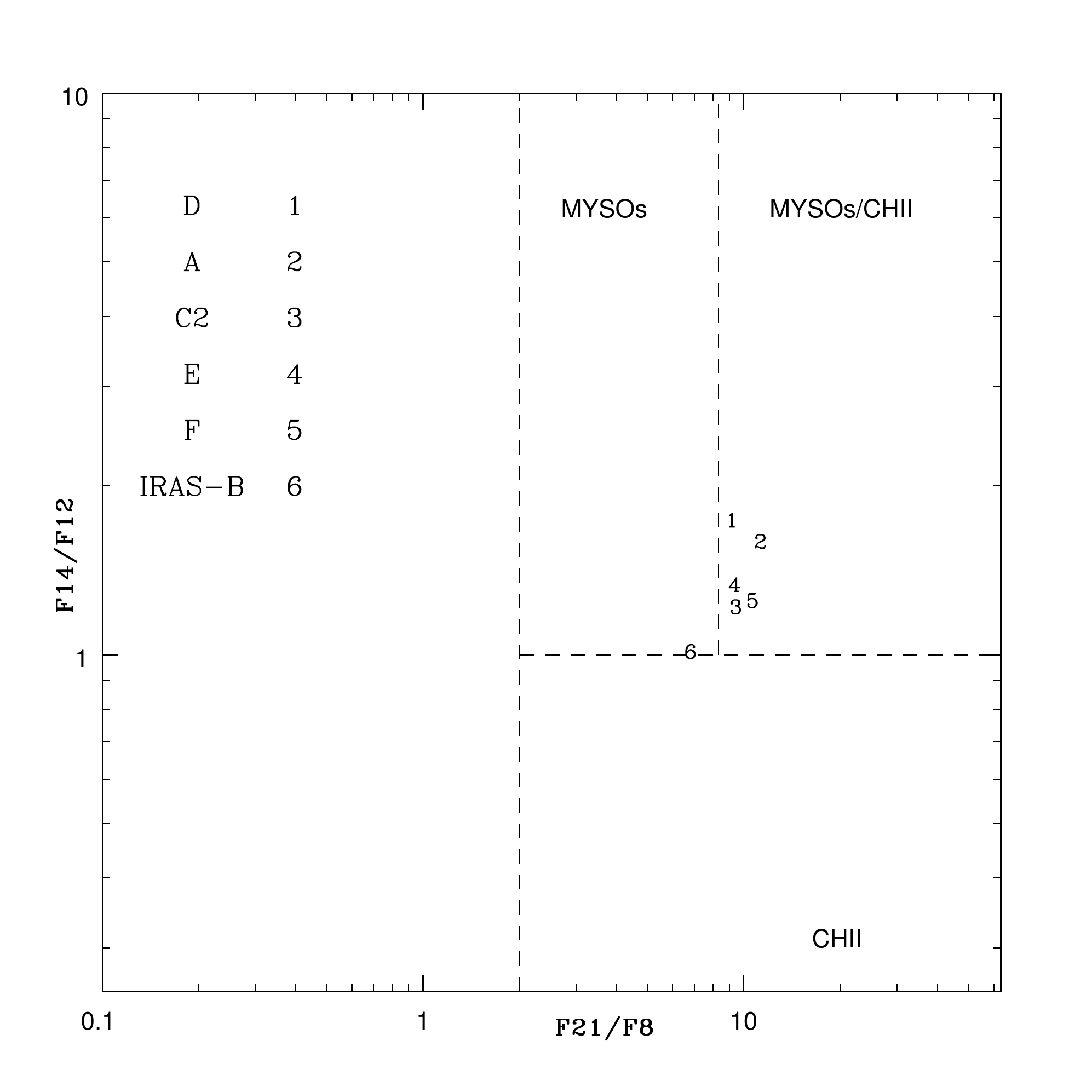}
      \caption{CC plot for the MSX point sources around Sh 2-100 region. The limits
          derived by Lumsden et al. (2002) for massive YSOs and \hii regions (dashed lines) are          
mentioned by their areas. The numbers represent the MSX  point sources
          associated with
         different \hii regions and are mentioned in the upper left corner of the figure.}
         \label{fig18}
   \end{figure}

 \begin{figure}
   \centering
   \includegraphics[width=8 cm, height=7 cm, trim = 00 30 00 90]{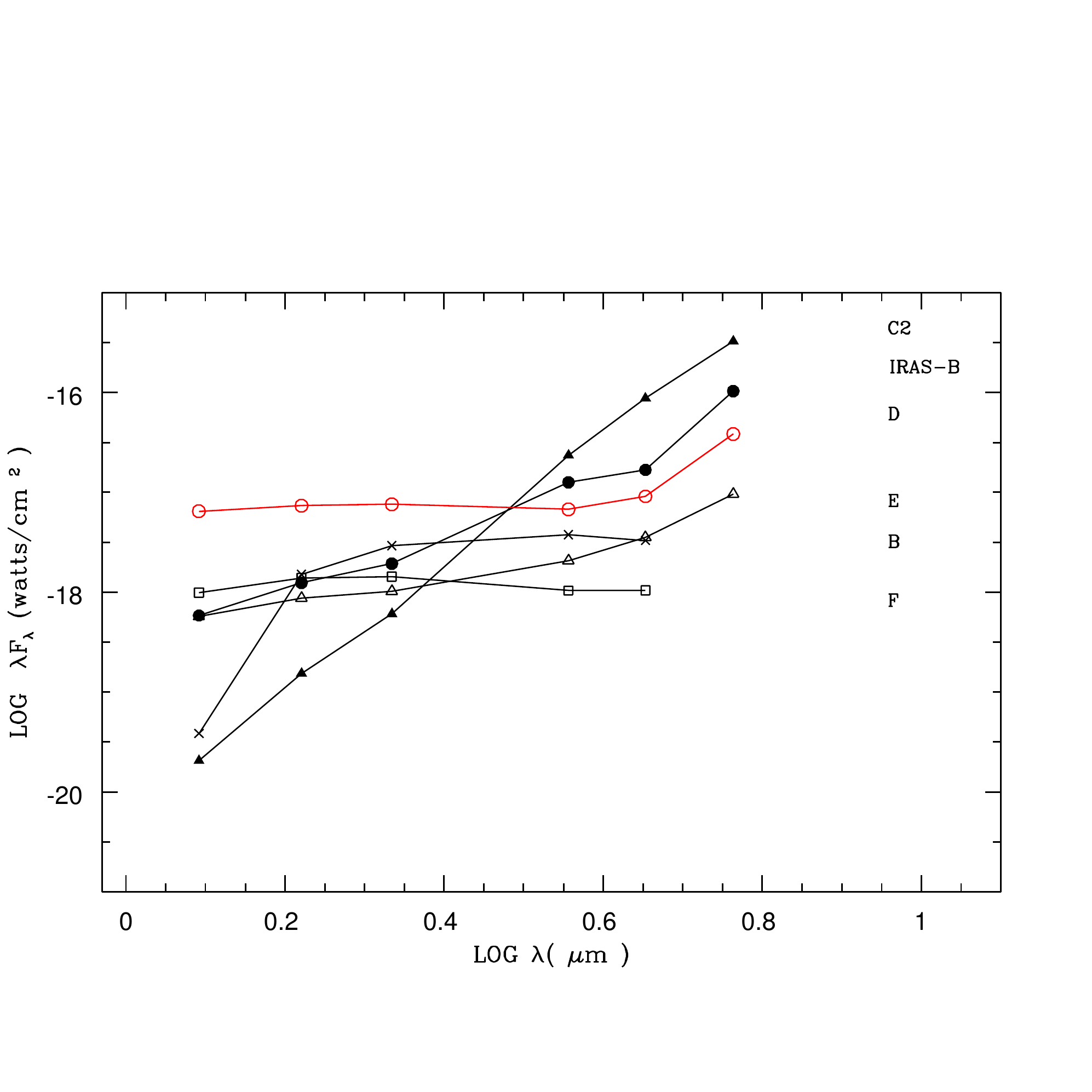}
      \caption{SEDS of bright NIR sources within
 individual \hii regions having IRAC counterparts.}         
 \label{fig19}
   \end{figure}

\begin{figure}
\centering
   \includegraphics[width=7 cm,height=8.5 cm,trim=150 100 170 120]{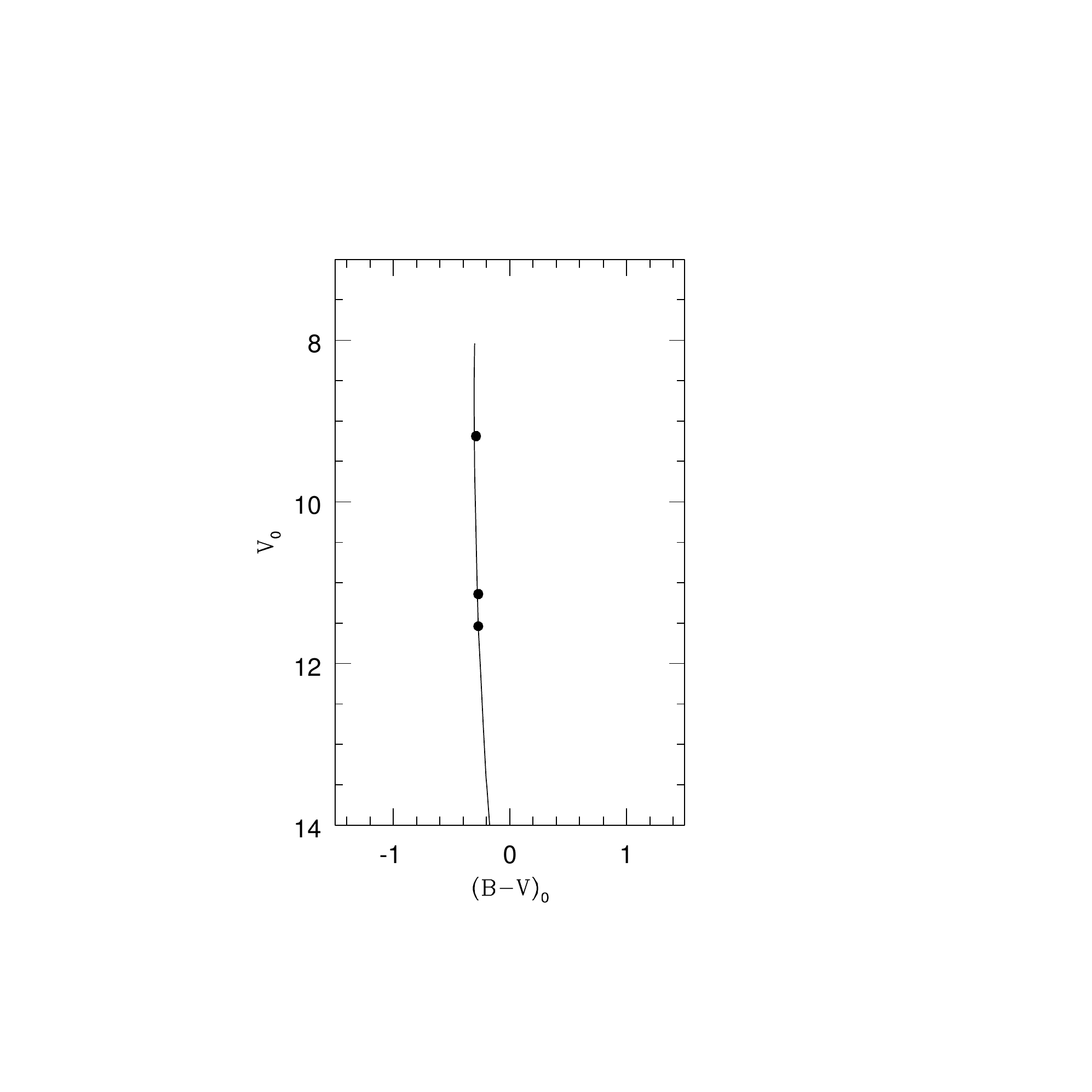}
      \caption{The intrinsic $V/B-V$ CM diagram for the ionizing stars
             of K3-50D, K3-50E and K3-50F regions. The solid line shows the theoretical isochrones
          taken from Girardi et al. (2002) for 1 Myr at a distance of 8.7 kpc.}
\label{fig20}
\end{figure}

\begin{figure}
   \centering
   \includegraphics[width=8 cm]{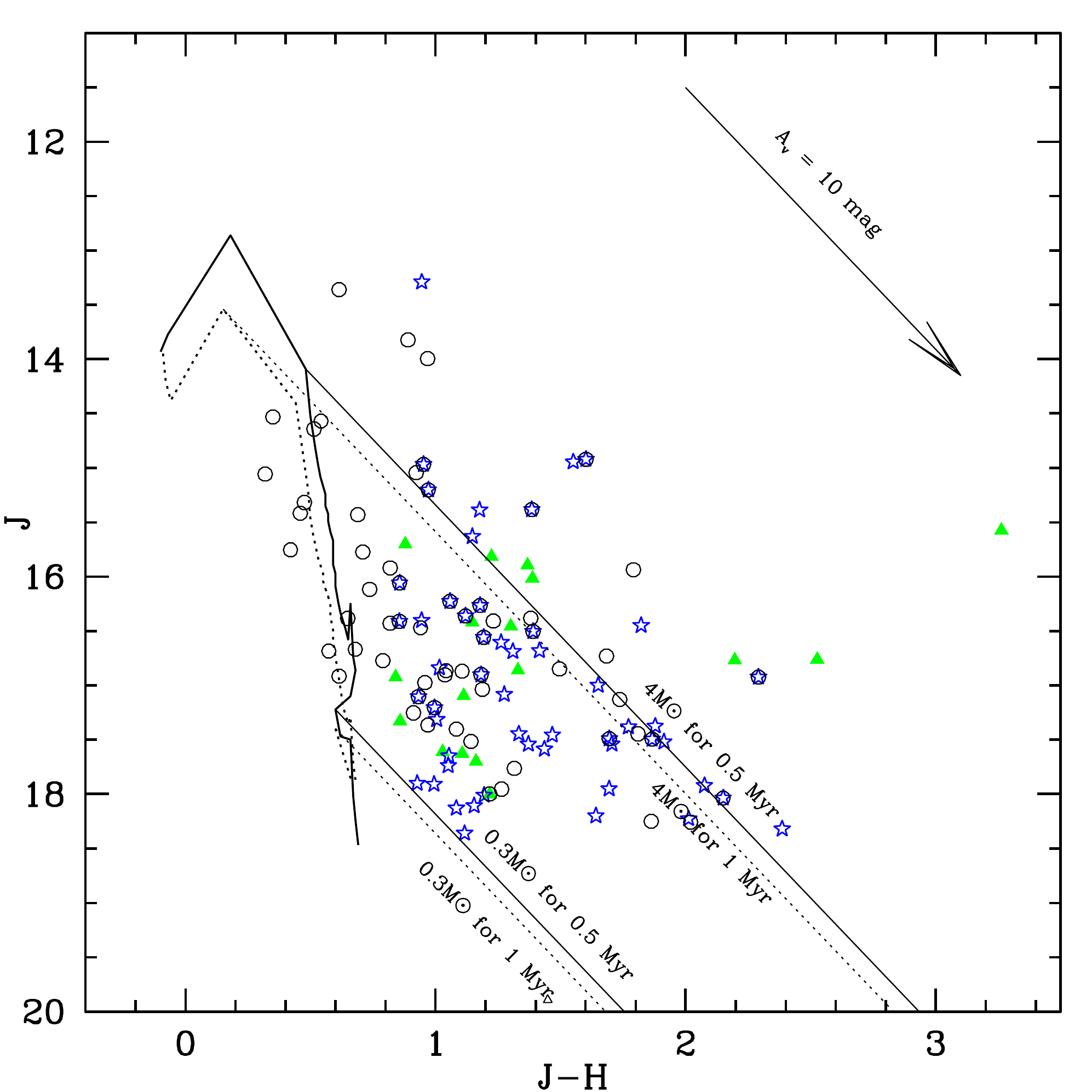}
      \caption{ CM diagram for YSO candidates in Sh 2-100 SFR.
The symbols are same as in Fig. 17.
The PMS isochrone of 0.5 Myr (solid curve) and 1 Myr (dotted curve)
from Siess et al. (2000) are drawn
 at a distance of  8.7 kpc and zero reddening.
The reddening vectors corresponding to 0.3$\msun$ and 4$\msun$,
are also shown in the figure. }
\label{fig21}
\end{figure}

\begin{figure*}
   \centering
  \includegraphics[width=12 cm]{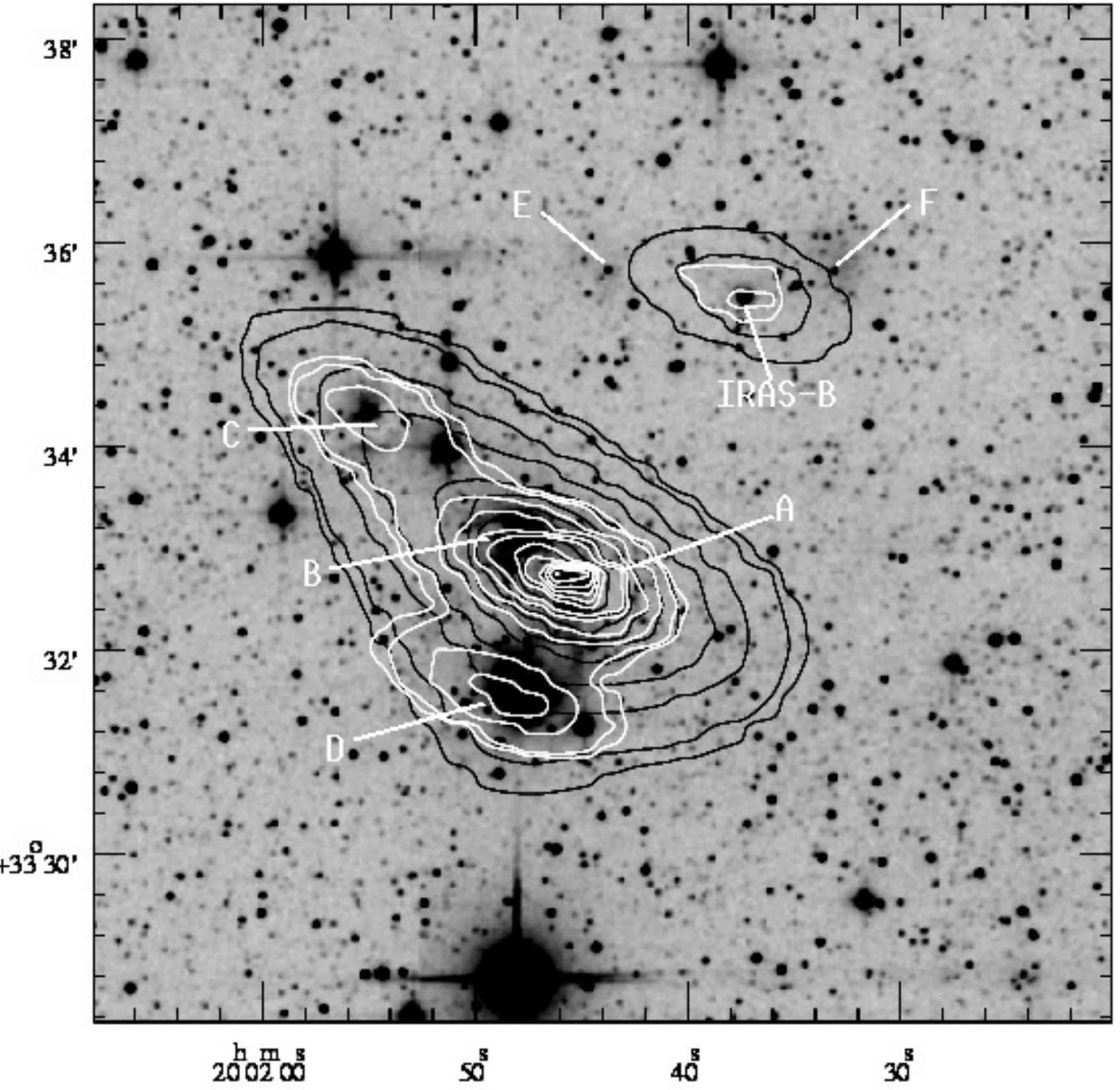}
      \caption{2MASS \ksb image overlaid with the warm dust contours at
 25 $\mu$m (black in color) and cold dust contours at 60 $\mu$m (white in color)
 taken from {\it HIRES} processed images of IRAS. The 25 $\mu$m contour levels are
at 1.5, 2, 3, 4, 5, 10, 20, 40, 60, 80, 90 \% of the peak value 43208 MJy sr$^{-1}$ 
and the 60 $\mu$m contour levels are at 1.5, 2, 3, 5, 7, 10, 15, 20, 25, 40, 60, 80, 90\% of the
peak value 44942 MJy sr$^{-1}$. Locations of the individual \hii regions are
also marked. The labelled axes are in J2000 coordinates.
}
\label{fig22}
\end{figure*}

\begin{figure*}
   \centering
   \includegraphics[width=10 cm]{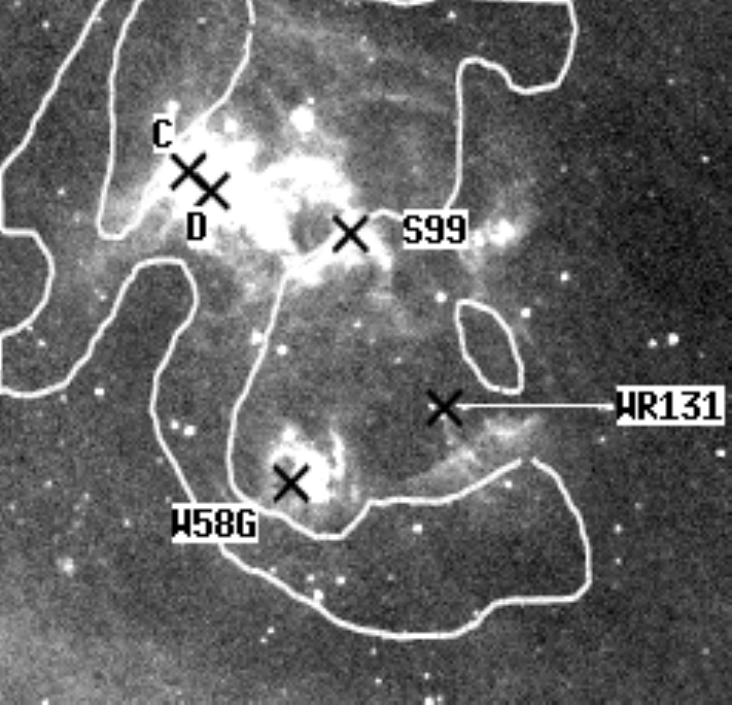}      
 \caption{MSX A-band image marked with area of \hi emission (thick solid line) 
taken from Israel (1980). The crosses
represent the positions of different sources in the W58 cloud complex (see the text).
North is up and east is to the left.}
\label{fig23}
\end{figure*}

\begin{figure*}
   \centering
   \includegraphics[width=6 cm, height=3 cm]{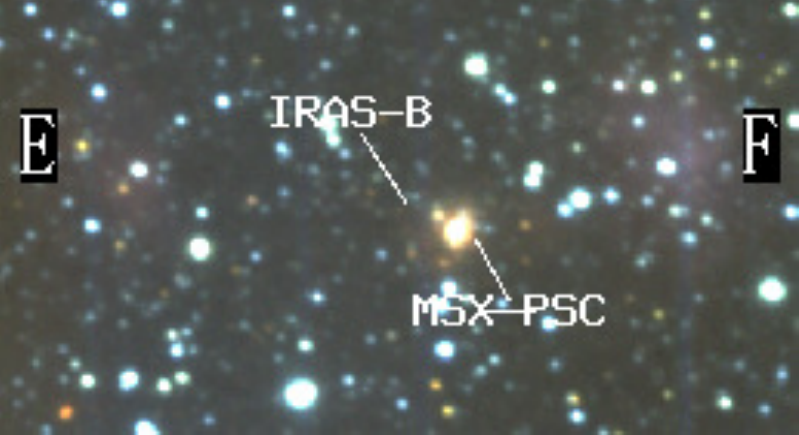}    \includegraphics[width=8 cm, height=6 cm]{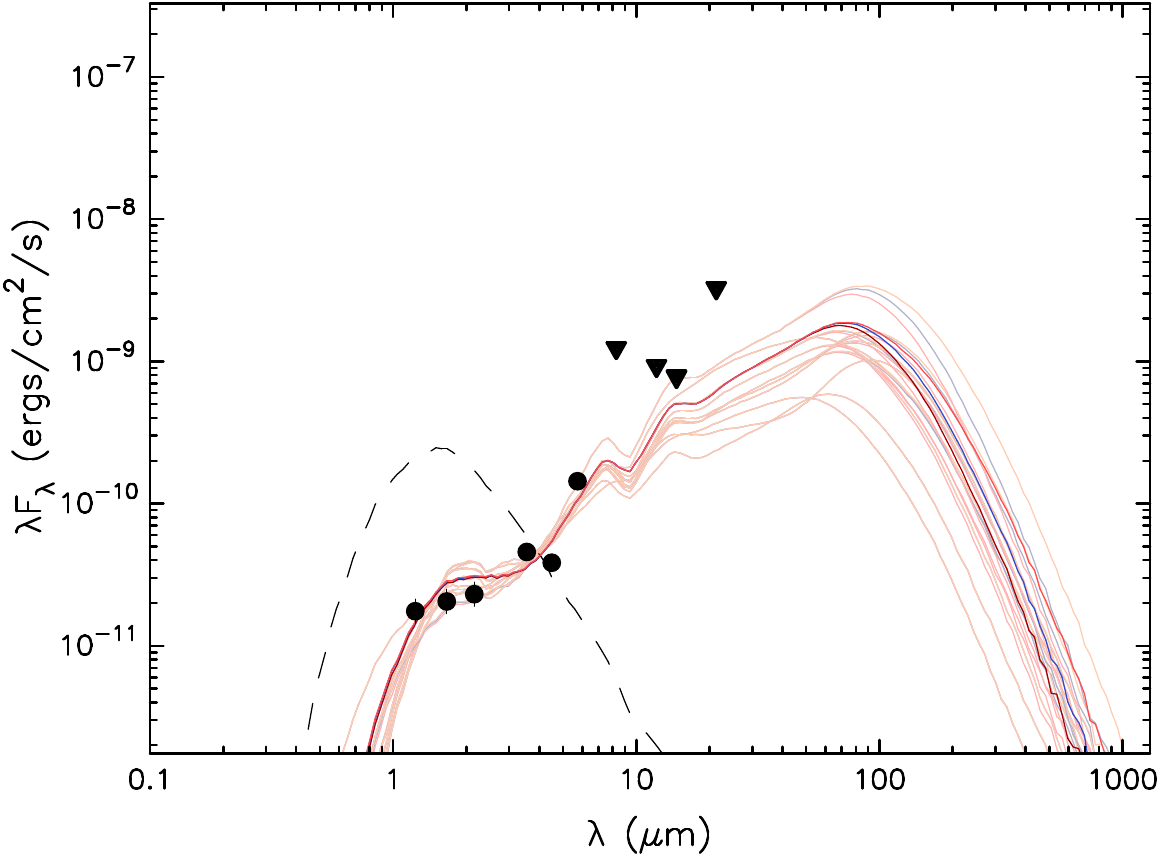}
      \caption{$Left$: $JHK$ color-composite image around IRAS-B region. $Right$: SED of the
luminous source associated with IRAS-B. The filled circles represent NIR and
{\it Spitzer} flux values, whereas the filled triangles denote upper limit of 
the MSX flux values.
The solid lines show all
models that fit the data reasonably with ${\chi}^2 - {\chi_{best}}^2 < 3$ .
 The dashed line shows the SED of the stellar photosphere in the best-fitting model. }
\label{fig24}
   \end{figure*}
\end{document}